%% file: aanda.tex
\documentclass{aa}

\usepackage[varg]{txfonts}
\usepackage{graphicx}
\usepackage[colorlinks=True, linkcolor=blue, citecolor=blue]{hyperref}

\def\paperversion{1}  
\input{header.tex}

\begin{document}

\title{The \noctua Suite of Simulations -- The Difficulty of Growing \\ Massive Black Holes in Low-Mass Dwarf Galaxies}
\subtitle{}

\titlerunning{The \noctua Suite of Simulations}
\authorrunning{J. Petersson et al.}

\author{Jonathan Petersson \orcid{0000-0001-7248-3898} \inst{1}\thanks{Email: jonathan.petersson@epfl.ch} \and
        Michaela Hirschmann \inst{1} \and
        Robin G. Tress \orcid{0000-0002-9483-7164}\inst{1} \and
        Marion Farcy \inst{1} \and
        Simon C. O. Glover \inst{2} \\
        Ralf S. Klessen \orcid{0000-0002-0560-3172} \inst{2,3,4,5} \and
        Thorsten Naab \orcid{0000-0002-7314-2558} \inst{6} \and
        Christian Partmann \inst{7} \and
        David J. Whitworth \inst{8}
        }

\institute{\galspec \and
           \zah \and
           \iwr \and
           \cfa \and
           \radcliffe \and
           \mpia \and
           \cca \and
           \unam
           }

\date{Received January 1, 2000; accepted January 1, 2000}
\date{}

\abstract
{}
{We study the individual and cumulative impact of stellar feedback processes on massive black hole (MBH) growth in a simulated low-mass dwarf galaxy. Furthermore, we explore the influence of the MBH initial mass ($10^{3-6} \Msol$) on the gas accretion, and whether or not artificially induced gas inflows can `boost' further gas accretion onto the MBH.}
{A suite of high-resolution radiation-hydrodynamic simulations called \noctua is performed, using the \areponoctua numerical framework for BHs in galaxy simulations. The chemical evolution of the gas is explicitly modelled in a time-dependent non-equilibrium way. Two types of stellar feedback are considered: individually-traced type II supernova (SNII) explosions, and radiatively transferred (on-the-fly) ionising stellar radiation (ISR) from OB stars. As part of the numerical framework, we develop and apply a novel physically-motivated model for MBH gas accretion, taking into account the angular momentum of the gas in the radiatively efficient regime, to estimate the gas accretion rate from the sub-grid accretion disc.}
{Without any stellar feedback, an initial $10^4 \Msol$ MBH is able to steadily grow over time, roughly doubling its mass after 800~Myr. Surprisingly, the growth of the MBH is more than doubled when only ISR feedback is considered, compared to the no stellar feedback run. This is due to the star formation rate (SFR) being highly suppressed (to a similar level or slightly above that when SNII feedback is considered), enabling a higher cumulative net gas inflow onto the MBH from not only the cold neutral- and molecular medium phases, but also the unstable- and warm neutral medium phases. With SNII feedback included, the gas accretion onto the MBH is episodic over time, and is suppressed by more than an order of magnitude already during the first 150~Myr. When combining SNII with ISR feedback, the growth of the MBH remains suppressed due to SNII feedback, but to a lesser extent compared to the SNII-only feedback run, due to a slightly lower SFR, and hence a reduced number of SNII events.}
{We conclude that SNII feedback is a strong regulator and suppressor of MBH growth, and that only an initial $10^5 \Msol$ MBH is able to consistently accrete gas in the radiatively efficient regime (in the presence of SNII feedback), while an initial $10^3 \Msol$ MBH barely is able to accrete any gas after 800~Myr. Combined with the fact that artificially induced gas inflows are unable to `boost' further gas accretion onto the MBH (even for an initial $10^6 \Msol$ MBH), this suggests that it is primarily the nearby gravitational potential around the MBH that determines how much the MBH can grow via gas accretion over time (at least in an isolated non-cosmological environment).}


\maketitle



\section{Introduction}
\label{sect:introduction}

The stellar mass, luminosity, and velocity dispersion of galaxies' spheroidal component are repeatedly inferred to strongly correlate with the mass of their residing central black hole (BH). These scaling relations are observed for supermassive BHs (SMBH) of $M_\mathrm{BH} \gtrsim 10^6 \Msol$ \citep{Silk1998, Magorrian1998, Tremaine2002, King2003, Haring2004, Graham2007, Gultekin2009, McConnell2013}, but also recently in the high-end intermediate mass BH (IMBH) regime of $M_\mathrm{BH} \sim 10^{4-6} \Msol$ \citep[see e.g.][]{Reines2015, Schutte2019}. As these massive BHs (MBH) grow via gas accretion, substantial amounts of energy are expected to be released into the surrounding medium, enough to power active galactic nuclei (AGN) and (in extreme cases) luminous quasars \citep{Salpeter1964, LyndenBell1969}. This so-called AGN feedback (composed of radiation, winds and jets), is thought to be able to regulate both the gas inflow and the amount of star formation in host galaxies, and thus be responsible for their observed co-evolution with the central BH \citep[at least in the SMBH regime, see e.g.][]{Fabian2012, Kormendy2013} -- even though non-causal origins for scaling relations have been discussed as well \citep{Hirschmann10, Jahnke11}. 

Numerical cosmological simulations provide a robust theoretical framework to study MBH accretion and AGN feedback in a galaxy evolution context \citep[e.g.][]{DiMatteo2008, Hirschmann2014, Choi2015, Schaye2015, Sijacki2015, Somerville2015, Volonteri2016,  Naab2017, Nelson2019}, and are proven to be successful when it comes to, for example, recreating many of the observed local scaling relations. However, as recently illustrated by \citet{habouzit2022}, these large-scale cosmological simulations struggle to reach consensus in the low-mass galaxy regime at high redshift ($z \geq 4$), and instead predict a wide range of MBH and AGN properties at a fixed host galaxy stellar mass $M_\star$. This discrepancy is ultimately due to the various (often ad-hoc) sub-grid models implemented for MBH seeding, accretion and AGN feedback (as a consequence of the limited spatial resolution). Furthermore, the physical modelling of galaxies' internal processes (e.g.\ stellar feedback) is thought to be able to strongly affect the time evolution and shape of the $M_\mathrm{BH} - M_\star$ relation as well, especially in the low stellar mass range $M_\star \leq 10^{10.5} \Msol$ \citep{Habouzit2021}. With the increasing number of active MBH candidates at high redshift over the last decade \citep[see e.g.][]{Willott2009, Willott2010, Mortlock2009, Mortlock2011, Kim2015b, Jiang2016, Reed2017, Reed2019, Matsuoka2016, Matsuoka2018, Matsuoka2019, Andika2020, Yang2020, Wang2021, Gloudemans2022, Banados2023}, accelerated by the advent of the James Webb Space Telescope \citep[see e.g.][]{Harikane2023, Kocevski2023, Kokorev2023, Larsson2023, Onoue2023, Greene2024, Lin2024, Maiolino2024, Matthee2024, Taylor2024}, the need for detailed theoretical models of MBH growth, in particular via gas accretion, is more pressing than ever. 

In parallel, the number of lower-mass MBH candidates at $10^4 < M_\mathrm{BH} < 10^6 \Msol$ in (typically low-metallicity) dwarf galaxies at low to intermediate redshifts has steadily increased as well \citep[see e.g.][]{Moran2014, Pardo2016, Nguyen2019, Baldassare2017, Baldassare2018, Baldassare2020, Mezcua2020, Reines2011, Reines2013, Reines2020, Mezcua2018, Mezcua2019b, Mezcua2023, Ubler2023, Sacchi2024}, suggesting that the dwarf galaxy regime might serve as a plausible channel for early MBH growth, provided that MBHs in local dwarf galaxies are leftovers from high redshift \citep[which is still debated, see e.g.][]{Mezcua2019a}. However, to what extent lower-mass MBHs can grow in mass when residing in this type of environment is unclear at the moment. While some theoretical studies suggest that early MBH growth is largely suppressed by strong supernova feedback at high redshift \citep{Dubois2015, Habouzit2017, Prieto2017, McAlpine2018, Trebitsch2018}, others argue that low-mass MBHs indeed can grow at a high enough rate to even invoke AGN feedback \citep[][but typically at slightly lower redshifts]{Barai2019, Sharma2020, Sharma2022, Koudmani2019, Koudmani2021, Koudmani2022}. Despite the high spatial resolution achievable in some of these works, many of them still rely on the same (or slightly modified) physical models for BH physics and stellar feedback, employed in large-scale cosmological simulations (see above).

Accurate modelling of MBH gas accretion is vital, because apart from governing the growth of the MBH itself, it also essentially determines the strength of the AGN feedback. Most large-scale (or zoom-in) cosmological simulations adopt a Bondi-Hoyle-Lyttleton formalism \citep{Hoyle1939, Bondi1944, Bondi1952}, assuming a homogeneous spherically-symmetric steady gas accretion flow (of zero angular momentum). However, this is far from always applicable to a galactic environment, in which the interstellar medium (ISM) is expected to be turbulent and composed of multi-phase gas \citep[see e.g.][]{Bellovary2013, Gaspari2013, Klessen2016}. Additionally, the characteristic length scale of Bondi-Hoyle-Lyttleton accretion is often not even resolved in state-of-the-art cosmological simulations. To account for this, many numerical models introduce a constant boost factor $\alpha$ to `boost' the estimated accretion rate \citep[see e.g.][]{Springel2005a}, with the motivation that at large radii, the density and temperature are under- and overestimated respectively (due to an over-pressurised ISM), meaning that the Bondi-Hoyle-Lyttleton accretion rate is overall underestimated \citep[with exceptions, e.g.][]{Weinberger2018}. The value of $\alpha$ is however very uncertain, and can range over several orders of magnitude \citep{Booth2009}. There are also large uncertainties in the numerical implementation of Bondi-Hoyle-Lyttleton accretion. \citet{Negri2017} find, for example, that differences in the spatial resolution of the simulation, and how the variables needed to estimate the accretion rate are computed, can lead to both over- and underestimated MBH growth. That said, we note that many studies try to get around these shortcomings by either modifying already existing Bondi-Hoyle-Lyttleton accretion schemes \citep[see e.g.][]{RosasGuevara2015, Steinborn2015}, or by developing their own BH accretion models, for instance, by the use of gravitational torques \citep{AnglesAlcazar2015} or by utilising a Bayesian forward-modelling approach \citep{Weinberger2025}. 

\begin{figure*}
    \centering
    \includegraphics[trim={0 12.5cm 0 12.5cm}, clip, width=17cm]{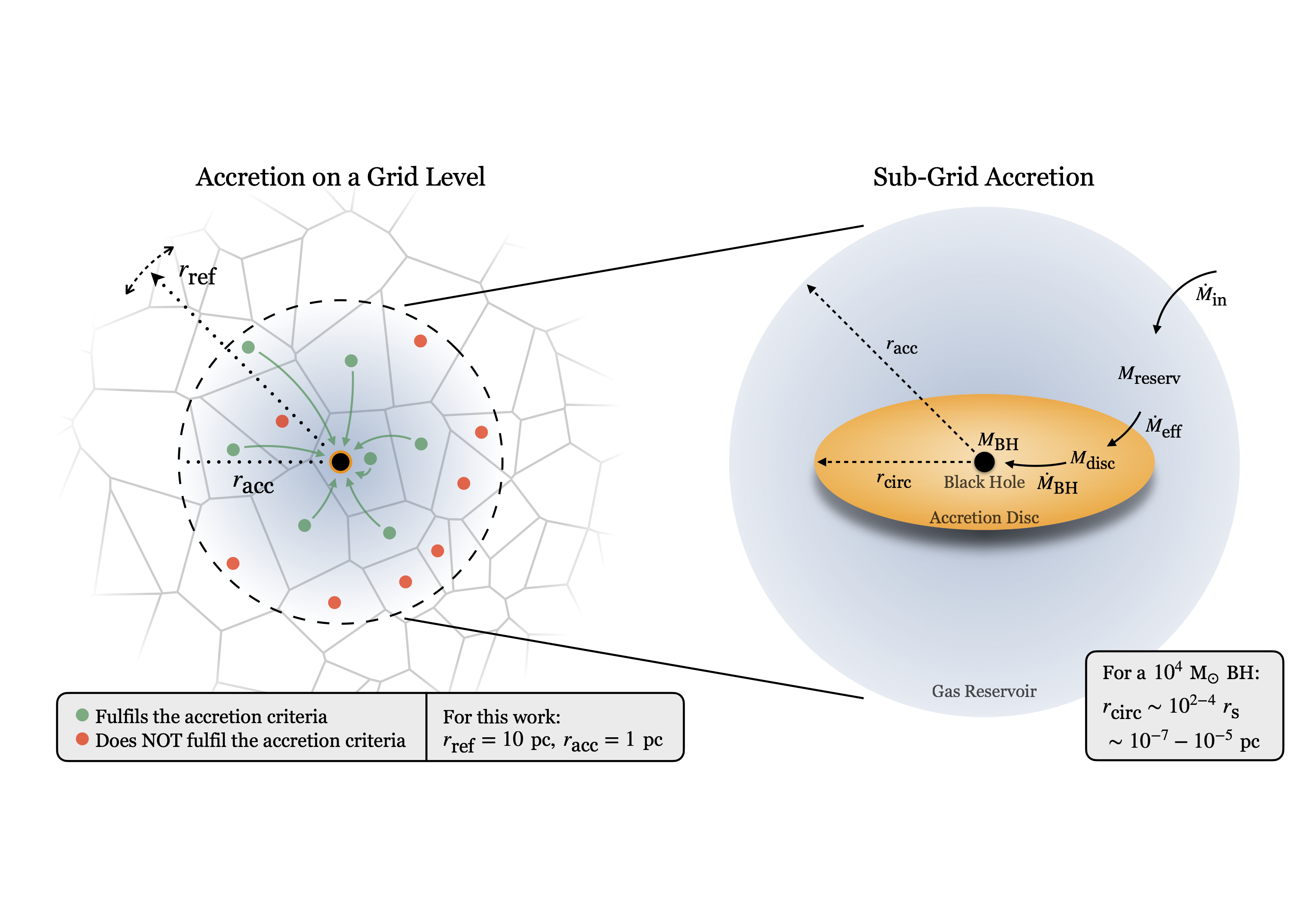}
    \caption{Sketch for how the MBH gas accretion model described in Sect.~\ref{sect:bh_accretion} behaves on a (sub-)grid level (with typical length scales adopted in this work provided). \textit{Left}: On a grid level, gas cells inside the refinement radius $r_\mathrm{ref}$ and the accretion radius $r_\mathrm{acc}$, are geometrically refined to ensure that the accretion region $r < r_\mathrm{acc}$ of the MBH is always resolved by at least a few tens of gas cells (see Sect.~\ref{sect:refinement} for more details). Whenever at least one active gas cell inside the accretion region fulfils the accretion criteria, the code first estimates the infalling gas accretion rate according to Eq.~(\ref{eq:infalling_gas_accretion_rate}), and then skim mass of individual gas cells following Eq.~(\ref{eq:skimming_mass_of_gas_cells}). \textit{Right}: If the infalling gas accretion rate $\dot{M}_\mathrm{in} > 0.02~\dot{M}_\mathrm{Edd}$ (or if $\dot{M}_\mathrm{BH}$ is already sufficiently high from the previous timestep), the skimmed gas is set to evolve on a sub-grid level. There, it is initially placed into a reservoir from which it is allowed to fall on a free-fall timescale down to the so-called circularisation radius $r_\mathrm{circ}$ (the radius at which an accretion disc is expected to form). Finally, the gas is accreted onto the MBH on a viscous timescale (see Eq.~\ref{eq:viscous_timescale}), assuming a geometrically thin, optically thick accretion disc. When $\dot{M}_\mathrm{in} < 0.02~\dot{M}_\mathrm{Edd}$, the accretion timescale is expected to be very short (compared to the viscous timescale), thus, no time-lag effects are considered. Instead, the skimmed gas is directly accreted onto the MBH (see Sect.~\ref{sect:bh_accretion_sub_grid} for more details).}
    \label{fig:accretion_model_sketch}
\end{figure*}

In order to more accurately model MBH gas accretion, while simultaneously modelling galaxies' internal processes in a physically-motivated way, a high spatial resolution is required. For this purpose, idealised galaxy simulations make a great case, thanks to the possibility to construct well-controlled initial conditions, and to run simulations at a relatively low computational cost. \citet{Sivasankaran2022} recently demonstrated using a set of idealised Milky Way- to Small Magellanic Cloud-like simulations, that MBH gas accretion can significantly alter and become highly variable over time when an explicit treatment of the ISM is considered \citep[compared to an effective equation of state, e.g.][]{Springel2003}. Likewise, idealised dwarf galaxy simulations have opened up new possibilities, allowing us to study in great detail the individual and cumulative impact of various physical processes important for galaxy evolution, and the influence they may have in shaping the overall global and local properties of dwarf galaxies \citep{Hu2016, Hu2017, Emerick2019, Agertz2020, Lahen2020, Gutcke2021, Smith2021, Whitworth2022, Whitworth2023, Andersson2023, Deng2024, Steinwandel2024}. Yet, so far only \citet{Partmann2025} have applied this type of numerical framework to study MBH accretion in the dwarf galaxy regime. In short, they find that stellar feedback can strongly suppress MBH gas accretion, and that it is only in the presence of a nuclear star cluster (NSC) that more gas is funnelled towards the central few parsecs, boosting both the gas accretion onto the MBH and the amount of star formation in the NSC. Similar to \citet{Sivasankaran2022}, \citet{Partmann2025} also find that MBH gas accretion (combined with nuclear star formation) is episodic over time, and that these episodes are strongly regulated by nearby supernova explosions. However, despite the inclusion of other stellar feedback processes in \citet{Partmann2025}, their separate impact on MBH growth is not assessed. Also, the interaction of ionising radiation from young massive stars with the ISM is not explicitly simulated, but instead modelled in an approximate way following a Str\"omgren sphere approach \citep[which can e.g.\ significantly reduce galactic outflows, see][]{Emerick2018}. Thus, the extent to which individual stellar feedback processes can regulate MBH gas accretion in the low-mass dwarf galaxy regime remains uncertain. 

To assess the regulating power of stellar feedback on MBH growth, we conduct a suite of high-resolution radiation-hydrodynamic simulations called \noctua\footnote{Named after the `little owl' -- \textit{Athene Noctua}, commonly adopted as a symbol for knowledge and wisdom, and where \textit{Noctua} is associated to the Latin word for `night'.}, where low-mass MBHs are allowed to accrete gas (without AGN feedback) in an isolated low-mass dwarf galaxy system, with and without different stellar feedback processes included. To achieve this, we introduce and apply the \areponoctua numerical framework for BHs in galaxy simulations, where the chemical evolution of the gas is explicitly modelled using a time-dependent non-equilibrium chemical network. A novel physically-motivated model for BH gas accretion is applied, thus avoiding the ambiguity of Bondi-Hoyle-Lyttleton accretion described above. For this study, two types of stellar feedback are considered: individually-traced type II supernova (SNII) explosions, and radiatively transferred (on-the-fly) ionising stellar radiation (ISR) from OB stars. By carefully analysing how the multi-phase ISM of the low-mass dwarf galaxy gradually changes when incorporating the different stellar feedback processes to the numerical set-up, the individual and cumulative impact of stellar feedback on MBH gas accretion is assessed in great detail. In a subset of simulations, we study how the gas accretion onto the MBH varies for different MBH initial masses, and whether or not an under- or overmassive BH is more or less favourable for sustained MBH growth (compared to our fiducial set-up). Lastly, we explore the effect of artificially induced gas inflows towards the galactic centre, mimicking strong gas inflows potentially happening in a cosmological setting and possibly at high redshift, and whether or not this can `boost' further gas accretion onto the MBH. 

The paper is structured as follows. Section~\ref{sect:methodology} describes the \areponoctua numerical framework used to conduct the \noctua suite of simulations. The main results of our simulations are presented in Sect.~\ref{sect:results}, while the results from the simulations of artificially induced gas inflows are instead introduced and discussed in Sect.~\ref{sect:discussion}. How our numerical models and results compare to previous studies are also discussed Sect.~\ref{sect:discussion}. Our main findings are summarised in Sect.~\ref{sect:conclusions}.

\section{Methodology}
\label{sect:methodology}

\begin{table*}
    \centering
    \caption{List of physical properties obtained for the low-mass dwarf galaxy system presented in Sect.~\ref{sect:initial_conditions}, with numerical details on the mass resolution and gravitational softening length of individual particles making up the different components. For the gas, an adaptive softening length $\epsilon_\mathrm{soft}$ that scales with the cell size is adopted, with a minimum value of 1.0 pc, while star particles formed during our simulations are given a fixed softening length of 2.0 pc instead.}
    \begin{tabular}{lccc}
        \hline
         & \textit{Dark Matter Halo} & \textit{Gaseous Disc} & \textit{Massive Black Hole} \\ \hline\hline
        \textit{Physical Properties} & & & \\
        Total mass [$\mathrm{M}_\odot$] &  $2.0 \times 10^{10}$ & $5.3 \times 10^7$ & $10^{3-6}$ \\
        Scale length [kpc] & 7.63 & 0.83 & -- \\
        Scale height [pc] & -- & 330 & -- \\ \hline
        \textit{Numerical Details} & & & \\
        Resolution [$\mathrm{M}_\odot$] & 500 & 20 & -- \\
        Softening length [pc] & 64 & $\mathrm{max}(\epsilon_\mathrm{soft},\ 1.0)$ & 0.2 \\
        Accretion radius [pc] & -- & -- & 1.0 \\ \hline 
    \end{tabular}
    \label{tab:initial_conditions}
\end{table*}

This section introduces the \areponoctua numerical framework for black holes (BH) in galaxy simulations. The hydrodynamics and physical framework for gas chemistry, star formation, and stellar feedback are all described in Sects.~\ref{sect:simulation_code}-\ref{sect:star_formation_and_stellar_feedback}. The implementation of BH gas accretion is presented in Sect.~\ref{sect:bh_accretion}, while the construction of initial conditions for the low-mass dwarf galaxy system is described in Sect.~\ref{sect:initial_conditions}. The full suite of simulations making up \noctua is presented in Sect.~\ref{sect:suite_simulations}.

\subsection{Simulation code}
\label{sect:simulation_code}

The simulations are performed using the moving-mesh hydrodynamic code \arepo \citep{Springel2010}, in which the conservation laws of hydrodynamics (assuming an ideal gas equation of state with an adiabatic index $\gamma=5/3$) are solved on a time-dependent mesh defined by the Voronoi tessellation of a set of discrete points. By allowing the mesh-generating points to move with the velocity of the local flow, a quasi-Lagrangian formulation is obtained, and the mesh is automatically and continuously adjusted in spatial resolution depending on the local density (i.e. cells will aim to retain a constant mass). The code also allows for adaptive mesh refinement, where cells are (de-)refined based on a set of (de-)refinement criteria (see Sect.~\ref{sect:refinement} for more details on the refinement strategy adopted in this work).

Dark matter (DM) and stars are treated as collisionless $N$-body particles and thus only interact with the gas (and each other) via gravity. The code is coupled to a $N$-body solver in which the gravitational field is calculated on a tree-based approach, utilising an adapted and improved version of the technique developed for \gadgettwo \citep{Springel2005b}.

\subsection{Chemical network}
\label{sect:chemical_network}

To model the chemistry and chemical evolution of the gas, the NL97 time-dependent non-equilibrium chemical network from \citet{Glover2012} is adopted \citep[originally implemented in \arepo by][]{Smith2014}. The network combines a treatment of hydrogen chemistry (including H$_{2}$ formation and destruction) based on the work by \citet{Glover2007a, Glover2007b}, with a simplified treatment of CO formation and destruction based on \citet{Nelson1997}. The network has been extensively applied to numerous previous studies in a wide range of physical set-ups, such as molecular clouds in a Galactic centre environment \citep{Bertram2015}, spiral arms of a Milky Way-like galaxy \citep{Smith2014}, giant star-forming molecular clouds in a M51-like galaxy \citep{Tress2020a}, as well as the entire central molecular zone \citep{Sormani2020, Tress2024}. Hence, only a short summary of the chemical network is given below. For more details, we refer to the previous studies already mentioned, and to the detailed description of the network in Section~3.4 of \citet{Sormani2018}. 

Changes in the chemical composition are tightly coupled to the heating and cooling of the gas. To compute the reaction terms of chemical species and the radiative- and chemical heating and cooling terms, some key quantities for the gas are initially required to be estimated. First, the dust grain number density is calculated assuming a constant dust-to-gas ratio of $10^{-3}$ \citep[equivalent to 10 \% of solar metallicity gas, see e.g.][]{Katia2001}. The dust grain temperature is determined via the same procedure as in Appendix~A of \citet{Glover2012}. Second, the intensity of the interstellar radiation field is calculated by first allowing it to be spatially constant at 10 \% of the solar neighbourhood value \citep{Draine1978}, and then attenuated via dust extinction and H$_2$ and CO self-shielding \citep[using the \treecol algorithm developed by][]{Clark2012}. Last, the intensity of the cosmic-ray ionisation rate is determined by assuming it to be spatially constant (zero attenuation) at $\zeta_\mathrm{H} = 3\times 10^{-18} \ \mathrm{s}^{-1}$ \citep[corresponding to 10 \% of the solar neighbourhood value inferred by][]{Goldsmith1978}. On this basis, the H and CO chemistry of each gas cell is calculated in a time-dependent non-equilibrium way. The associated heating and cooling rates are simultaneously and self-consistently inferred for $T < 10^4 \K$ gas. For higher temperatures, the contribution of H excitation cooling (`Lyman-$\alpha$ cooling') is computed using the non-equilibrium HI abundance provided by the chemical network, whereas for the contributions made by the permitted lines of He and metals, we instead use tabulated cooling rates from \citet{Gnat2012}, computed assuming collisional ionisation equilibrium. We note that in the current implementation of the chemical network, a temperature floor of 20 K is imposed. This is to avoid numerical instabilities occasionally arising when gas cells close to the resolution limit are undergoing strong adiabatic cooling. For the simulations of this study (with a target mass resolution of $20 \Msol$ per gas cell), this is equivalent to a Jeans length \citep{Jeans1902} of $\sim 0.3 \pc$. Since this is comparable or below the maximum spatial resolution of the gas (see Fig.~\ref{fig:resolution} in Appendix~\ref{app:resolution}), we do not expect this temperature floor to have any strong influence on the overall results of our simulations.

\subsection{Star formation \& stellar feedback}
\label{sect:star_formation_and_stellar_feedback}

\begin{table*}
    \centering
    \caption{The \noctua suite of simulations. The simulations are divided into different subsets depending on the scientific question they aim to address (described in more detail in Sect.~\ref{sect:suite_simulations}): `\textit{Stellar Feedback}' for when gradually adding stellar feedback processes to the numerical set-up; `\textit{Black Hole Mass}' for when varying the MBH initial mass between $10^{3-6} \Msol$; `\textit{Artificial Inflows}' for when manually reducing the angular momentum of the gas in the low-mass dwarf galaxy by $L_\mathrm{reduce}$, to induce artificial inflows towards the galactic centre and the central MBH (see Sect.~\ref{sect:art_inflows} for more details on this).}
    \begin{tabular}{l|ccccc}
        \hline Simulation & Run Time & $M_\mathrm{BH}$ [$\mathrm{M}_\odot$] & SNII & ISR & $L_\mathrm{reduce}$ [\%] \\ \hline\hline
        \textit{Stellar Feedback} & & & & & \\
        \texttt{MBH4\_noFB} & 800 Myr & $10^4$ & --           & --           & -- \\
        \texttt{MBH4\_SNII} & 800 Myr & $10^4$ & $\checkmark$ & --            & -- \\
        \texttt{MBH4\_ISR}  & 800 Myr & $10^4$ & --           & $\checkmark$ & -- \\
        \texttt{MBH4\_full} & 800 Myr & $10^4$ & $\checkmark$ & $\checkmark$  & -- \\ \hline
        \textit{Black Hole Mass} & & & & & \\
        \texttt{MBH3} & 800 Myr & $10^3$ & $\checkmark$ & -- & -- \\
        \texttt{MBH5} & 800 Myr & $10^5$ & $\checkmark$ & -- & -- \\
        \texttt{MBH6} & 300 Myr & $10^6$ & $\checkmark$ & -- & -- \\ \hline
        \textit{Artificial Inflows} & & & & & \\
        \texttt{MBH4\_L20} & 300 Myr & $10^4$ & $\checkmark$ & -- & 20 \\
        \texttt{MBH4\_L50} & 300 Myr & $10^4$ & $\checkmark$ & -- & 50 \\
        \texttt{MBH6\_L20} & 300 Myr & $10^6$ & $\checkmark$ & -- & 20 \\ \hline
    \end{tabular}
    \label{tab:simulations}
\end{table*}

Star formation is modelled on a sub-grid level following the star particle approach by \citet{Goller2025}. To determine whether an active gas cell is star-forming or not, the local Jeans mass $M_\mathrm{J}$ of the cell is calculated as follows
\begin{equation}
    M_\mathrm{J} = \left(\frac{3}{4\pi \rho}\right)^{1/2} \left(\frac{5e_\mathrm{in}(\gamma - 1)}{G}\right)^{3/2}, 
\end{equation}
\noindent where $\rho$ and $e_\mathrm{in}$ are the mass density and specific internal energy respectively of the gas cell, and $G$ is the gravitational constant. If the mass of the cell $M_\mathrm{cell} > M_\mathrm{J}/N$, where $N=8$ \citep[following the criterion by][]{Truelove1997}, the cell is flagged as `possibly star-forming'. The probability $p$ for a `possibly star-forming' gas cell to form a star particle is given by the following expression
\begin{equation}
    p = \left(1 - e^{-\lambda}\right), \quad \mathrm{with} \quad \lambda = \mathrm{SFR} \frac{\Delta t}{M_\mathrm{cell}}, 
    \label{eq:probability_sf}
\end{equation}
\noindent where $M_\mathrm{cell}$ is the mass of the cell, $\Delta t$ is the current timestep of the simulation, and SFR is the local star formation rate in the cell, defined as
\begin{equation}
    \mathrm{SFR} = \epsilon_\mathrm{ff}\frac{M_\mathrm{cell}}{t_\mathrm{ff}}, \quad \mathrm{with} \quad t_\mathrm{ff} = \sqrt{\frac{3\pi}{32G\rho}}, 
\end{equation}
\noindent in which the local star formation efficiency per free-fall time $\epsilon_\mathrm{ff} = 0.01$. At star formation, star particles are populated by individual massive stars ($8 < M_\star < 120 \Msol$) sampled from the high-mass end of the Kroupa initial mass function \citep[IMF, ][]{Kroupa2001}, following the technique of \citet{Sormani2017}. Consequently, star particles do not represent individual stars, but instead small stellar clusters (who on average fully sample the IMF) that evolve and interact with the ISM via stellar feedback. For this study, two types of stellar feedback are considered: type II supernovae (SNII) and ionising stellar radiation (ISR), each described separately below. Stellar mass loss and metal injection are not considered.

\subsubsection{Type II supernova feedback}
\label{sect:supernova_feedback}

Whenever a sampled massive star ($>8 \Msol$) reaches the end of its lifetime \citep[inferred from Table 25.6 in][]{Maeder2009}, a SNII explodes at the location of its host star particle, injecting energy into the surrounding medium following the prescription of \citet{Tress2020a}. In short, the injection region of a SNII is determined by the radius $R_\mathrm{inj}$, which is defined as the smallest possible radius of a sphere containing 40 gas cells around the host particle. Based on the mean density of gas cells within the injection region, the radius $R_\mathrm{ST}$ of a SN remnant at the end of its Sedov-Taylor phase is calculated (assuming a SN energy of $10^{51} \erg$). If the Sedov-Taylor phase of the SN remnant is resolved, i.e. if $R_\mathrm{ST} > R_\mathrm{inj}$, a thermal energy of $10^{51} \erg$ is isotropically injected into the region ($R<R_\mathrm{inj}$), and all gas is fully ionised. On the other hand, if the Sedov-Taylor phase is unresolved, i.e. if $R_\mathrm{ST} < R_\mathrm{inj}$, direct injection of thermal energy is unreliable due to numerical `overcooling', making it impossible to generate a strong shock and deposit the correct amount of kinetic energy into the ISM. Instead, the SNII is modelled in a momentum conserving way, where the interior of the injection region is heated to $10^4 \K$ and momentum is injected following the analytical solution of the so-called snow-plow phase \citep[see e.g.][]{Gatto2015, Martizzi2015}. Due to the relatively small amount of stellar mass in our low-mass dwarf galaxy, and hence a low expected rate of type Ia supernovae \citep[SNIa, see e.g.][]{Maoz2012}, no SNIa feedback is included for this study.

\subsubsection{Ionising stellar radiation feedback}
\label{sect:ionising_stellar_radiation}

Star particles hosting at least one sampled O/B star are treated as ionising sources. The radiated photons are advected through the simulation domain by solving the radiative transfer (RT) equation on-the-fly using the \sweep code \citep{Peter2023}. The number of ionising photons injected per unit time into the domain by each star particle, is determined by the sum of ionising photon rates for every O and B star assigned to each particle. The rates of ionising photons for individually sampled OB stars are calculated assuming black body radiation, using the surface temperatures and stellar radii (at their main sequence phase) of \citet{Ekstrom2012}. A single frequency bin is adopted for energies above 13.6~eV. The advected photons are passed as an input to the thermochemistry routine, in which they can ionise H and heat the gas. With the current implementation, no radiation pressure is considered.

\subsection{Black hole gas accretion}
\label{sect:bh_accretion}

\begin{figure*}
    \centering
    \includegraphics[trim={0 3.5cm 0 3.5cm}, clip, width=17cm]{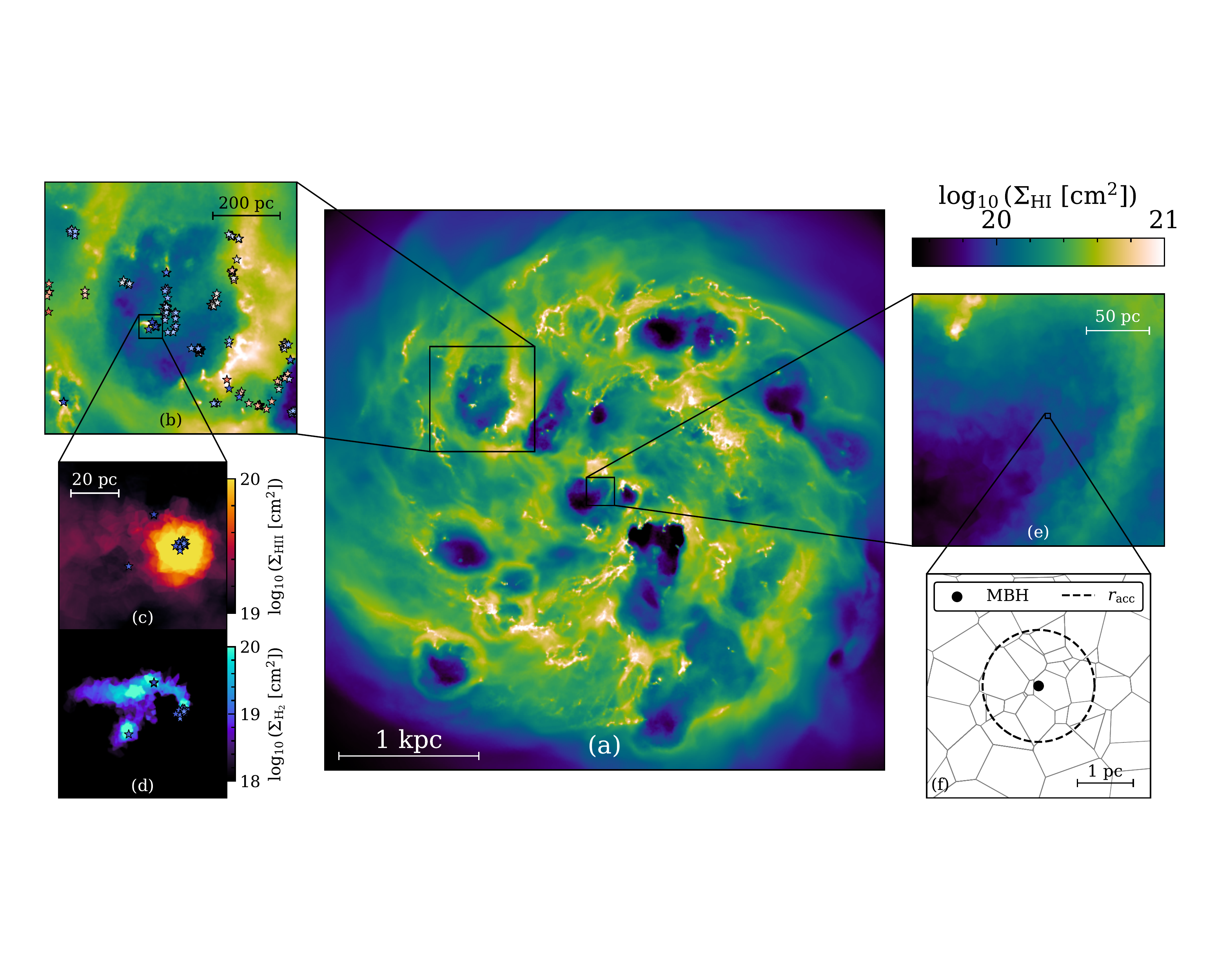}
    \caption{Visual demonstration of the complete physical framework described in Sect.~\ref{sect:methodology}. \textit{Panel (a)}: Face-on projected column density map of the HI surface density in the low-mass dwarf galaxy at $t \sim 500 \Myr$ in the \texttt{MBH4\_full} simulation (see Table~\ref{tab:simulations}). \textit{Panel (b)}: Zoom-in on a recently active star-forming region where SNII explosions have injected energy into the surrounding gas, pushing it in all directions shaping a `bubble' in the ISM. Star particles formed during the simulation are indicated and colour-coded according to their stellar age (normalised between 0 and 100~Myr, where blue represents the youngest and red the oldest stellar populations respectively). \textit{Panel (c--d)}: Further zoom-in on a group of young star particles from panel (b). By displaying face-on projected column density maps of the HII and H$_2$ surface densities, side-by-side, we can distinguish the molecular cloud from which the star particles were born from (panel~d), and how the ionising stellar radiation (modelled using radiative transfer on-the-fly) from individually sampled O and B stars propagates and ionises the surrounding gas (panel~c). \textit{Panel (e)}: Zoom-in on the region close to the central MBH. \textit{Panel (f):} Further zoom-in on the MBH and its accretion radius $r_\mathrm{acc}$ (black dashed circle), with a slice (aligned in the $z$-direction with the MBH) through the gas displaying its Voronoi structure. At this scale, individual gas cells can be seen inside the accretion region ($r<r_\mathrm{acc}$), highlighting the refinement strategy outlined in Sect.~\ref{sect:refinement}.}
    \label{fig:poster}
\end{figure*}

The central massive BH (MBH) is treated as a collisionless sink particle, i.e.\ beyond gravitationally interacting with the DM, stars, and gas, it is also allowed to skim mass from nearby gas cells. With the high gas resolution achievable in the simulations of this study (see Table~\ref{tab:initial_conditions}), reaching sub-parsec maximum spatial resolution (enough to resolve the Bondi radius of cold and warm gas around a $10^4 \Msol$ MBH, see Fig.~\ref{fig:resolution} in Appendix~\ref{app:resolution}), the gas accretion onto the MBH is modelled in a physically-motivated mass-flux approach, instead of the commonly adopted analytical solution of Bondi-Hoyle-Lyttleon accretion (which has several numerical uncertainties as described in Sect.~\ref{sect:introduction}). 

In Fig.~\ref{fig:accretion_model_sketch}, we illustrate how the gas accretion onto the MBH is modelled on a grid- and sub-grid level (left and right side of the figure respectively). For the MBH to skim mass on a grid level, individual gas cells are required to fulfil a set of accretion criteria:
\begin{enumerate}[(i)]
    \item The position $r$ of the gas cell is inside the accretion radius $r_\mathrm{acc}$ of the MBH.
    \item The gas cell is gravitationally bound to the MBH, i.e. $U > (E_\mathrm{k} + E_\mathrm{in})$, where $U = GM_\mathrm{BH}m_\mathrm{cell}/r$ is the cell's gravitational binding energy to the MBH (for a gas cell of $m_\mathrm{cell}$ in mass), $E_\mathrm{k} = m_\mathrm{cell}\Delta v^2 / 2$ is the kinetic energy associated to its motion (with respect to the MBH), and $E_\mathrm{in} = m_\mathrm{cell}e_\mathrm{in}$ is its internal energy.
    \item The flow of the gas represented by the cell is converging onto the MBH. This is achieved by ensuring that the divergence of the velocity and acceleration vector (with respect to the velocity and acceleration of the MBH) is negative (i.e. $\vec{\nabla} \cdot \vec{v} < 0$ and $\vec{\nabla} \cdot \vec{a} < 0$ respectively). 
\end{enumerate}
\noindent In the event of one or more gas cells fulfilling the accretion criteria (marked by green dots in Fig.~\ref{fig:accretion_model_sketch}), the infalling gas accretion rate onto the MBH is estimated as
\begin{equation}
    \dot{M}_\mathrm{in} = 4 \pi r_\mathrm{acc}^2 \langle \rho \rangle \langle v_r \rangle,
    \label{eq:infalling_gas_accretion_rate}
\end{equation}
\noindent where $\langle \rho \rangle$ and $\langle v_r \rangle$ is the non-weighted mean density and mass-weighted mean radial velocity, respectively, of only the gas cells fulfilling the accretion criteria. For the same set of gas cells, mass is then skimmed on a cell-by-cell basis (indicated by green arrows in Fig.~\ref{fig:accretion_model_sketch}) as follows
\begin{equation}
    \dot{M}_\text{cell} = \dot{M}_\text{in} f_\text{cell}, \quad \text{with} \quad f_\text{cell} = M_\text{cell} \left/ \sum\limits_{n=1}^N M_n \right., 
    \label{eq:skimming_mass_of_gas_cells}
\end{equation}
\noindent where $N$ is the total number of gas cells fulfilling the accretion criteria. The amount of mass skimmed from a gas cell is therefore proportional (but capped at 90 \%) to the cell's total initial mass. To ensure that the momentum of the system is conserved over time, skimmed gas cells are required to transfer an equivalent amount of momentum (with respect to the amount of mass skimmed) to the MBH. We note that when skimming mass of individual gas cells, strong variations in the gas density inside $r_\mathrm{acc}$ can occasionally arise and cause unwanted effects. To avoid numerical artifacts of this kind, we manually `smooth' the density of skimmed gas cells to their (non-weighted) average value after every accretion event.

\subsubsection{Sub-grid gas accretion model}
\label{sect:bh_accretion_sub_grid}

\begin{figure}
    \resizebox{\hsize}{!}{\includegraphics{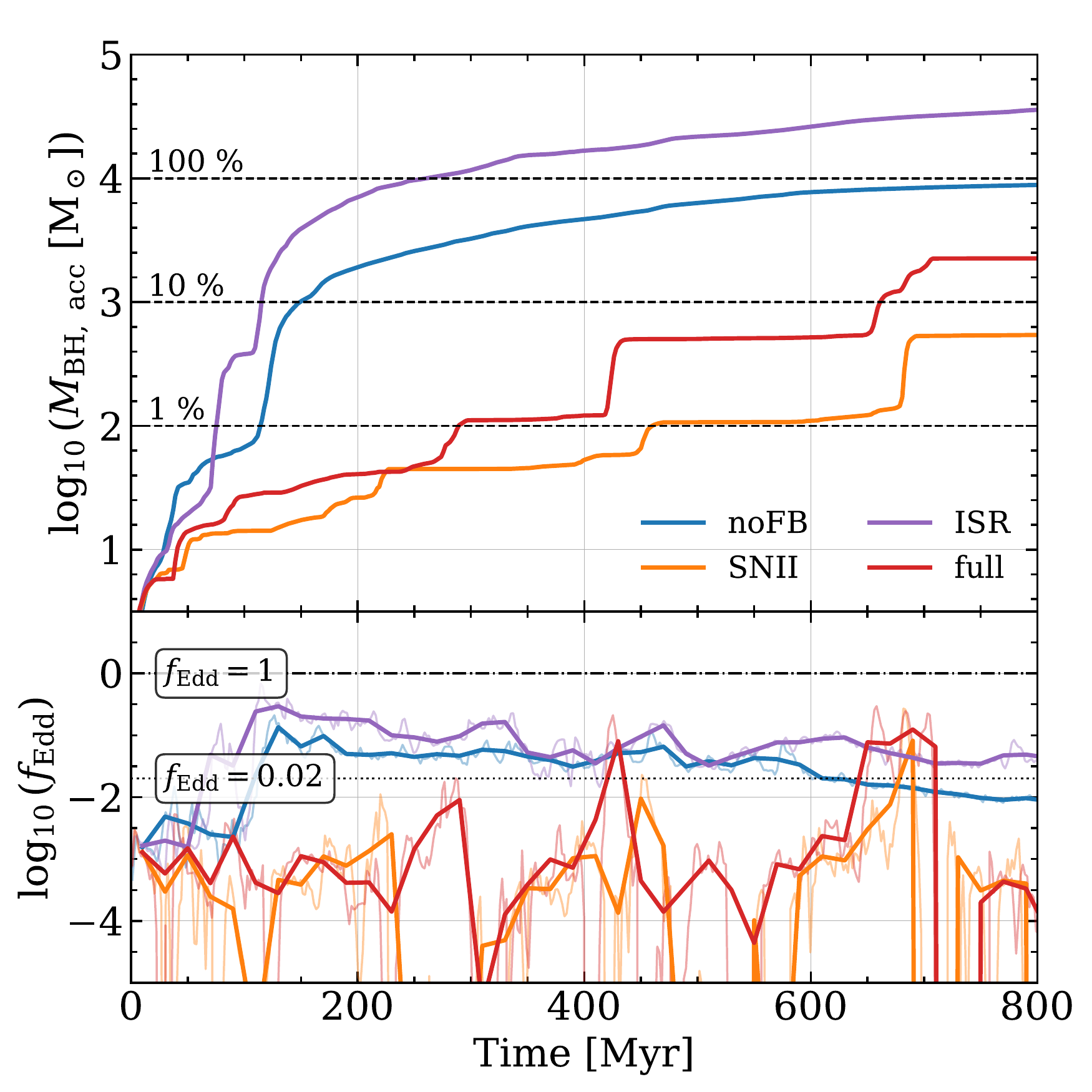}}
    \caption{\textit{Top}: Amount of mass accreted over time by the MBH in our `\textit{Stellar Feedback}' simulations (see Table~\ref{tab:simulations}). Dashed lines indicate the percentage growth of an initial $10^4 \Msol$ MBH. \textit{Bottom}: Corresponding time-evolution of the MBH gas accretion rate (in terms of the Eddington fraction $f_\mathrm{Edd} = \dot{M}_\mathrm{BH} / \dot{M}_\mathrm{Edd}$). Transparent background lines correspond to the obtained gas accretion rate of individual snapshots, while opaque lines mark the running mean value over 20 Myr.}
    \label{fig:bh_growth}
\end{figure}

Below the resolution limit, a sub-grid treatment for the gas accretion onto the MBH similar to \citet{Yuan2018} is adopted. For the radiatively efficient regime, i.e. when $\dot{M}_\mathrm{BH} > 0.02~\dot{M}_\mathrm{Edd}$\footnote{Inferred observationally from the transition between the hard and soft state in BH X-ray binaries \citep{McClintock2006}. In the equation for $\dot{M}_\mathrm{Edd} = (4\pi G M_\mathrm{BH} m_\mathrm{p})/(\epsilon_\mathrm{r}\sigma_\mathrm{T} c)$, where $m_\mathrm{p}$ is the proton mass, $\sigma_\mathrm{T}$ is the Thomson cross-section and $c$ the speed of light, we assume a radiative efficiency of $\epsilon_\mathrm{r} = 0.1$ \citep{Inayoshi2020}.}, a geometrically thin, optically thick accretion disc is expected to arise at the so-called circularisation radius $r_\mathrm{circ}$ of the infalling gas, and fuel the MBH via the loss of angular momentum in the gas due to viscous forces in the disc. To model the evolution of this, time-lag effects are introduced to delay the gas accretion from newly skimmed gas onto the MBH \citep[an approach previously adopted in numerous one- and two-dimensional simulations performed by e.g.][]{Ciotti2007, Ciotti2009, Novak2011, Gan2014, Ciotti2017}. 

Based on the infalling gas accretion rate at $r_\mathrm{acc}$ according to Eq.~(\ref{eq:infalling_gas_accretion_rate}), the time evolution of the effective gas accretion rate onto the accretion disc can be written as
\begin{equation}
    \begin{aligned}
         &\frac{\mathrm{d}\dot{M}_\mathrm{eff}}{\mathrm{d}t} = \frac{\dot{M}_\mathrm{in}(r_\mathrm{acc}) - \dot{M}_\mathrm{eff}}{\tau_\mathrm{ff}}, \quad \mathrm{where} \quad \tau_\mathrm{ff} = \frac{r_\mathrm{acc} - r_\mathrm{circ}}{v_\mathrm{ff}} \\ &\mathrm{and} \quad v_\mathrm{ff} = \left( \frac{2GM_\mathrm{BH}}{r_\mathrm{acc}}\right)^{1/2},
    \end{aligned}
    \label{eq:effective_accretion_rate}
\end{equation}
\noindent assuming a free-fall timescale\footnote{This indirectly assumes that all gas skimmed by the MBH has a relatively low specific angular momentum, which is most likely not always true. However, for the purpose of this paper, the free-fall timescale is a sufficient approximation. Trying to constrain the timescale(s) on which gas is transported from $r_\mathrm{acc}$ down to $r_\mathrm{circ}$ is beyond the scope of this paper, but presents a potential topic for future work.} for the gas transport from $r_\mathrm{acc}$ down to $r_\mathrm{circ}$ (each extent indicated by a black dashed line in the right hand sketch of Fig.~\ref{fig:accretion_model_sketch}). Gas that is not directly fuelled onto the accretion disc in a given timestep is placed into a `reservoir' ($M_\mathrm{reserv}$ in Fig.~\ref{fig:accretion_model_sketch}), from which the gas is allowed to continue to fuel the accretion disc on a free-fall timescale, but at later times as the simulation progresses. If the supply of infalling gas suddenly ends (temporarily or for a longer period of time), that is, when $\dot{M}_\mathrm{in}(r_\mathrm{acc}) = 0$, Eq.~(\ref{eq:effective_accretion_rate}) implies that the rate of which gas will continue to fuel the accretion disc (assuming there is already an existing reservoir of gas), will exponentially decrease over time, until the reservoir is depleted. 

The accretion disc ($M_\mathrm{disc}$ in Fig.~\ref{fig:accretion_model_sketch}) fuels the MBH on the instantaneous viscous timescale \citep{Kato2008}:
\begin{equation}
    \tau_\mathrm{vis} \approx 1.2 \times 10^6 \ \mathrm{yr} \ \left(\frac{\alpha}{0.1}\right)^{-1} \left(\frac{r_\mathrm{circ}}{100r_\mathrm{s}}\right)^{7/2} \left(\frac{M_\mathrm{BH}}{10^9 \ \mathrm{M}_\odot}\right), 
    \label{eq:viscous_timescale}
\end{equation}
\noindent assuming a geometrically thin, optically thick accretion disc, where $\alpha$ is the viscosity parameter \citep{Shakura1973} and $r_\mathrm{s}$ is the Schwarzschild radius ($r_\mathrm{s} \equiv 2GM_\mathrm{BH}/c^2$). The circularisation radius $r_\mathrm{circ}$ is essentially a free parameter which can be scaled for instance with the Schwarzschild radius of the MBH \citep{Yuan2018}. However, in this work we adopt a different approach, by estimating $r_\mathrm{circ}$ based on the angular momentum of the infalling gas (but without tracking the individual spin evolution of the MBH). For more details regarding this, see Appendix~\ref{app:rcirc}. The inflowing gas accretion rate onto the MBH from the accretion disc is estimated as
\begin{equation}
    \dot{M}_\mathrm{disc, \ inflow} = \frac{M_\mathrm{disc}}{\tau_\mathrm{vis}}, 
\end{equation}
\noindent where $M_\mathrm{disc}$ is the total mass of the accretion disc ($r < r_\mathrm{circ}$ in Fig.~\ref{fig:accretion_model_sketch}). The final MBH gas accretion rate follows as
\begin{equation}
    \dot{M}_\mathrm{BH} = \dot{M}_\mathrm{disc, \ inflow} - \dot{M}_\mathrm{w}, 
\end{equation}
\noindent where $\dot{M}_\mathrm{w}$ is the mass flux of winds expected to arise as the gas spirals inward towards the MBH. However, since feedback from the MBH is neglected in this study\footnote{Feedback from the MBH will instead be considered and studied in an upcoming dedicated paper, following the study of Farcy et al., (submitted).}, $\dot{M}_\mathrm{w}$ is set to zero, meaning that $\dot{M}_\mathrm{BH} = \dot{M}_\mathrm{disc, \ inflow}$. 

For the radiatively inefficient regime, that is, when $\dot{M}_\mathrm{BH} < 0.02~\dot{M}_\mathrm{Edd}$, there is no well-defined circularisation radius \citep{Bu2014} for the infalling gas, and the geometrically thin disc approximation mentioned above is therefore no longer valid. Instead, the gas accretion flow is expected to be geometrically thick, optically thin, and advection-dominated. Additionally, the accretion timescale is expected to be very short compared to the viscous timescale in the radiatively efficient regime \citep{Narayan1995}. Time-lag effects are therefore neglected, and the newly skimmed gas is instead directly accreted onto the MBH. We note that it is both the infalling gas accretion rate $\dot{M}_\mathrm{in}(r_\mathrm{acc})$ and the instantaneous $\dot{M}_\mathrm{BH}$ from the previous timestep that determines which regime the gas accretion onto the MBH should be modelled according to, in other words, if $\mathrm{max(\dot{M}_\mathrm{in}, \ \dot{M}_\mathrm{BH})} > 0.02~\dot{M}_\mathrm{Edd}$, newly skimmed gas is set to follow the radiately efficient regime's sub-grid model (described above), otherwise, it follows the radiately inefficient regime's direct accretion approach. This ensures that the gas accretion from an already massive accretion disc can keep the MBH in the radiatively efficient regime (regardless of $\dot{M}_\mathrm{in}$).

\subsubsection{Refinement strategy}
\label{sect:refinement}

\begin{figure*}
    \centering
    \includegraphics[width=17cm]{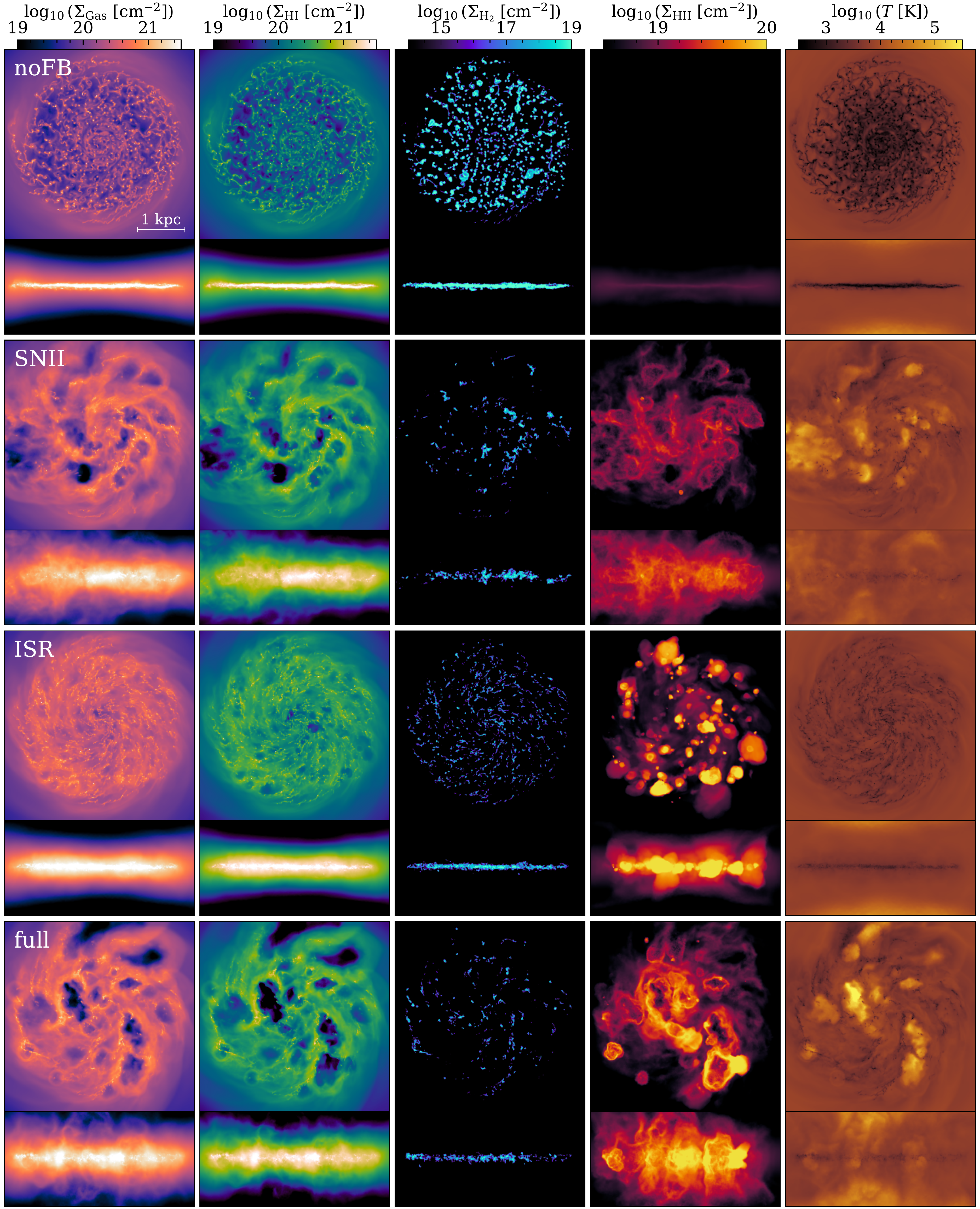}
    \caption{Collection of projected column density maps for the total gas (1st column), HI (2nd column), H$_2$ (3rd column) and HII (4th column) surface densities, as well as density-weighted gas temperature maps (5th column), of the low-mass dwarf galaxy at $t\sim 380 \Myr$ in our `\textit{Stellar Feedback}' simulations (see Table~\ref{tab:simulations}). For every map, both the face-on (top) and edge-on (bottom) projected view of the galaxy are included.}
    \label{fig:comparison}
\end{figure*}

\begin{table*}
    \centering
    \caption{Definitions of ISM phases adopted from \citet{Kim2023}, and their respective gas temperature and hydrogen abundance ranges. The hydrogen abundances are obtained following the relation: $\chi_\mathrm{HI} = 1 - 2\chi_{\mathrm{H}_2} - \chi_\mathrm{HII}$, meaning that $\chi_{\mathrm{H}_2} = 0.5$ for gas cells where all of the hydrogen is in molecular form.}
    \begin{tabular}{lccc}
        \hline
        ISM Phase Definition & Abbreviation & Temperature [K] & Hydrogen Abundance \\ \hline\hline
        Hot Ionised Medium & HIM & $T > 5 \times 10^5$ & -- \\
        Warm-Hot Ionised Medium & WHIM & $3.5 \times 10^4 < T < 5.0 \times 10^5$ & -- \\
        Warm Collisionally Ionised Medium & WCIM & $1.5 \times 10^4 < T < 3.5 \times 10^4$ & $\chi_\mathrm{HII} > 0.5$ \\
        Warm Photoionised Medium & WPIM & $6.0 \times 10^3 < T < 1.5 \times 10^4$ & $\chi_\mathrm{HII} > 0.5$ \\
        Warm Neutral Medium & WNM & $6.0 \times 10^3 < T < 3.5 \times 10^4$ & $\chi_\mathrm{HI} > 0.5$ \\
        Unstable Ionised Medium & UIM & $T < 6.0 \times 10^3$ & $\chi_\mathrm{HII} > 0.5$ \\
        Unstable Neutral Medium & UNM & $ 5.0 \times 10^2 < T < 6.0 \times 10^3$ & $\chi_\mathrm{HI} > 0.5$ \\
        Cold Neutral Medium & CNM & $ T < 5.0 \times 10^2$ & $\chi_\mathrm{HI} > 0.5$ \\
        Cold Molecular Medium & UMM & $ T < 6.0 \times 10^3$ & $\chi_{\mathrm{H}_2} > 0.25$ \\ \hline
    \end{tabular}
    \label{tab:ism_phase_definitions}
\end{table*}

Based on the standard (de-)refinement criteria in \arepo, the code will at any given time try to keep the mass of all individual gas cells as close as possible (within a factor two) to the designated target mass resolution ($20 \Msol$ in our case) of the simulation. On top of this, we employ two additional geometrical refinement criteria around the central MBH. By applying a first layer of maximum volume limit on the gas cells inside the refinement radius $r_\mathrm{ref} = 10 \pc$ (see left side of Fig.~\ref{fig:accretion_model_sketch}), we ensure that gas cells close to the MBH are comparable in size (or smaller) to the size of the accretion region ($r_\mathrm{acc} = 1 \pc$). We then apply a second layer of maximum volume limit on the gas cells inside $r_\mathrm{acc}$, to ensure that the accretion region ($r < r_\mathrm{acc}$) is always resolved by at least a few tens of gas cells. By gradually refining the gas cells as they get close to the MBH via these two layers of geometrical refinement, we make sure that gas cells will continue to enter the accretion region as the simulation progresses. Additionally, the code will at any given time try to avoid having neighbouring gas cells of too large volume difference (to ensure numerical stability), thus, in the situation of two neighbouring gas cells having a volume ratio greater than eight, the larger gas cell will be refined.

\subsection{Initial conditions}
\label{sect:initial_conditions}

\begin{figure}
    \resizebox{\hsize}{!}{\includegraphics{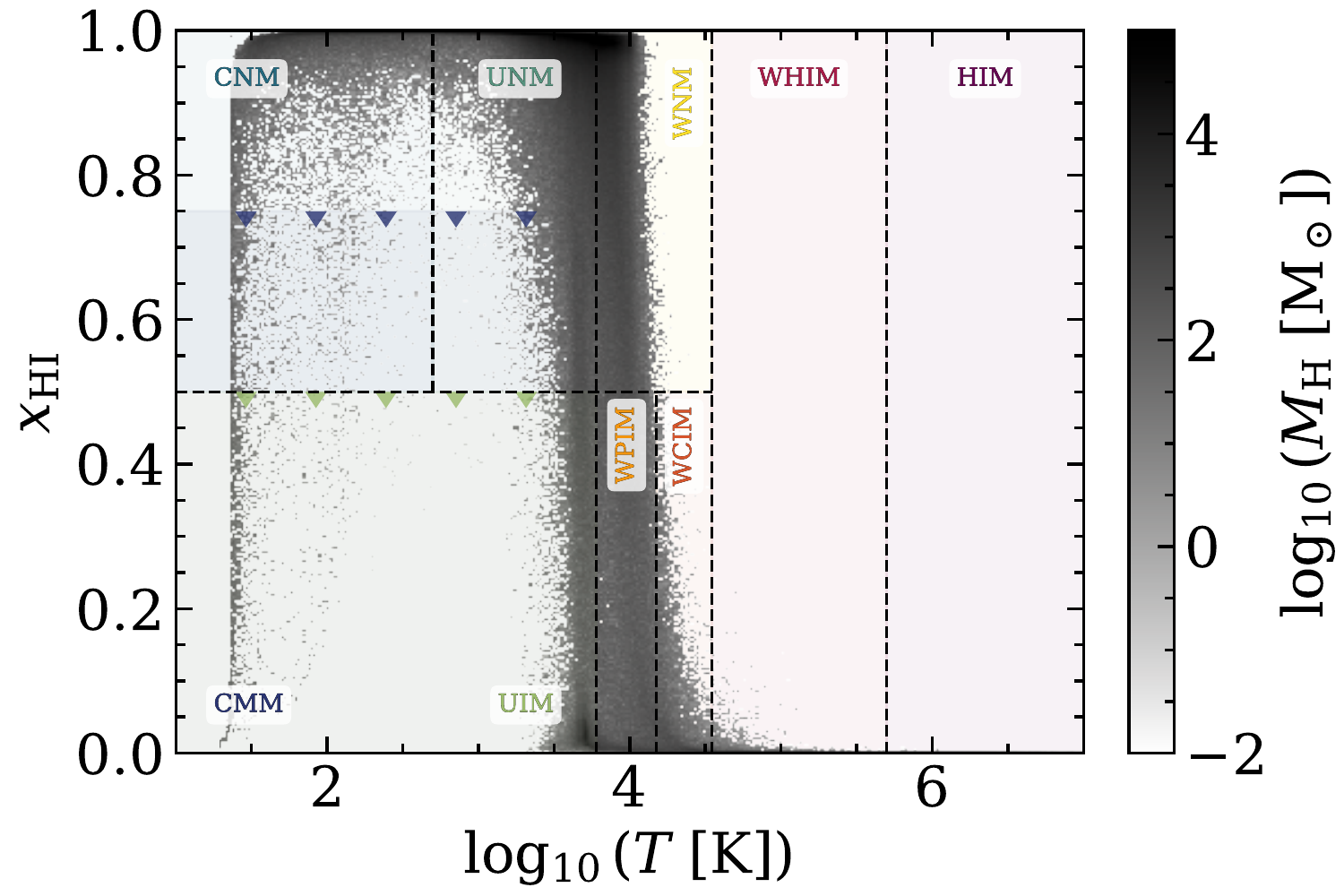}}
    \caption{Abundance of HI as a function of gas temperature for the ISM in our \texttt{MBH4\_full} simulation (at $t\sim 380 \Myr$), mass-weighted by the total amount of H per bin. Following the ISM definitions of \citet{Kim2023}, summarised in Table~\ref{tab:ism_phase_definitions}, the ISM is split into different phases based on its HI, HII or H$_2$ abundance, and gas temperature. The CMM and UIM are categorised by their H$_2$ and HII abundance respectively. For the CMM, $\chi_{\mathrm{H}_2} > 0.25$, meaning that $\chi_\mathrm{HI} < 0.75$, while for the UIM, $\chi_\mathrm{HII} > 0.5$, implying that $\chi_\mathrm{HI} < 0.5$. These upper limits are indicated by blue and green upside-down triangles for the CMM and UIM respectively.}
    \label{fig:ism_phases_definitions}
\end{figure}

\begin{figure}
    \resizebox{\hsize}{!}{\includegraphics{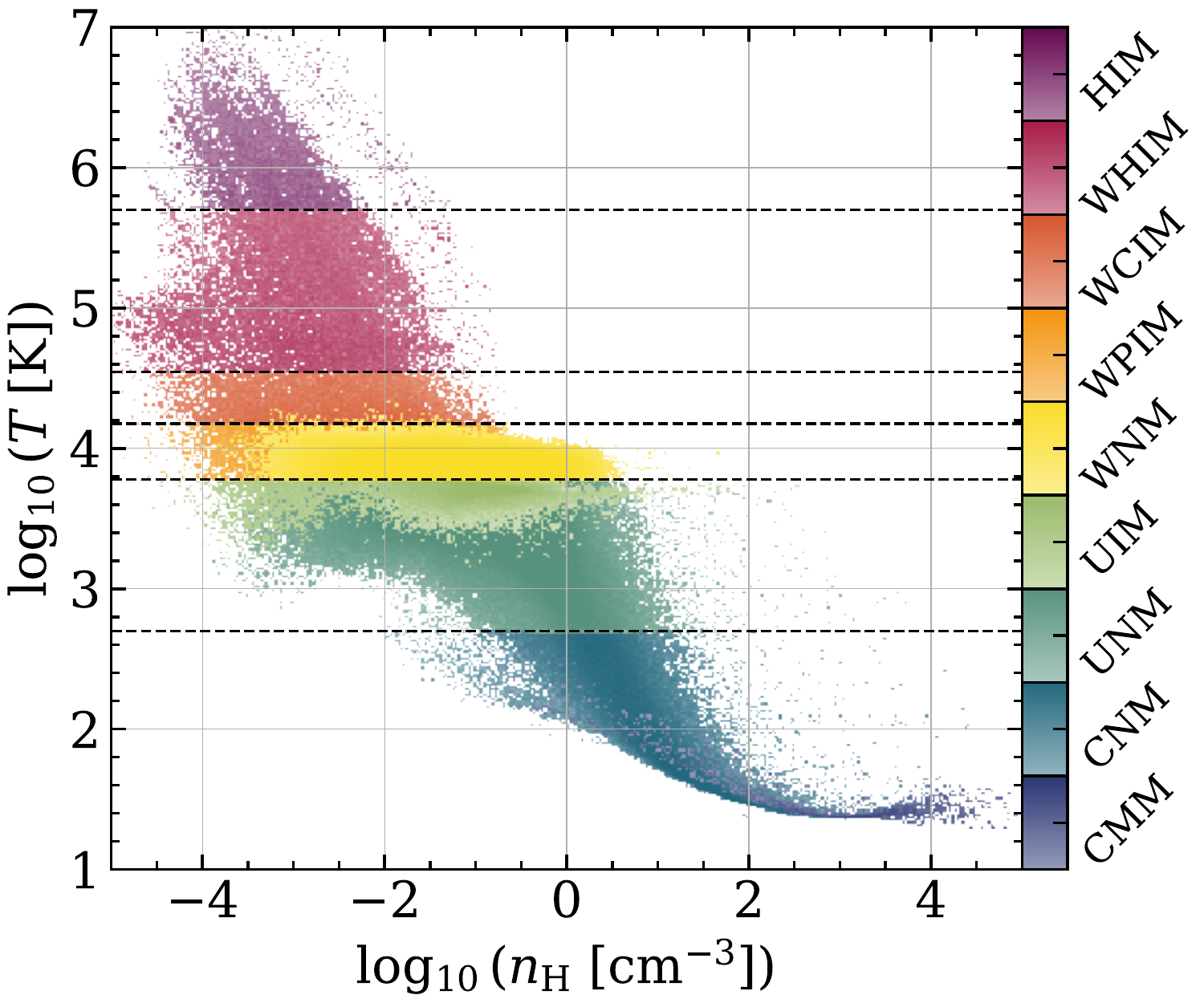}}
    \caption{Gas temperature as a function of H number density for the ISM in our \texttt{MBH4\_full} simulation (at $t\sim 380 \Myr$), mass-weighted by the total amount of H per bin, and colour-coded according to the ISM phase definitions of Table~\ref{tab:ism_phase_definitions} (temperature ranges are indicated by dashed lines).}
    \label{fig:ism_phases_diagram}
\end{figure}

\begin{figure*}
    \centering
    \includegraphics[width=17cm]{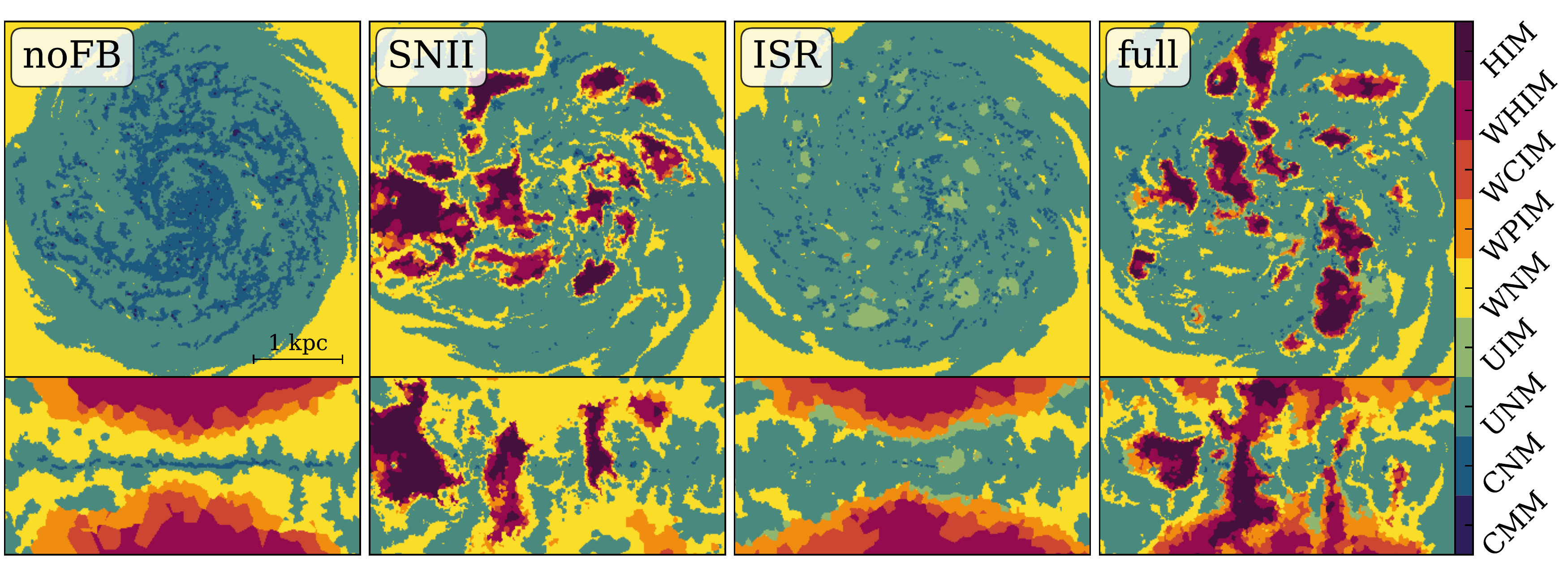}
    \caption{Slices through the ISM in our `\textit{Stellar Feedback}' simulations (see Table~\ref{tab:simulations}), with a face-on (top) and edge-on (bottom) view (at $z=0$ and $y=0$ respectively) of the low-mass dwarf galaxy at $t \sim 380 \Myr$, colour-coded according to the ISM phase definitions in Table~\ref{tab:ism_phase_definitions}.}
    \label{fig:ism_phases_comparison}
\end{figure*}

The low-mass dwarf galaxy system used to conduct the simulations of this study, is generated in isolation using the code and method developed by \citet{Springel2005a}. Two components initially make up the system: a DM halo ($2 \times 10^{10} \Msol$) and a gaseous disc ($5.3 \times 10^{7} \Msol$). The mass distribution of the DM halo follows a \citet{Hernquist1990b} profile with an enforced NFW-equivalent \citep{Navarro1996} density profile at the centre, whereas the gaseous disc is modelled according to an exponential surface density profile. To minimise the effect of numerical dynamical heating on the MBH due to the limited mass resolution of DM particles \citep{Hernquist1990a}, while trying to keep the additional computational cost minimal, the DM mass resolution is set to $500 \Msol$ per particle. When generating the initial conditions (IC), parameter values are chosen in such way that the resulting physical properties of the system broadly agree with previous studies of simulated dwarf galaxies \citep{Hu2016, Whitworth2022}. The obtained physical properties, together with the mass resolution and gravitational softening length of each component, are summarised in Table~\ref{tab:initial_conditions} \citep[the small baryonic mass fraction is in agreement with the abundance matching results of][]{Moster2010, Moster2013}. After generating the ICs, an MBH (in the mass range of $10^{3-6} \Msol$) is manually added to the galactic centre. The initial metallicity of the gas is set to $0.1~\mathrm{Z}_\odot$ \citep[assuming the same relative abundance pattern as inferred for the Sun, see][]{Asplund2009} uniformly across the gaseous disc. Prior to any analysis, the system is initially simulated for $\sim 800 \Myr$ at a coarse target mass resolution ($500 \Msol$ per gas cell), enough to experience an initial burst of star formation and reach a steady-state. This is enabled by reducing the stellar lifetimes of massive stars by a factor of 10, speeding up the injection of energy into the ISM from SNII (ISR is not yet included). As a final preparation, the system is run for an additional $\sim 200 \Myr$ without the reduction of stellar lifetimes (but with no risk of running out of gas since the total gas depletion time of the system is $> 10 \Gyr$), during which the target mass resolution is gradually improved to the desired value for the analysis ($20 \Msol$ per gas cell). This is to ensure that smaller-scale structures, typically arising at higher resolution, are already formed and well-relaxed at the start of the analysis. During this first Gyr, the MBH is not allowed to skim mass, but can dynamically evolve (e.g.\ with nearby star particles).

\subsection{Suite of simulations}
\label{sect:suite_simulations}

Using the numerical and physical set-up described above, a suite of high-resolution radiation-hydrodynamic simulations called \noctua is performed. Throughout the full suite, the target mass resolution and the MBH accretion radius are set to $20 \Msol$ per gas cell and 1~pc respectively (motivated by the resolution study presented in Appendix~\ref{app:resolution}). The simulations are divided into three subsets depending on the scientific question they aim to address. For the first subset of simulations, stellar feedback processes are gradually added to the numerical set-up in order to examine their individual and cumulative impact on MBH gas accretion. Within this subset, the initial mass of the MBH is set to $10^4 \Msol$, in agreement to recent detections of MBH candidates in dwarf galaxies \citep[e.g.][]{Sacchi2024}, and in broad alignment with what is expected from local scaling relations \citep[see e.g.][]{Reines2015}. For the second subset of simulations, the influence of the MBH initial mass on the gas accretion is investigated, by varying the initial mass between $M_\mathrm{BH}=10^{3-6} \Msol$. In the last subset of simulations, we explore whether or not it is possible to `boost' the gas accretion onto the MBH via induced gas inflows towards the galactic centre, by artificially reducing the angular momentum of the gas in the galaxy (mimicking the effect of strong gas inflows potentially happening in a cosmological setting). The full list of subsets and their respective simulations, which together make up the \noctua suite of simulations, are summarised in Table~\ref{tab:simulations}. 

To demonstrate the physical framework used to conduct the \noctua suite of simulations, Fig.~\ref{fig:poster} shows an example simulation (\texttt{h1b4\_full} in Table~\ref{tab:simulations}) where all of the above mentioned stellar feedback processes are included. In panel (a), we show the face-on projected column density map of the HI surface density in our low-mass dwarf galaxy at $t\sim 500 \Myr$. By zooming in on one of the many ISM `bubbles' recently created by SNII explosions (panel b), and further zooming in on one of the groups of recently formed star particles (panel c), we can see the effect of the ionising stellar radiation produced by the newly-formed O and B stars. We can also distinguish the molecular cloud from which the newly-formed star particles were born (panel d). Lastly, panel (e) shows a zoom-in on the region close to the MBH, with a further zoom-in on the MBH accretion region ($r < r_\mathrm{acc}$) in panel (f). As intended by the refinement strategy outlined in Sect.~\ref{sect:refinement}, the region inside the accretion radius $r_\mathrm{acc}$ is resolved by at least a few tens of gas cells. Without this refinement, there is no guarantee that the MBH will always have gas cells to skim mass from, in other words, there is a risk that the code might miss skimming mass from potential gas cells.

\section{Results}
\label{sect:results}

We proceed by showing the main results from the \noctua suite of simulations in the following way. Section~\ref{sect:bh_growth} presents the impact of stellar feedback on MBH growth from the `\textit{Stellar Feedback}' simulations (see Table~\ref{tab:simulations}). This is followed by a detailed analysis of the interstellar medium (ISM) in Sect.~\ref{sect:multiphase_ism}, which is further extended in Sect.~\ref{sect:inflow_outflow_properties}, by exploring the properties of inflowing and outflowing gas close to the MBH. Lastly, in Sect.~\ref{sect:bh_initial_mass}, we address to what extent the MBH initial mass influences the gas accretion (the `\textit{Black Hole Mass}' simulations in Table~\ref{tab:simulations}).

\subsection{Stellar feedback regulated massive black hole growth}
\label{sect:bh_growth}

\begin{figure*}
    \centering
    \includegraphics[width=17cm]{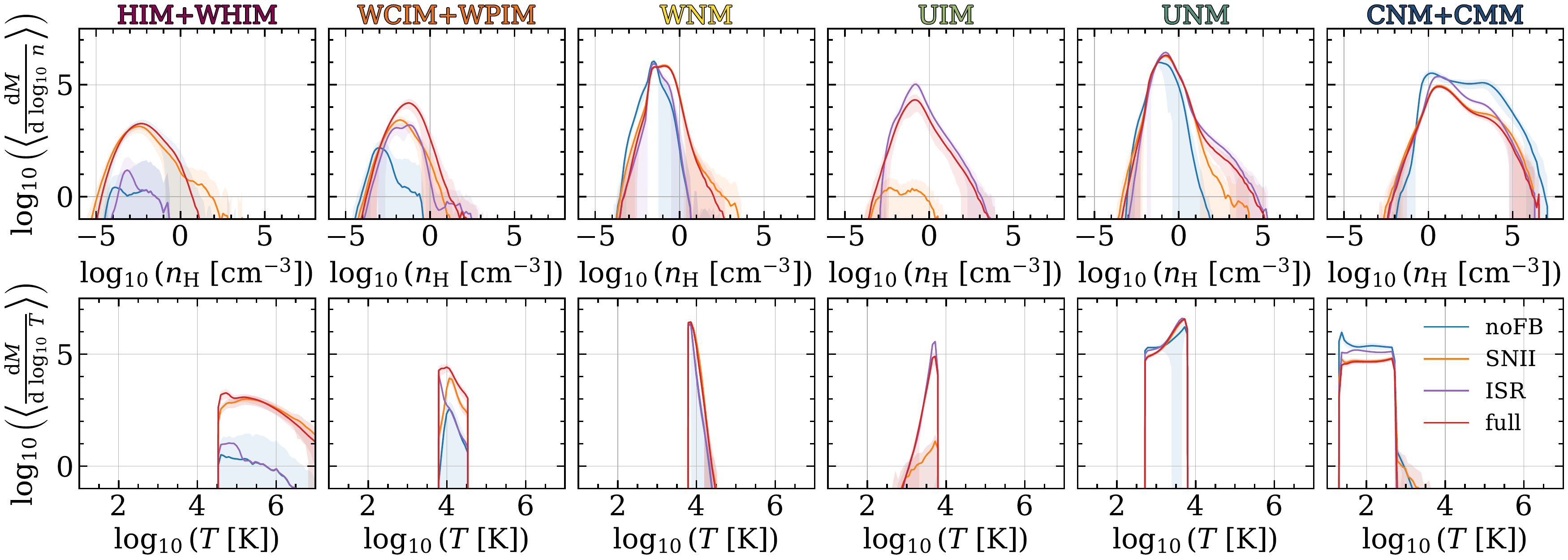}
    \caption{Mass-weighted probability density functions (integrated and averaged over 800~Myr) of the gas density (top row) and gas temperature (bottom row) for different phases of the ISM (limited to gas inside $R < 2.5 \kpc$ and $|z| < 300 \pc$ of the low-mass dwarf galaxy mid-plane) in our `\textit{Stellar Feedback}' simulations (see Table~\ref{tab:simulations}). To reduce the complexity of tracking each individual ISM phase of Table~\ref{tab:ism_phase_definitions}, some of the most hot and cold phases are paired together into single phases (see Sect.~\ref{sect:multiphase_ism} for more details). Shaded areas correspond to one standard deviation ($\pm \sigma$). The sharp cuts in the gas temperature distribution functions are linked to how the different ISM phases are defined (see Tabel~\ref{tab:ism_phase_definitions}).}
    \label{fig:ism_phases_pdf}
\end{figure*}

In the top panel of Figure~\ref{fig:bh_growth}, we show the amount of mass accreted over time by an initial $10^4 \Msol$ MBH in the `\textit{Stellar Feedback}' simulations. Without any stellar feedback included (blue line), the MBH grows steadily over time, increasing its mass by $\sim 100~\%$ of its initial mass after 800~Myr. With the addition of type II supernova feedback (SNII, orange line), the growth of the MBH is suppressed by more than an order of magnitude (already during the first 150~Myr). However, when only considering ionising stellar radiation feedback (ISR, purple line), the opposite effect can be seen, where the growth of the MBH is instead enhanced at all times compared to the no stellar feedback run. Yet, when SNII is combined with ISR (SNII+ISR, red line), the growth of the MBH remains significantly suppressed, but with an overall slightly higher MBH growth throughout the full simulation compared to the SNII-only feedback run. The bottom panel of Figure~\ref{fig:bh_growth} shows the corresponding time evolution of the MBH Eddington ratio ($f_\mathrm{Edd} = \dot{M}_\mathrm{BH} / \dot{M}_\mathrm{Edd}$) in each of our simulations, illustrating that SNII feedback not only reduces $f_\mathrm{Edd}$ (i.e.\ the rate at which gas accretes onto the MBH), compared to the no stellar feedback run, but also makes it more episodic over time with occasional time periods of no gas accretion at all. The addition of SNII feedback moves the MBH from having a continuous steady gas accretion rate at $f_\mathrm{Edd} \gtrsim 0.02$ in the radiatively efficient regime, to instead more consistently accrete at $f_\mathrm{Edd} < 0.02$ in the radiately inefficient regime. The effect of ISR feedback on $f_\mathrm{Edd}$ is less striking than when SNII feedback is included, but is still apparent, as it slightly increases $f_\mathrm{Edd}$ at close to all times, regardless of whether SNII feedback is already included or not. 

With the gas being the only source of mass for the MBH to accrete from (at least in this study), a detailed view of the ISM in the low-mass dwarf galaxy is needed to robustly understand the regulating power of stellar feedback on MBH growth. In Fig.~\ref{fig:comparison}, we therefore present a series of face-on and edge-on projected column density maps displaying the total gas, HI, H$_2$, and HII surface densities, as well as the gas temperature (columns left to right), in each of the `\textit{Stellar Feedback}' simulations (top row: \texttt{MBH4\_noFB}, second row: \texttt{MBH4\_SNII}, third row: \texttt{MBH4\_ISR}, bottom row: \texttt{MBH4\_full}) at $t\sim 380 \Myr$. From visual inspection, it is evident that SNII feedback strongly disrupts the collapse of cold dense gas, hindering the formation of H$_2$ by heating and partially ionising the gas. However, it is only with ISR feedback that prominent and well-confined HII regions start to appear throughout the galaxy. When only ISR feedback is included (i.e. no SNII feedback), the ISM is visually less clumpy compared to the no stellar feedback run, but not as diffuse and hot as when SNII feedback is present. In fact, by estimating the fraction of mass enclosed in the least volume-filling gas, taking into account the total gas mass and volume of the low-mass dwarf galaxy (confined to $R<2.5 \kpc$ and $|z| < 100 \pc$, centred around the mid-plane of the galaxy), indeed the ISM is most `clumpy' when no stellar feedback is considered. For the no stellar feedback run, $\sim 60~\%$ of the total gas mass is concentrated to the 10~\% least volume-filling gas. This is followed by the ISR-only feedback run, where $\sim 40~\%$ of the mass is concentrated to the same volume fraction, and least so in the simulations of SNII feedback, where the mass fraction in both simulations are $\sim 20~\%$. With a less clumpy ISM, a higher fraction of the total gas mass is expected to reside in the ambient (more diffuse) medium, enabling more gas to be accreted onto the MBH. This partly explains why the MBH grows the most when only ISR feedback is included (see Fig.~\ref{fig:bh_growth}), despite the visually larger amount of H$_2$ (expected to be a main fuel source for the MBH) seen in the no stellar feedback run. Nevertheless, this is not sufficient to explain the suppression of MBH growth seen when SNII feedback is present, meaning that other mechanisms are also likely to take part in regulating MBH gas accretion.

\subsection{A multi-phase interstellar medium}
\label{sect:multiphase_ism}

\begin{figure*}
    \centering
    \includegraphics[width=17cm]{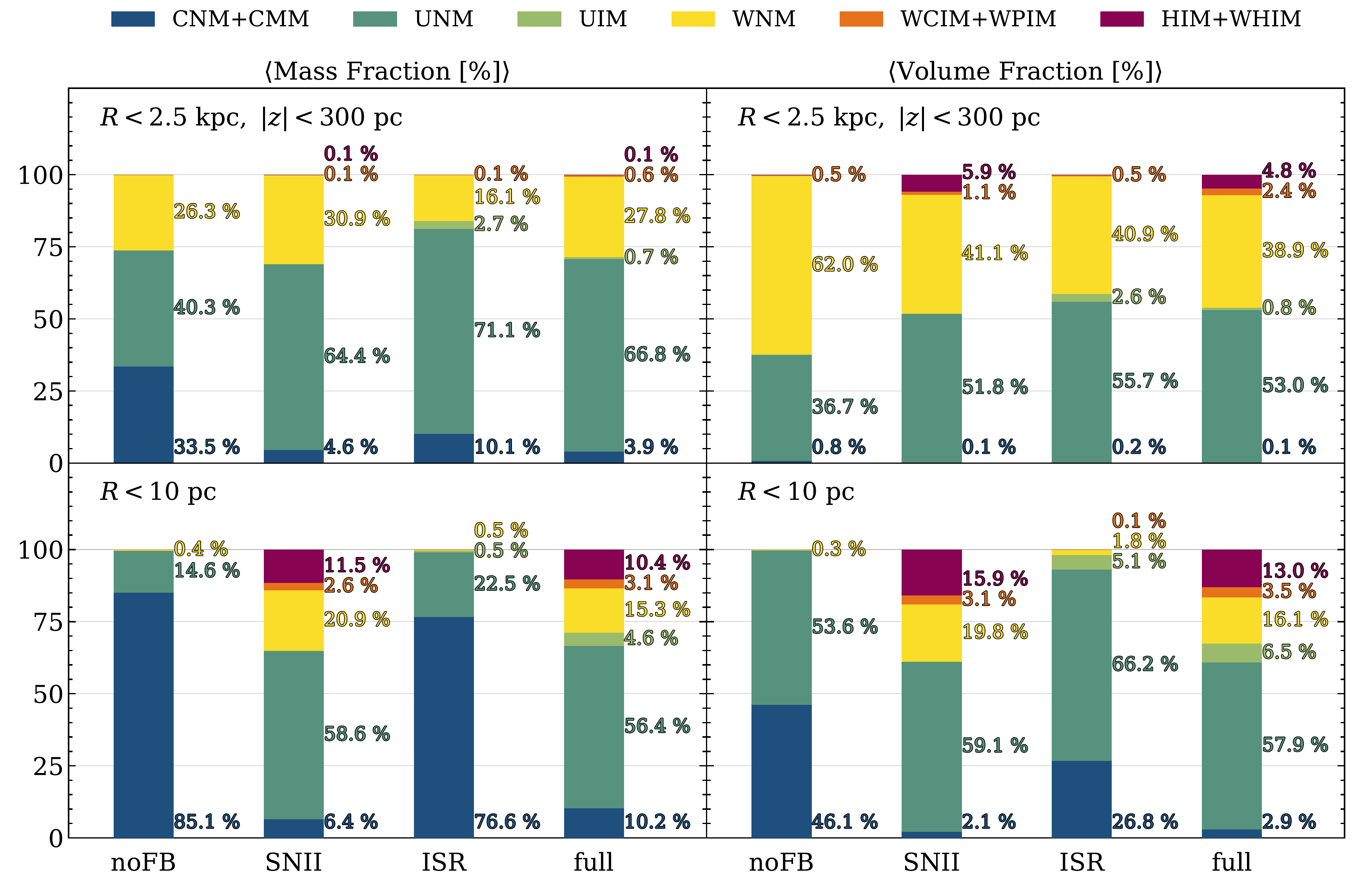}
    \caption{Time-integrated (over 800~Myr) average mass- and volume fractions (left and right column respectively) for different ISM phases in our `\textit{Stellar Feedback}' simulations, both on a global scale ($R < 2.5 \kpc$ and $|z| < 300 \pc$ around the mid-plane of the low-mass dwarf galaxy, top panels) and on a local scale close to the MBH ($R < 10 \pc$, bottom panels). Fractions $< 0.1~\%$ are not displayed.}
    \label{fig:ism_phases_fraction}
\end{figure*}

We extend our analysis from Sect.~\ref{sect:bh_growth} in trying to link the discrepancy in MBH growth (due to which stellar feedback processes are included or not) with the ISM composition (and distribution) of our low-mass dwarf galaxy, by categorising individual gas cells according to their ISM phase. As already demonstrated in Fig.~\ref{fig:comparison}, the projected hydrogen column densities (second to fourth column) are distinctly different depending on the stellar feedback process(es) included. By classifying individual gas cells according to both their temperature and hydrogen abundances, following the ISM phase definitions of \citet{Kim2023}, we can therefore more thoroughly disentangle the individual and cumulative impact of SNII and ISR feedback on the ISM composition (and distribution) of our low-mass dwarf galaxy in the `\textit{Stellar Feedback}' simulations. The adopted ISM phases of \citet{Kim2023} and their respective definitions, are summarised in Table~\ref{tab:ism_phase_definitions}. We illustrate how the gas is split into different phases in Fig.~\ref{fig:ism_phases_definitions} (using the definitions in Table~\ref{tab:ism_phase_definitions}), by showing the HI abundance as a function of gas temperature for the ISM (at $t\sim 380 \Myr$) in the \texttt{MBH4\_full} simulation, mass-weighted by the total amount of H per bin, and indicate different regions according to their ISM phase (using the abbreviations in Table~\ref{tab:ism_phase_definitions}). A substantial amount of H can be located in the unstable neutral and ionised medium phases (UNM and UIM respectively), as well as the warm neutral and photoionised medium phases (WNM and WPIM respectively). This agrees well with \citet{Hu2016}, who find that the ISM in a similar dwarf galaxy set-up is dominated by $100 \ \K \leq T \leq 3 \times 10^4 \K$ gas. In Fig.~\ref{fig:ism_phases_diagram}, we show the corresponding phase diagram (gas temperature as a function of H number density) for the same snapshot of the ISM in the \texttt{MBH4\_full} simulation as in Fig.~\ref{fig:ism_phases_definitions}, mass-weighted by the total amount of H per bin, and colour-coded according to the ISM phase definitions in Table~\ref{tab:ism_phase_definitions}. Compared to Fig.~\ref{fig:ism_phases_definitions}, where the most visually populated ISM phases are clearly disentangled from one another, Fig.~\ref{fig:ism_phases_diagram} shows instead a significant overlap for the UNM, UIM, WNM, and WPIM phases. Thus, by only applying temperature ranges to define ISM phases (which is traditionally done), chemically different ISM phases are at risk of being treated the same, despite being associated to vastly different physical processes (as discussed below). 

Using the ISM phase definitions in Table~\ref{tab:ism_phase_definitions}, Fig.~\ref{fig:ism_phases_comparison} shows colour-coded face-on (top row) and edge-on (bottom row) slices (at $z=0$ and $y=0$ respectively) through the ISM in each of our `\textit{Stellar Feedback}' runs (different columns as indicated). The UNM is the visually most dominating phase of the ISM in all of the physical set-ups. Extended regions of the cold neutral medium phase (CNM) with patches of the cold molecular medium phase (CMM) can primarily be seen in the no feedback run. However, extended regions of the CNM phase can also be distinguished in the ISR-only feedback run (but to a lesser extent compared to the no feedback run). It is only when SNII feedback is included that the CNM phase is strongly suppressed, together with the emergence of prominent regions of the hot and warm-hot ionised medium phases (HIM and WHIM respectively). Similarly, the UIM phase is only prominent when ISR feedback is included. 

\begin{figure*}
    \centering
    \includegraphics[width=17cm]{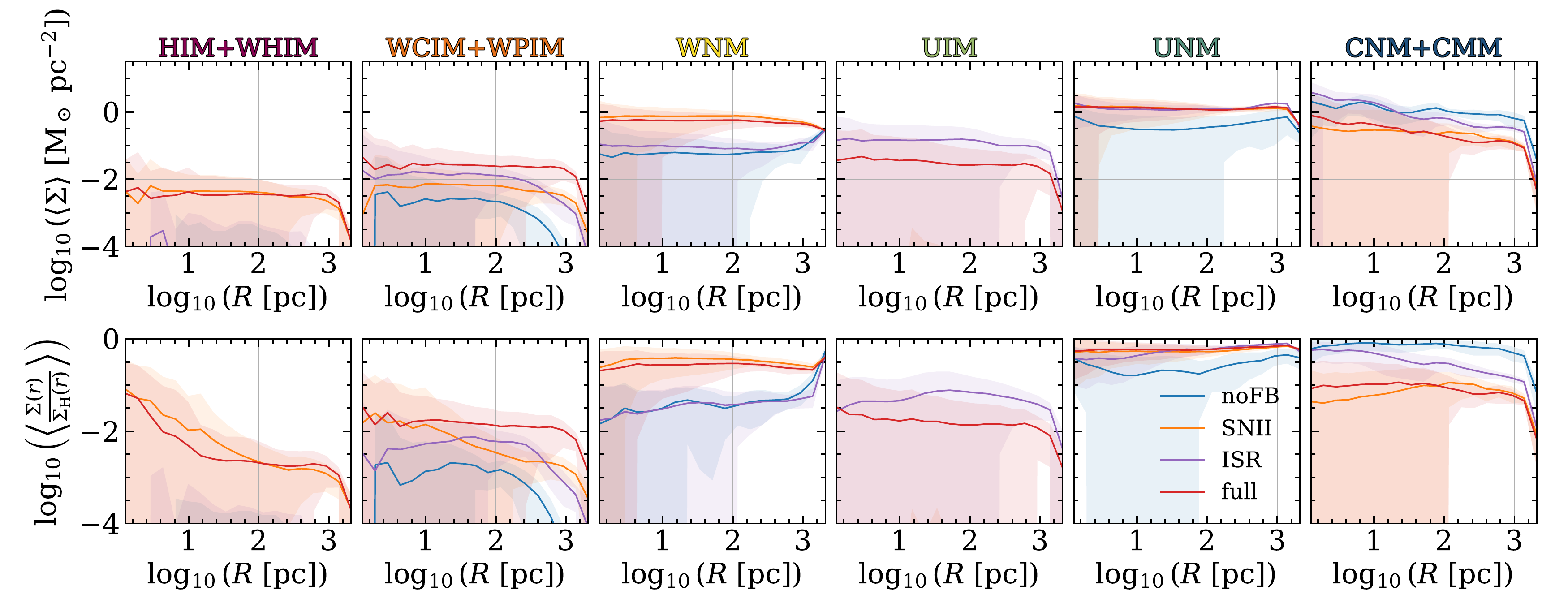}
    \caption{\textit{Top row}: Radial surface density profiles (integrated and averaged over 800~Myr) for different ISM phases in our `\textit{Stellar Feedback}' simulations (see Table~\ref{tab:simulations}). The radius $R$ is with respect to the position of the MBH, and the calculation of surface densities are limited to gas inside $|z| < 300 \pc$ of the MBH. \textit{Bottom row}: Corresponding mass fraction as a function of radius for each ISM phase in the top row. Shaded areas correspond to one standard deviation ($\pm \sigma$).}
    \label{fig:ism_phases_surfdens}
\end{figure*}

\begin{figure*}
    \centering
    \includegraphics[width=17cm]{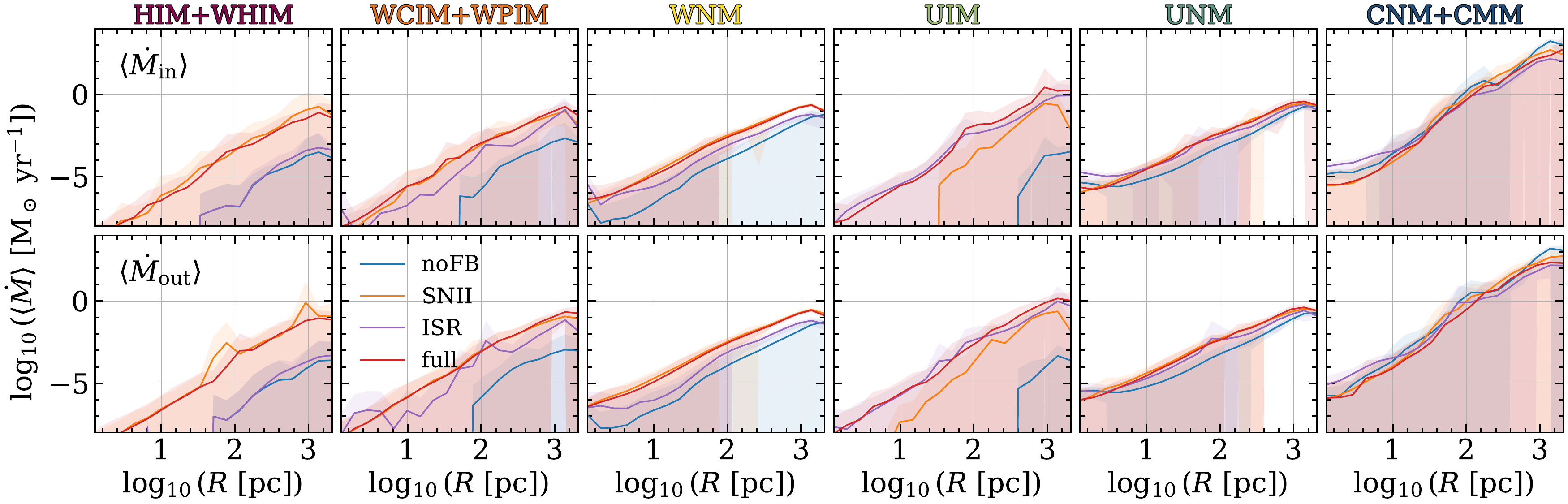}
    \caption{Time-integrated (over 800~Myr) average radial profiles of the gas inflow (top row) and outflow (bottom row) rate for different ISM phases in our `\textit{Stellar Feedback}' simulations (see Table~\ref{tab:simulations}). The radius $R$ is with respect to the position of the MBH, and the average gas inflow and outflow rate in each radial bin is calculated for a spherical shell around the MBH (see Sect.~\ref{sect:inflow_outflow_properties} for more details). Shaded areas correspond to one standard deviation ($\pm \sigma$).}
    \label{fig:time_int_inflow_outflow}
\end{figure*}

To further quantify the effect of SNII and ISR feedback on the ISM composition of our low-mass dwarf galaxy, in Fig.~\ref{fig:ism_phases_pdf}, we show mass-weighted probability density functions (integrated and averaged over 800~Myr) of the gas density (top row) and temperature (bottom), for different ISM phases (different columns as indicated) in each of our `\textit{Stellar Feedback}' runs (different coloured lines). Tracking all of the above mentioned ISM phases (see Table~\ref{tab:ism_phase_definitions}) individually, and their respective relation (or not) to MBH growth, is very demanding and time-consuming. Hence, to reduce the complexity for the remaining part of the analysis, some of the hottest and coldest phases of the ISM are paired together into single phases in Fig.~\ref{fig:ism_phases_pdf} (without causing any relevant loss of information). Since the CNM and CMM phases are both expected to be associated to star formation, and seemingly closely coupled to each other (see Fig.~\ref{fig:ism_phases_comparison}), we combine these two phases into one: CNM+CMM. Similarly, the HIM and WHIM phases seem to both be strongly correlated to SNII explosions, thus, they are combined into a single phase: HIM+WHIM. Lastly, since both the warm collisionally ionised medium (WCIM) and WPIM phases tend to appear in the outskirts of the HIM+WHIM phase (see Fig.~\ref{fig:ism_phases_comparison}), we also combine these two phases into one: WCIM+WPIM. 

As expected, Fig.~\ref{fig:ism_phases_pdf} shows that SNII feedback indeed is the main contributor to generating the HIM+WHIM phase \citep[orange and red lines, in agreement to][]{Hu2017}. For building the WCIM+WPIM phase, both SNII and ISR feedback are seemingly strong contributors (individually and cumulatively), while for the UIM phase, only ISR feedback is able to significantly generate a substantial amount of gas in this phase (purple and red lines). For the WNM phase, SNII feedback seem to sustain a larger amount of gas at high number densities (red and orange curves in the WNM column). The same effect can be seen for the UNM phase, but when ISR feedback is included (purple and red curves in the UNM column). Unsurprisingly, the largest amount of CNM+CMM phases can found in the no stellar feedback run (blue line), followed by the ISR-only feedback run (purple line), and lastly the simulations where SNII feedback is included.

\begin{figure*}
    \centering
    \includegraphics[width=17cm]{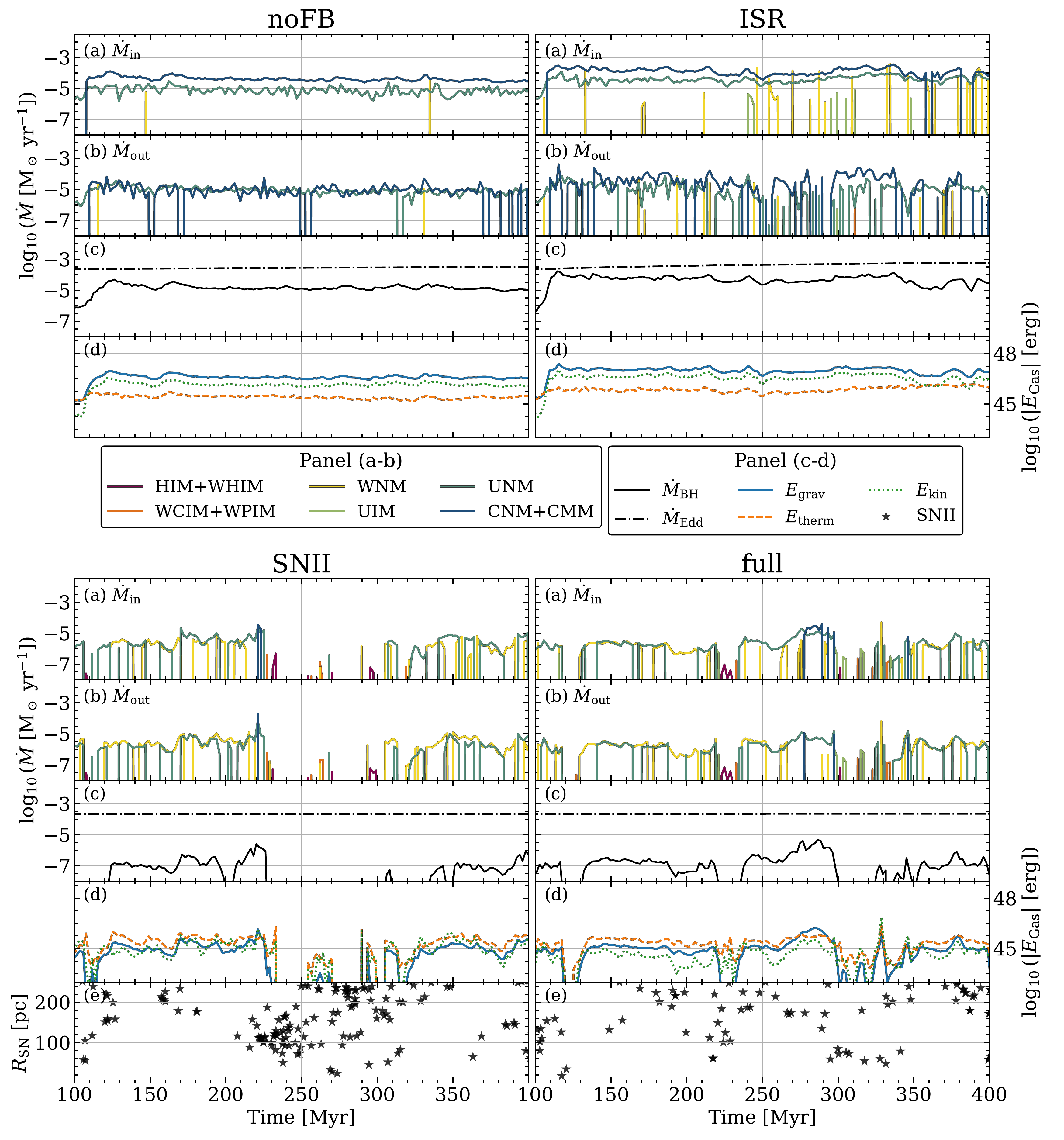}
    \caption{\textit{Panel (a-b)}: Cutout of the time evolution (between 100 and 400~Myr) for the gas inflow (panel a) and outflow (panel b) rate at the vicinity of the MBH (through a spherical shell at  $R = [1, 3] \pc$) for different ISM phases (coloured lines), in our `\textit{Stellar Feedback}' simulations (see Table~\ref{tab:simulations}). \textit{Panel (c)}: Corresponding time evolution of the MBH gas accretion rate and its Eddington accretion limit (black solid and dash-dotted line respectively). \textit{Panel (d)}: As a function of time, the total gravitational binding- (blue solid line), thermal- (orange dashed line), and kinetic energy (green dotted line) of all the inflowing and outflowing gas between $R = [1, 3] \pc$. \textit{Panel (e)}: Radial distances (with respect to the position of the MBH) to nearby SNII explosions as a function of time.}
    \label{fig:time_evol_inflow_outflow}
\end{figure*}

To make a fair comparison of how the different ISM phases in Fig.~\ref{fig:ism_phases_pdf} relate to one another (irrespective of the amount of gas residing in the galaxy, which may vary due to differences in the star formation rate), Fig.~\ref{fig:ism_phases_fraction} shows the time-integrated (over 800~Myr) average mass- and volume fraction (left and right column respectively) of each ISM phase, both globally ($R < 2.5 \kpc$ and $|z| < 300 \pc$, top panels) and locally at the vicinity of the MBH ($R < 10 \pc$, bottom panels), in our `\textit{Stellar Feedback}' runs. The hot-, unstable-, and warm ionised medium phases (UIM, WCIPM+WPIM, and HIM+WHIM) only marginally contribute to the total amount of mass in the ISM on a global scale across all runs (top left panel), but display significantly higher mass fractions close to the MBH when SNII feedback is included (bottom left panel). That said, these phases contribute more to the ISM in terms of volume (right column), but are still minor compared to the strong volume-filling of the WNM (especially on a global scale, top right panel). However, the UNM phase is overall the most mass- and volume-dominating phase of the ISM in our low-mass dwarf galaxy, both globally and close to the MBH, regardless of the stellar feedback process(es) included. This is true except for the mass fractions close to the MBH in the runs without SNII feedback included (bottom left panel), where the CNM+CMM phase is completely dominating the mass contribution to the ISM. We also note that the WNM is the most volume-filling phase of the ISM on a global scale in the no stellar feedback run (top right panel). This is because of the strong collapse of the gaseous disc in the $z$-directions (see e.g. the edge-on slice of the ISM in Fig.~\ref{fig:ism_phases_comparison} for the no stellar feedback run), enabling more of the surrounding WNM phase to be taken into account when calculating the volume fractions (since we consider all gas within $|z| < 300 \pc$). 

\begin{figure*}
    \centering
    \includegraphics[width=17cm]{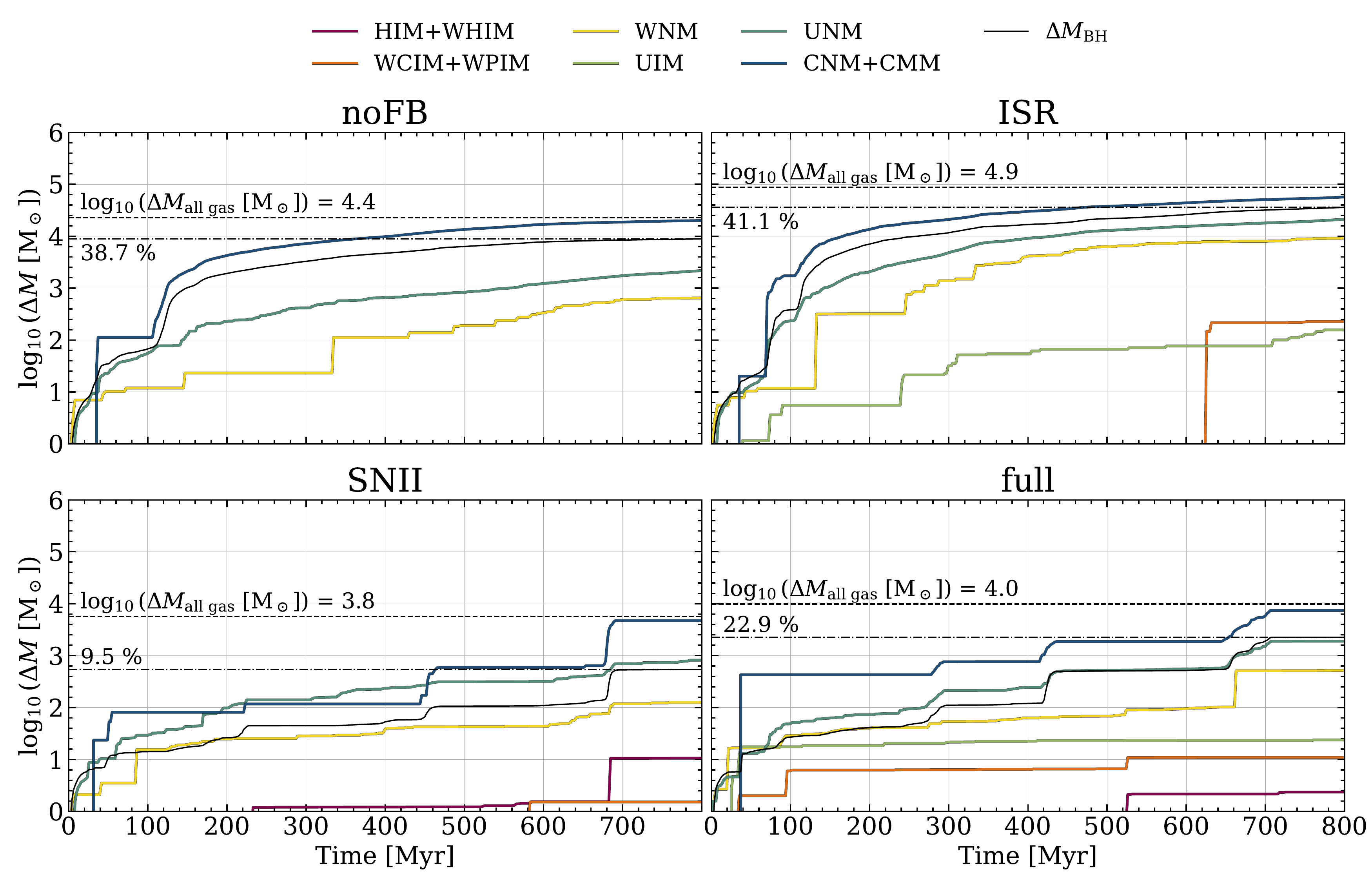}
    \caption{Cumulative net gas inflow over time at the vicinity of the MBH (through a spherical shell at $R = [1, 3] \pc$) for different ISM phases in our `\textit{Stellar Feedback}' simulations (see Table~\ref{tab:simulations}). For each run, the total cumulative net gas inflow (including all of the different ISM phases) is provided and indicated by a black dashed line. The growth of the MBH (black solid line for reference), in terms of percentage of the total cumulative net gas inflow, is provided and indicated by a black dash-dotted line.}
    \label{fig:cummulative_inflow_outflow}
\end{figure*}

The discrepancy in mass fractions on a global scale versus close to the MBH in Fig.~\ref{fig:ism_phases_fraction} (e.g.\ in the CNM+CMM phase), suggests that the ISM composition of our low-mass dwarf galaxy is radially dependent as we approach the central MBH, and that it may vary depending on the stellar feedback process(es) included. To explore this further, Fig.~\ref{fig:ism_phases_surfdens} shows radial surface density profiles (integrated and averaged over 800~Myr, top row) and associated mass fraction profiles (bottom row), for different ISM phases (different columns as indicated), in each of our `\textit{Stellar Feedback}' runs (different coloured lines). The estimated uncertainty of both profiles for each ISM phase is indicated by colour- (according to the simulation) shaded areas (corresponding to one standard deviation). No strong radial dependence for the surface densities can be distinguished among the different ISM phases, except for the WNM and CNM+CMM phases. The surface density of the CNM+CMM phase increases as we get close to the MBH, mostly in the runs without SNII feedback included (blue and purple lines), but also slightly when SNII feedback is considered (orange and red lines). With the surface density of the UNM phase seemingly being constant at all radii (for all the runs and with only small variations between them), this explains why the mass fractions of the CNM+CMM phase is the highest on a local scale close to the MBH, compared to on a global scale in Fig.~\ref{fig:ism_phases_fraction} (especially for the runs without SNII feedback included). For the WNM phase, the surface density initially increases and decreases at large radii ($\sim 1 \kpc$) for the runs with and without SNII feedback included respectively, but then remains relatively stable as we approach the MBH. The mass fraction of the WNM phase is consistently higher at all radii in the runs with SNII included than in the runs without (bottom panel in the WNM column). Since the opposite effect can be seen for the CNM+CMM phase, where the mass fraction of the CNM+CMM phase is consistently higher at all radii in the runs without SNII included than with (bottom panel in the CNM+CMM column), we conclude that SNII feedback sustains a higher mass fraction of the ISM in the WNM phase, at the expense of mainly the CNM+CMM phase\footnote{The UNM phase is not primarily considered, since the mass fraction profiles are only marginally different between the runs. The drop in mass fraction for the UNM phase in the no stellar feedback run (blue curve) is due to the extreme increase in mass fraction of the CNM+CMM phase (see the blue curve in the bottom panel of the CNM+CMM column in Fig.~\ref{fig:ism_phases_surfdens}).}. Lastly, we note that mass fractions for the HIM+WHIM and WCIM+WPIM phases increase as we get close to the MBH, possibly due to higher star formation rate (as a consequently of the higher surface density of the CNM+CMM phase). However, we also note that the majority of ISM phases in Fig.~\ref{fig:ism_phases_surfdens} exhibit a large scatter (colour-shaded areas) in their respective surface density and mass fraction profiles at small radii, including the HIM+WHIM and WCIM+WPIM phases. This is likely due to the stochastic nature of star formation and stellar feedback. At large radii, the volume within which these processes act is substantially larger than at small radii, and is thus less affected by the stochasticity. At small radii however, a few stochastic events can significantly alter the ISM phase composition, explaining for example the slight increase of mass fractions in the HIM+WHIM and WCIM+WPIM phases seen in Fig.~\ref{fig:ism_phases_fraction} (when SNII feedback is included), which is possibly due to a few SNII events occurring very close to the MBH.

\subsection{Properties of inflowing and outflowing gas}
\label{sect:inflow_outflow_properties}

\begin{figure}
    \resizebox{\hsize}{!}{\includegraphics{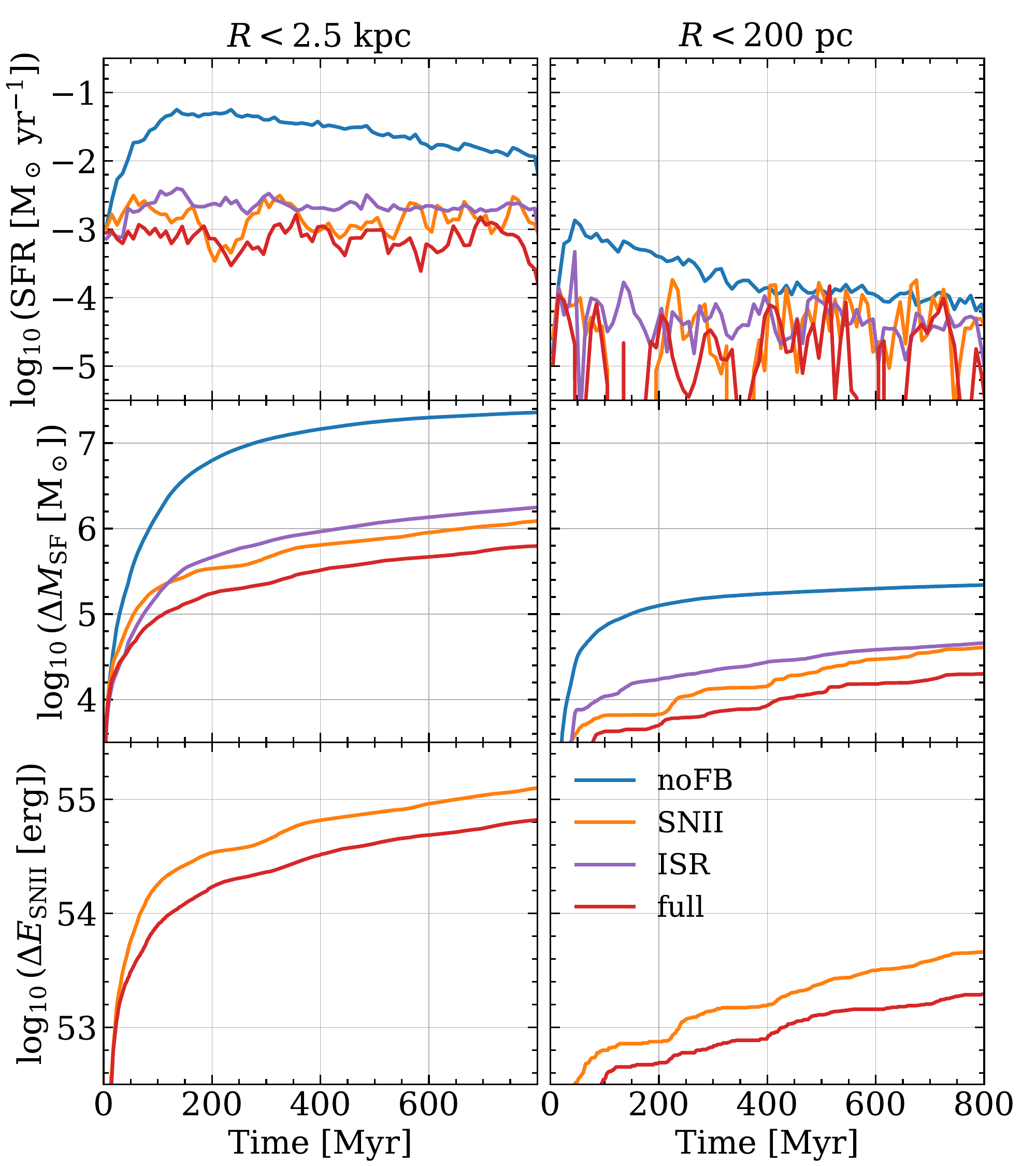}}
    \caption{Time evolution of the SFR (in bins of 10~Myr, top row), cumulative amount of stars formed (middle row), and cumulative amount of energy injected into the ISM from SNII explosions (bottom row), on a global scale ($R < 2.5 \kpc$, left column) and close to the MBH ($R < 200 \pc$, right column), in our `\textit{Stellar Feedback}' simulations (see Table~\ref{tab:simulations}).}
    \label{fig:cumulative_star_formation}
\end{figure}

As stated in Sect.~\ref{sect:multiphase_ism}, the composition of the ISM does not only distinctly differ between our `\textit{Stellar Feedback}' simulations, but can also spatially vary within the low-mass dwarf galaxy. How this relates to the growth of the MBH is however not trivial. To address this, Fig.~\ref{fig:time_int_inflow_outflow} shows the average gas inflow (top row) and outflow (bottom row) rate as a function of radius (with respect to the position of the MBH) for different ISM phases (different columns as indicated) in our `\textit{Stellar Feedback}' simulations (different coloured lines). For each ISM phase, the gas inflow and outflow rate is estimated as: $\dot{M} = 4 \pi r^2 \langle | v_r | \rangle \langle \rho \rangle$, where $\langle | v_r | \rangle$ and $\langle \rho \rangle$ are the mean absolute radial velocity (with respect to the MBH) and mean density respectively, of inflowing ($v_r < 0$) and outflowing ($v_r > 0$) gas through a spherical shell of radius $r$ and thickness $\mathrm{d}r$. The estimated uncertainty of gas inflow and outflow rates for each ISM phase is indicated by colour- (according to the simulation) shaded areas (corresponding to one standard deviation). As expected, whenever SNII feedback is included, a greater exchange of inflowing and outflowing gas in the hot and warm phases (HIM+WHIM, WCIM+WPIM, and WNM) can be distinguished at all radii, compared to the runs without SNII feedback included. The same is true for the UIM phase when ISR feedback is included. For the UNM and CNM+CMM phases, the average inflow and outflow rates are close to identical between the different runs, except close to the MBH. At this scale, the average inflow and outflow rates of the UNM and CNM+CMM phases tend to be slightly higher when no SNII feedback is considered (blue and purple curves), compared to the SNII feedback runs (orange and red curves). This suggests that the UNM and CNM+CMM phases are mainly responsible for the enhanced MBH growth seen in Fig.~\ref{fig:bh_growth}, when no SNII feedback is included. However, the estimated uncertainty of the gas inflow outflow rates in Fig.~\ref{fig:time_int_inflow_outflow} is substantial, especially at small radii close to the MBH. In addition to that, the gas inflow rate at a certain radius is typically matched by a roughly equal amount of gas outflow, meaning that it is difficult to identify any significant net-inflows towards the MBH. 

To more precisely study how the inflowing and outflowing gas of various ISM phases correlate with the gas accretion onto the MBH, Fig.~\ref{fig:time_evol_inflow_outflow} shows a cutout of the time evolution (between 100 and 400~Myr) of the gas inflow (panel~a) and outflow (panel~b) rate at the vicinity of the MBH (through a sphere with an inner and outer boundary of $R = [1, 3] \pc$ around the MBH) for different ISM phases (different coloured lines), in our `\textit{Stellar Feedback}' runs. Figure~\ref{fig:time_evol_inflow_outflow} also shows the corresponding time evolution of the MBH gas accretion rate $\dot{M}_\mathrm{BH}$ and its Eddington accretion limit $\dot{M}_\mathrm{Edd}$ (panel~c), the time evolution of the total gravitational binding- (blue solid line), thermal- (orange dashed line), and kinetic energy (green dotted line) of all the inflowing and outflowing gas (between $R = [1, 3] \pc$, panel~d), and the radial distance (with respect to the position of the MBH) to nearby SNII explosions as a function of time (for the runs with SNII feedback included, panel~e).

By comparing the gas inflow and outflow rates of each ISM phase (Fig.~\ref{fig:time_evol_inflow_outflow}, panel~a and~b) with the time evolution of the MBH gas accretion rate (panel~c), we can see that when no stellar feedback is considered (top left panels), a continuous net gas inflow of the CNM+CMM phase is the main contributor to the gas accretion onto the MBH. The same is true for the ISR-only feedback run (top right panels). However, compared to the no stellar feedback run, this scenario seems to enable a more significant net gas inflow of the UNM phase as well. With SNII feedback included (bottom row of panels), the continuous net gas inflow of the CNM+CMM phase is completely suppressed (with only occasional inflows and outflows over time), leaving the UNM phase (which is far more episodic over time compared to the no SNII feedback runs) and the WNM phase as the main sources for the MBH to skim mass from. 

For the gravitational binding energy of all the inflowing and outflowing gas (Fig.~\ref{fig:time_int_inflow_outflow}, panel~c), it is consistently higher than both the thermal and kinetic energy over time, in the runs without SNII feedback included (top row panels), suggesting that a high fraction of the net-inflowing CNM+CMM phase fulfils the accretion criteria (defined in Sect.~\ref{sect:bh_accretion}). With the inclusion of SNII feedback however (bottom row panels), the gravitational binding energy is instead comparable or below that of the thermal and kinetic energy, meaning that the likelihood for the net-inflowing UNM and WNM phases to fulfil the accretion criteria (gravitationally bound to- and a converging flow onto the MBH) is substantially lower. Consequently, the fraction of gas cells from which the MBH can skim mass also gets reduced. 

Since the energy injection of a SNII explosion ($10^{51} \erg$) is much higher than the typical gravitational binding energy of the inflowing and outflowing gas close to the MBH ($\sim 10^{45} \erg$ according to Fig.~\ref{fig:time_int_inflow_outflow}, panel d), SNII do not necessary have to explode in the direct vicinity of the MBH to regulate the gas accretion, but can also occur at larger distances and still indirectly influence the gas inflow towards the MBH. For the runs with SNII feedback included, we can see in panel (e) that whenever there is high number of relatively nearby ($\sim 100-200 \pc$) SNII explosions, or only a few very nearby (10s of pc) ones, the gas inflow drops (panel a), and the MBH gas accretion rate decreases (panel c). This also explains why the gas inflow (and thus the MBH gas accretion rate) is more episodic over time, than in the runs without SNII included. 

Despite Fig.~\ref{fig:time_evol_inflow_outflow} occasionally showing strong gas inflows (e.g.\ in the runs without SNII feedback included, panel~a), the majority of inflows are typically followed by outflows of similar strength (panel~b), suggesting that the MBH only skims mass from a minor fraction of the total inflowing gas. To quantify this, Fig.~\ref{fig:cummulative_inflow_outflow} shows the cumulative net gas inflow over time at the vicinity of the MBH (through a spherical shell at $R = [1, 3] \pc$ around the MBH) for different ISM phases (different coloured lines) in our `\textit{Stellar Feedback}' runs (different panels as indicated). By comparing the total cumulative net gas inflow of all the different ISM phases (black dashed line), with the total amount of mass accreted by the MBH (black solid line), a conversion factor for how much of the total cumulative net gas inflow is contributed to the growth of the MBH is estimated (black dash-dotted line), for each run. Based on this, the conversion factor is the highest in the runs without SNII feedback included ($\sim 40~\%$), followed by the SNII+ISR feedback run ($\sim 20~\%$), and the least in the SNII-only feedback run ($\sim 10\%$). This agrees well with the gravitational binding energy being more comparable to the thermal and kinetic energy of the inflowing and outflowing gas close to the MBH when SNII feedback is included, compared to when no SNII feedback is considered (Fig.~\ref{fig:time_int_inflow_outflow}, panel e), thus making it more difficult for the inflowing gas to fulfil the accretion criteria.

Figure~\ref{fig:cummulative_inflow_outflow} also clearly shows that the total amount of cumulative net gas inflow is significantly higher without SNII feedback (top row), compared to with SNII feedback (bottom row) included. The highest total cumulative net-inflow is achievable in the ISR-only feedback run (top right), followed by the no stellar feedback run (top left). For both runs is the total cumulative net gas inflow primarily driven by a strong net gas inflow of the CNM+CMM phase, however, for the ISR-only feedback run, a substantial net gas inflow of the UNM phase (and to some extent the WNM phase) can be seen as well (in agreement with Fig.~\ref{fig:time_evol_inflow_outflow}, top right panel~a and~b). The same applies to the runs where SNII feedback is included, that is, with the cumulative net gas inflow of the CNM+CMM phase strongly suppressed, the cumulative net gas inflows of the UNM and WNM phases are of great importance to grow the MBH \citep[in agreement to][]{Partmann2025}. Even though the overall gas accretion onto the MBH itself is suppressed by a lower total cumulative net gas inflow, compared to the no SNII feedback runs. 

\begin{figure}
    \resizebox{\hsize}{!}{\includegraphics{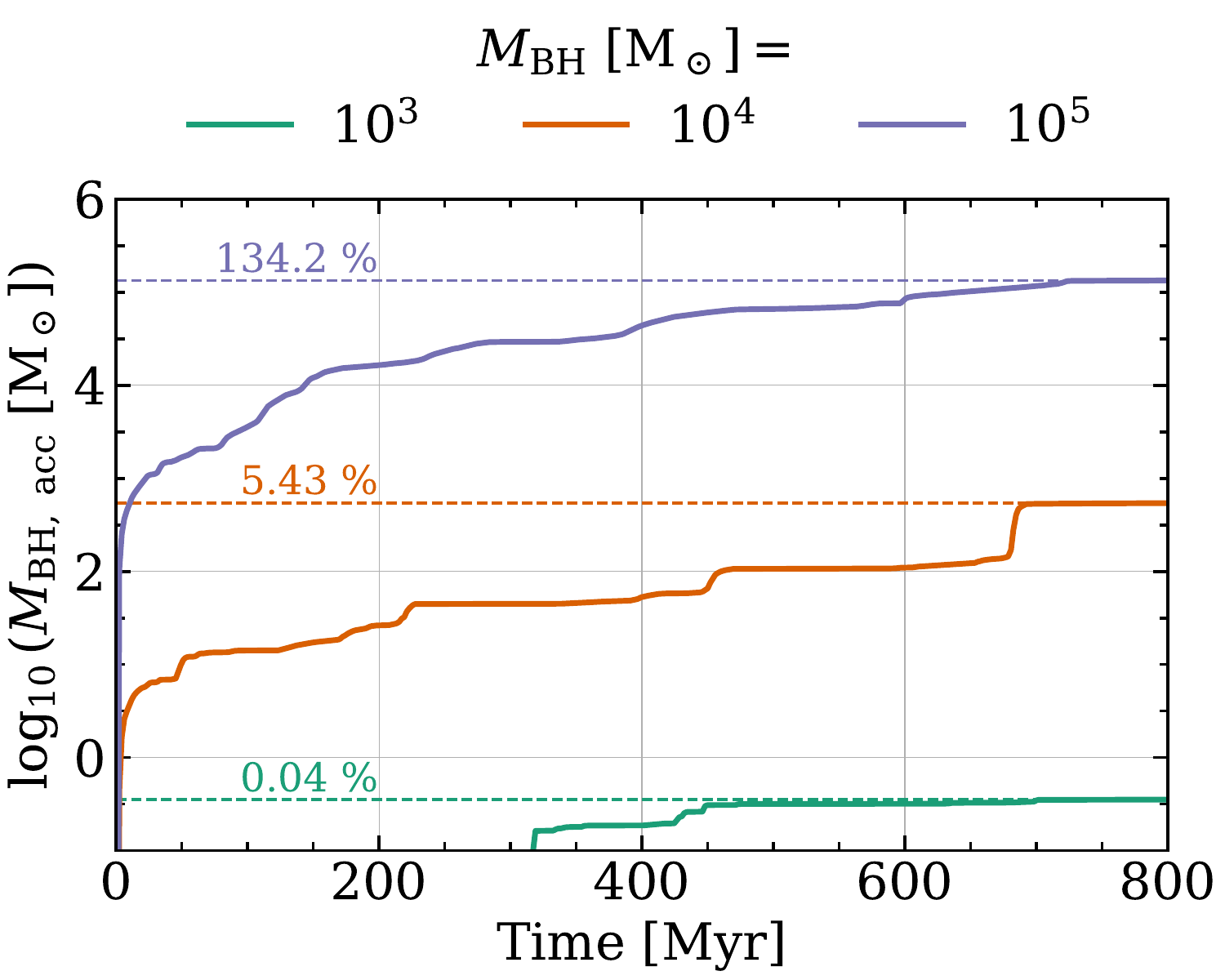}}
    \caption{Amount of mass accreted over time by MBHs of different initial masses. For the initial $10^4 \Msol$ MBH, the \texttt{MBH4\_SNII} is displayed, while for the initial $10^3 \Msol$ and $10^5 \Msol$ MBHs, the \texttt{MBH3} and \texttt{MBH5} simulations in the `\textit{Black Hole Mass}' subset (see Table~\ref{tab:simulations}) are respectively shown. Coloured dashed lines indicate the percentage growth of each MBH (with respect to their initial masses).}
    \label{fig:bh_growth_initial_mass}
\end{figure}

Lastly, in Fig.~\ref{fig:cumulative_star_formation} we check on the time evolution of the star formation rate (SFR, top row), the cumulative amount of stars formed (middle row), and the cumulative amount of energy injected into the ISM from SNII explosions (bottom row), in our `\textit{Stellar Feedback}' simulations (different coloured lines), both on a global scale ($R < 2.5 \kpc$, left column) and close to the MBH ($R < 200 \pc$, right column). The SFR is the lowest when SNII feedback is combined with ISR feedback, resulting in the least amount of stars formed both on a global scale throughout the low-mass dwarf galaxy, and close to the MBH. Consequently, the cumulative amount of energy injected into the ISM from SNII explosions is also the smallest (on both scales). Based on our analysis above, where we show that nearby SNII explosions are able to suppress the gas accretion onto the MBH, this explains the slightly higher total cumulative net gas inflow and conversion factor in the SNII+ISR feedback run in Fig.~\ref{fig:cummulative_inflow_outflow} (bottom right), compared to the SNII-only feedback run (bottom left), and hence the slight increase in MBH growth seen in Fig.~\ref{fig:bh_growth} (red curve). For the runs without SNII feedback included, the SFR is by far the highest in the no stellar feedback run, while in the ISR-only feedback run, it is instead comparable or slightly higher than in the SNII-only feedback run \citep[in agreement to e.g.][]{Rosdahl2015}. The risk of depleting the gas via star formation is therefore significantly higher when no stellar feedback is considered (depletion timescale of $\sim 1 \Gyr$), compared to when either (or both) SNII and ISR feedback is included (depletion timescales of $\gtrsim 10 \Gyr$). Thus, in the ISR-only feedback run, the larger gas content (due to the longer depletion time compared to the no stellar feedback run) is most likely a contributing factor to the higher total cumulative net gas inflow seen in Fig.~\ref{fig:cummulative_inflow_outflow}, making the MBH grow the most among all our `\textit{Stellar Feedback}' simulations.

\subsection{Gas accretion onto under- and overmassive black holes}
\label{sect:bh_initial_mass}

\begin{figure}
    \resizebox{\hsize}{!}{\includegraphics{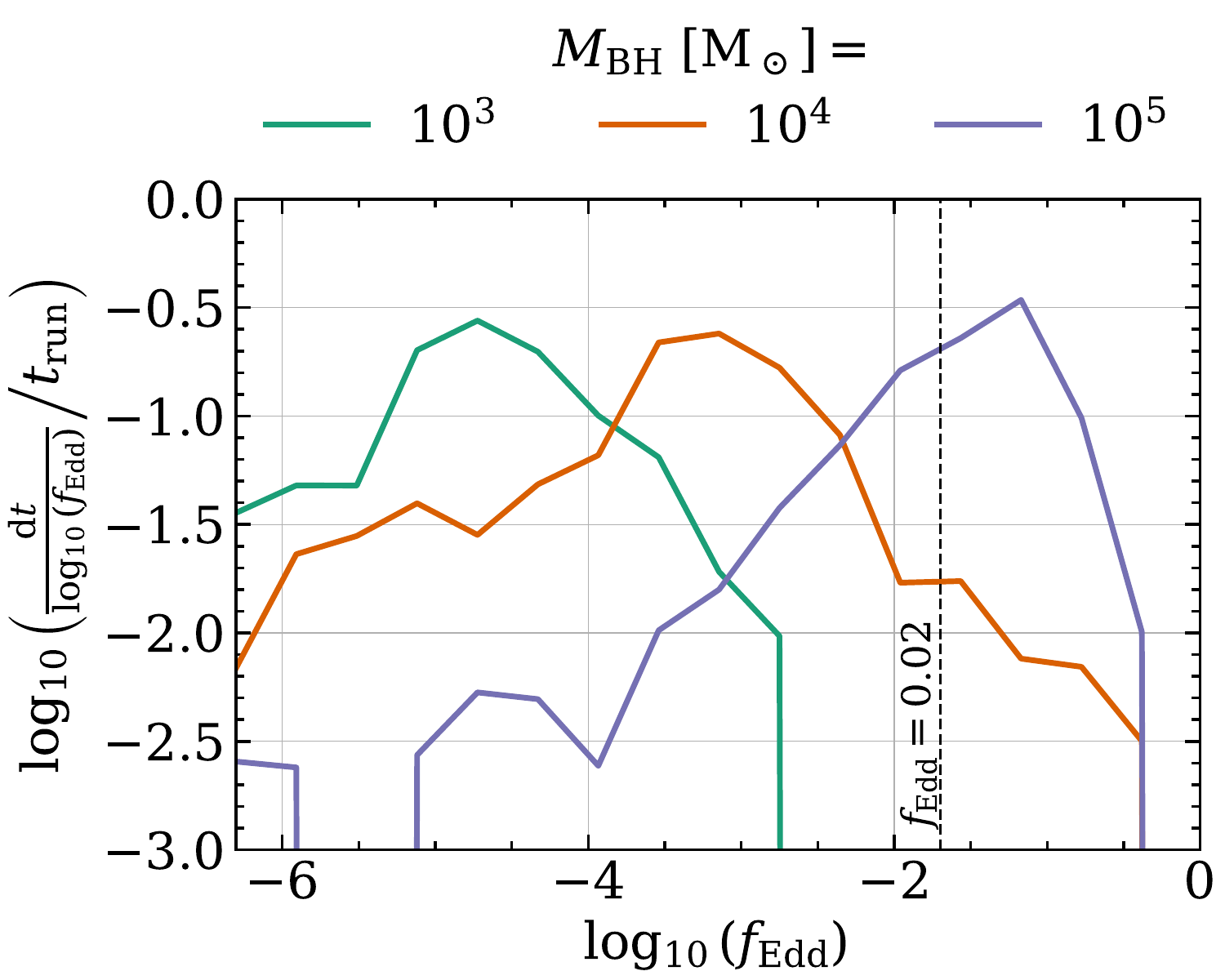}}
    \caption{Fraction of time spent at various Eddington fractions ($f_\mathrm{Edd} = \dot{M}_\mathrm{BH}/\dot{M}_\mathrm{Edd}$) for MBHs of different initial masses (same simulations as in Fig.~\ref{fig:bh_growth_initial_mass}). As a reference, the limit for the radiatively efficient regime $f_\mathrm{Edd} = 0.02$ (see Sect.~\ref{sect:bh_accretion_sub_grid} for more details) is indicated by a black dashed line.}
    \label{fig:eddington_fraction_dist}
\end{figure}

Based on our analysis above, SNII feedback seems to be the limiting factor in regulating MBH growth. To assess the validity of this for MBHs of different initial masses, in Fig.~\ref{fig:bh_growth_initial_mass}, we show the amount of mass accreted over time by an initial under- and overmassive BH ($10^3$ and $10^5 \Msol$ respectively), assuming the \texttt{MBH4\_SNII} simulation of an initial $10^4 \Msol$ MBH to be a reference point. To spare the computational cost, ISR feedback is not considered, since it only marginally boosts the gas accretion onto the MBH when combined with SNII feedback (according to our analysis in Sect.~\ref{sect:bh_growth})\footnote{We note that the gas accretion onto an undermassive BH with a shallow potential well might be more easily affected by early stellar feedback processes, than onto an overmassive BH with a deep potential well. However, this type of parameter exploration goes beyond the scope of this paper, and we therefore leave it for a future study instead.}. The initial $10^5 \Msol$ MBH does not only accrete more mass compared to our initial $10^4 \Msol$ MBH, but also grows more with respect to its initial mass ($> 130~\%$ compared to $\sim 5~\%$ for our initial $10^4 \Msol$ MBH). The opposite can be seen for the initial $10^3 \Msol$ MBH, which accretes $< 1 \Msol$ (corresponding to only $\sim 0.04~\%$ of its initial mass). Hence, SNII feedback seems to be more effective in suppressing gas accretion as the MBH initial mass decreases (due to the weaker gravitational potential).

In Fig.~\ref{fig:eddington_fraction_dist}, we show the fraction of time spent accreting at various Eddington fractions for each MBH of different initial masses in Fig.~\ref{fig:bh_growth_initial_mass}. The distribution of Eddington fractions are distinctly different between the runs, with for example only the initial $10^5 \Msol$ MBH peaking in the radiatively efficient regime ($f_\mathrm{Edd} > 0.02$). As the initial mass of the MBH decreases, so does the peak of the distribution. For the initial $10^3 \Msol$ MBH, this goes to the extreme with $f_\mathrm{Edd} \sim 10^{-5}$, completely halting the growth of the MBH as seen in Fig.~\ref{fig:bh_growth_initial_mass}.

\section{Discussion}
\label{sect:discussion}

Having shown in Sect.~\ref{sect:results} our main results from the \noctua suite of simulations, in this section we continue by discussing the numerical caveats of our study (Sect.~\ref{sect:simulation_caveats}), and present the results on massive black hole (MBH) growth when artificial gas inflows are induced, mimicking the effect of strong gas inflows potentially happening in a cosmological setting (Sect.~\ref{sect:art_inflows}). We then contextualise our results to the study of \citet{Partmann2025}, at present the only comparable study to our work, and discuss the implications it may have for MBH growth in the low-mass dwarf galaxy regime (Sect.~\ref{sect:comparison_studies}).

\subsection{Simulation Caveats}
\label{sect:simulation_caveats}

\subsubsection{Resolution}
\label{sect:caveats_resolution}

One of the major limitations when simulating the physical interplay between MBHs and their host galaxies, is the extreme dynamical range required. Trying to resolve the gas accretion flow from galactic scales (several kpc), all the way down to the Schwarzschild radius of a $10^4 \Msol$ MBH ($\sim 10^{-9} \pc$), is computationally infeasible at the moment. That said, some studies try to overcome this via `hyper-refinement techniques' \citep{Hopkins2024}, but at the expense of only being able to simulate supermassive BHs ($\sim 10^7 \Msol$) for $\sim 10^4 \yr$ at the highest spatial resolution ($10^{-5}-10^{-4} \pc$). However, for the purpose of this study, where a time evolution of at least a few 100 Myr is needed to properly investigate the impact of stellar feedback on MBH growth, a sub-grid treatment for the gas accretion onto the MBH below the resolution limit is therefore still needed. 

In our adopted MBH gas accretion model (see Sect.~\ref{sect:bh_accretion}), the MBH is only allowed to skim mass of individual gas cells inside its accretion radius $r_\mathrm{acc}$. At smaller scales, the dynamics of the gas is too uncertain to accurately follow the gas inflow towards the MBH. The value of $r_\mathrm{acc}$ is thus essentially determined by the highest spatial resolution achievable for a resolved gas accretion flow onto the MBH. With a target gas mass resolution of $20 \Msol$ and sub-parsec maximum spatial resolution (see Fig.~\ref{fig:resolution} in Appendix~\ref{app:resolution}), $r_\mathrm{acc}$ is therefore set to 1 pc throughout the full suite. At this scale, the gas accretion onto the MBH is well converged (as demonstrated in Fig.~\ref{fig:bh_growth_acc_radius} of Appendix~\ref{app:resolution}), and agrees well with \citet{Partmann2025}, who despite using a different hydrodynamic and physical framework, also find convergence at $\leq 1 \pc$ for MBH accretion in a similar dwarf galaxy set-up. 

Even though the dedicated refinement strategy adopted in Sect.~\ref{sect:refinement} ensures that the accretion region ($r < r_\mathrm{acc}$) of the MBH is always resolved by a few tens of gas cells, the global target mass of gas cells can still influence the gas accretion onto the MBH -- either via gas inflows from galactic scales, or indirectly via star formation and stellar feedback. Nevertheless, our resolution study presented in Fig.~\ref{fig:bh_growth_resolution} of Appendix~\ref{app:resolution}, illustrates that a target gas mass of $20 \Msol$ is enough to reach convergence for MBH growth (in the current numerical framework).

\subsubsection{Black hole physics}
\label{sect:caveats_bh_physics}

Since the main purpose of this work is to study MBH growth in the presence of stellar feedback due to type II supernova (SNII) explosions and ionising stellar radiation (ISR) from OB stars, no feedback from the MBH is considered (this is instead deferred to a forthcoming study). The obtained gas accretion rate in each simulation should therefore be treated as an `upper limit' on how much the MBH can potentially grow in that respective set-up. 

We note that the sub-grid model for gas accretion onto the MBH is a simplified version of reality. For the radiatively efficient regime ($f_\mathrm{Edd} > 0.02$), the time for which gas is transported from the accretion radius $r_\mathrm{acc}$ down to the circularisation radius $r_\mathrm{circ}$ is most likely more complex than assuming a free-fall timescale. However, it could be argued that this is compensated by allowing $r_\mathrm{circ}$ to vary with the angular momentum of the infalling gas (and by such altering the viscous timescale in Eq.~\ref{eq:viscous_timescale}), but this introduces other uncertainties, such as the efficiency factor for angular momentum transport between $r_\mathrm{acc}$ and $r_\mathrm{circ}$ (see Appendix~\ref{app:rcirc} for more details). In addition to that, no time-lag effects are considered at all for the radiatively inefficient regime ($f_\mathrm{Edd} < 0.02$). \citet{Koudmani2024} recently presented an `unified' accretion disc model where the unresolved gas accretion disc of the MBH is explicitly modelled for the radiatively inefficient regime in great detail. While it would be certainly interesting to investigate by how much such a detailed gas accretion scheme compares to our results, this clearly goes beyond the scope of our study. We speculate though that our main results regarding how stellar feedback impacts the gas inflow towards the MBH would hardly change, since a more detailed sub-grid accretion scheme most seemingly would only delay the gas accretion, but not change the total amount of gas accreted by the MBH. This could, however, change with the inclusion of active galactic nucleus (AGN) feedback, which very likely also will be delayed in time (and adjusted in strength). During this time delay, existing gas structures from the accretion event may have time to evolve, suggesting that the anticipated AGN feedback might interact with the nearby gas in a sightly different way (compared to if the AGN feedback is instantaneously injected into the ISM), possibly boosting or suppressing further gas accretion onto the MBH.

\subsubsection{Missing galaxy physics}
\label{sect:caveats_stellar_feedback_models}

In this section, we discuss some of the missing physical processes in our low-mass dwarf galaxy set-up, starting with stellar feedback. Assuming that other early stellar feedback processes may have a similar effect on MBH growth as the ISR feedback (in e.g.\ enhancing the growth of the MBH compared to no stellar feedback, see Sect.~\ref{sect:bh_growth}), stellar winds could potentially therefore have a positive influence on MBH gas accretion (even though at $0.1 \ Z_\odot$ stellar winds are expected to be less significant than e.g. in the Milky Way). Not because of the additional injection of mass (our system is already very gas rich), but because of the early dispersion of giant molecular clouds, suppressing star formation and therefore the number of SNII (which we show in Sect.~\ref{sect:results} is a strong suppressor of MBH growth). Other stellar feedback processes that may have an influence on the MBH gas accretion include: type Ia supernovae (even though the rate at which they occur in our low-mass dwarf galaxy model is expected to be very small due to the relatively low stellar mass content), cosmic rays (CR), and photoelectric heating (PE). In the current physical framework, PE is included as part of the spatially constant interstellar radiation field, and hence does not properly capture the effects close to regions of active star formation. \citet{Hu2017} recently demonstrated that spatially and temporally varying PE heating alone can be as efficient as SNII in suppressing star formation, and would thus be interesting to further explore, but in the context of MBH gas accretion. Since radiation pressure is not included in the current implementation of ISR feedback (see Sect.~\ref{sect:ionising_stellar_radiation} for more details), this also opens up an additional feedback channel potentially important for regulating MBH growth. To address some of these missing aspects, we aim to conduct an additional suite of simulations, where for example the impact of CRs on MBH gas accretion is investigated. \citet{Farcy2022} find that CRs can reduce the star formation rate in simulated dwarf galaxies (as well as the clumpiness of star formation), which based on our analysis in Sect.~\ref{sect:results}, would naively favour further MBH growth (due to a lower rate of SNII explosions). That said, CRs might also be able to drive cold gas out from the galaxy, and thus reduce the gas accretion onto the MBH. 

We also note that magnetic fields are not included in our suite of simulations. On a global scale, \citet{Whitworth2023} find that magnetic fields in low-metallicity dwarf galaxies are unable to suppress star formation, thus, based on our analysis in Sect.~\ref{sect:results}, we do not expect magnetic fields to have any significant effect on the MBH gas accretion. However, magnetic fields may be able to remove angular momentum from the inflowing gas and deposit it further out, thus enhancing further MBH growth. On smaller scales, magnetic fields are expected to be more important, by stabilising the gas from collapsing and forming stars due to the magnetic field pressure. This is especially true as we get close to the accretion disc of the MBH \citep{Hopkins2024}, but this is far from achievable in the simulations of this work. Still, to investigate the role of magnetic fields in MBH gas accretion (in the low-mass dwarf galaxy regime), we aim to conduct high-resolution radiation-magneto-hydrodynamic simulations in a future study.

\subsubsection{The dwarf galaxy environment}
\label{sect:caveats_environment}

The simulations of our low-mass dwarf galaxy system are conducted in an isolated environment, allowing us to run multiple high-resolution simulations of various physical set-ups (due to their relatively low computational cost compared to cosmological runs), and isolate the internal physical processes of the low-mass dwarf galaxy from external events and study their individual and cumulative impact on MBH gas accretion. However, by construction then, the system lacks the cosmic environment needed for continuous gas inflows from the cosmic web, dynamical interactions with other galaxies, and mergers between BHs. Even though it remains unclear, these missing processes could be of great importance for MBH growth, especially at high redshift. For instance, by the use of cosmological zoom-in simulations, a handful of recent studies have been able to infer sustained super-Eddington gas accretion onto MBHs of $10^{4-5} \Msol$ at high redshift $(z \gtrsim 7)$, growing the MBH by several orders of magnitudes in $<$ a few 100 Myr \citep[see e.g.][]{Lupi2024, Husko2025}\footnote{However, see also \citet{Massonneau2023}, who is able to demonstrate super-Eddington gas accretion onto a $10^6 \Msol$ MBH in an isolated low-mass galaxy at redshift $z=4$}, but typically in more massive systems than our simulated low-mass dwarf galaxy. That said, idealised custom-made simulations (with varying resolution) of giant molecular clouds \citep{Shi2023}, and extremely gas-rich low-mass galaxies \citep{Mehta2024}, both meant to represent possible high-redshift environments, have also been able demonstrate substantial MBH growth, enabling BH seeds of $\sim 1 - 10^4 \Msol$ to even grow into supermassive BHs. But only for a limited time evolution of $\leq 10 \Myr$, not enough to fully capture the impact of stellar feedback on MBH growth, and without any feedback from the BHs themselves.

\subsection{The effect of artificially induced gas inflows on massive black hole growth}
\label{sect:art_inflows}

\begin{figure}
    \resizebox{\hsize}{!}{\includegraphics{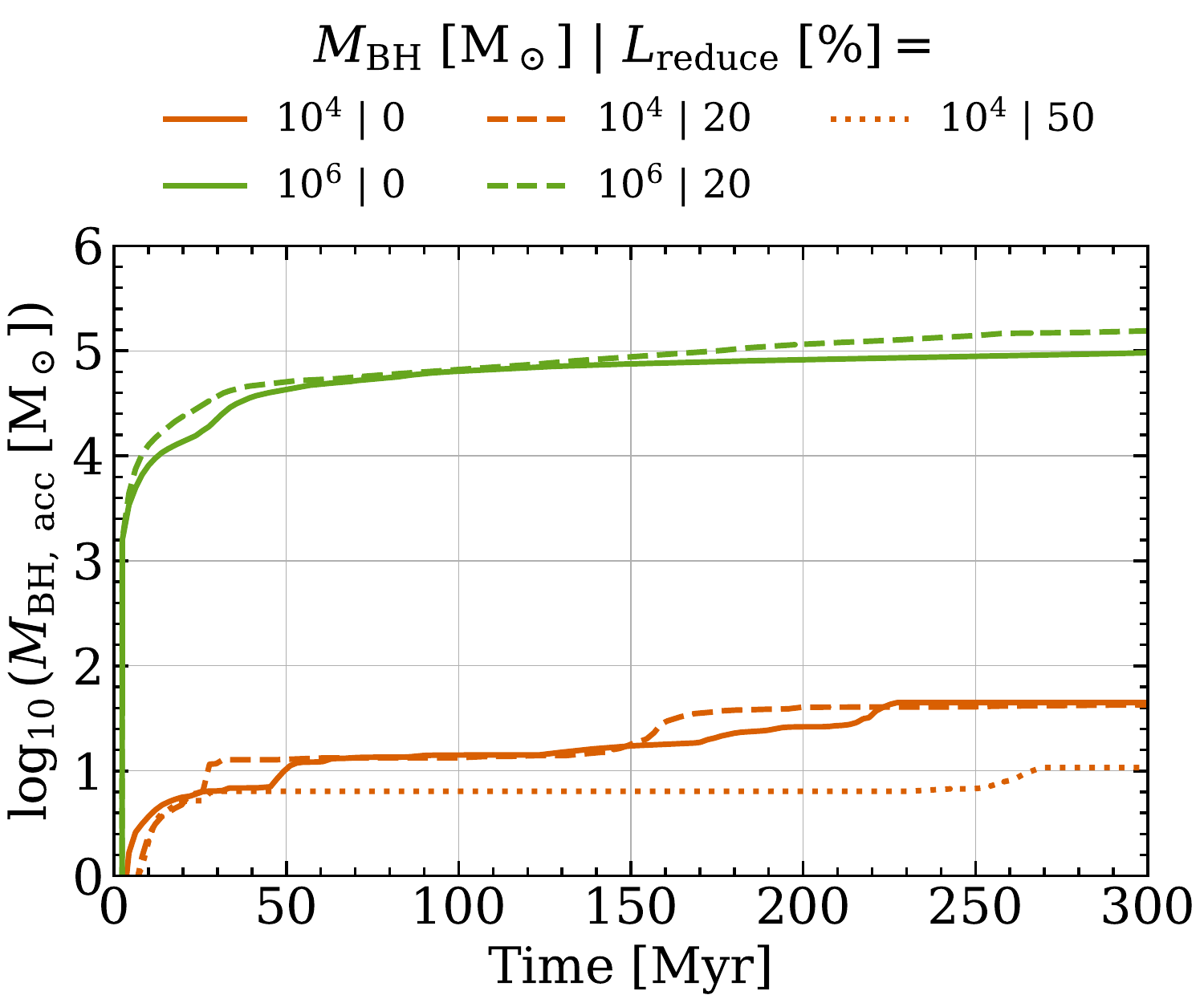}}
    \caption{Amount of mass accreted over time for an initial $10^4 \Msol$ MBH (orange lines) with and without uniformly reducing the angular momentum of the gas in the low-mass dwarf galaxy by $20$ and $50$ \% (\texttt{MBH4\_L20} and \texttt{MBH4\_L50} in Table~\ref{tab:simulations} respectively). No ISR feedback is considered. The simulation of an initial $10^{4} \Msol$ MBH with $L_\mathrm{reduce} = 0 \ \%$ corresponds to the \texttt{MBH4\_SNII} simulation (see Table~\ref{tab:simulations}). A pair of simulations of an initial $10^6 \Msol$ MBH (green lines) with and without $L_\mathrm{reduce} = 20 \ \%$ is also conducted (\texttt{MBH6\_L20} and \texttt{MBH6} in Table~\ref{tab:simulations} respectively).}
    \label{fig:bh_growth_art_inflow}
\end{figure}

To explore the effect of substantial gas inflows potentially happening in a cosmological setting (especially at high redshift) on MBH growth, for example by continuous gas inflows from the cosmic web or galaxy mergers \citep[see e.g.][]{Mayer2019, Zwick2023, Mayer2024}, we conduct a subset of simulations where the angular momentum of the gas is uniformly reduced by a certain percentage $L_\mathrm{reduce}$ across the gaseous disc of the low-mass dwarf galaxy (i.e. the `\textit{Artificial Inflows}' simulations in Table~\ref{tab:simulations}). In Fig.~\ref{fig:bh_growth_art_inflow}, we show the amount of mass accreted over time by an initial $10^4 \Msol$ MBH (orange lines) when $L_\mathrm{reduce}$ is equal to 20 and 50 \% (dashed and dotted line respectively), and compare this to the fiducial set-up where $L_\mathrm{reduce} = 0 \ \%$, more specifically, the \texttt{MBH4\_SNII} simulation in Table~\ref{tab:simulations} (ISR feedback is not considered). As illustrated in both Fig.~\ref{fig:bh_growth} and~\ref{fig:bh_growth_initial_mass}, the change in MBH growth for different physical set-ups is typically already clearly visible in the first few 100 Myr. To save computational time, we therefore only run our simulations of artificial gas inflows in Fig.~\ref{fig:bh_growth_art_inflow} for 300 Myr. No clear difference in MBH growth can be seen for the initial $10^4 \Msol$ MBH, with only marginally less mass accreted when $L_\mathrm{reduce} = 50 \%$. We explain this via an increased star formation rate in the galactic centre, subsequently boosting the number of SNII occurring, suppressing the gas accretion onto the MBH despite the substantial gas inflow. A similar pair of simulations for an initial $10^6 \Msol$ MBH with and without $L_\mathrm{reduce} = 20 \%$ is conducted as well. As expected, the MBH of $10^6 \Msol$ grows substantially more than its $10^4 \Msol$ counterpart, but once again without any clear difference between the runs with and without artificial inflows. This suggests that the gravitational potential around the MBH is most vital in determining the gas accretion, while larger-scale inflows seem to be of minor importance. However, we want to emphasise that these simulations are a numerical experiment, and that the impact of gas inflows in a cosmic environment most likely is more nuanced. We therefore defer a more in-depth investigation of this to future a study.

\subsection{Comparison to Partmann et al. (2025)}
\label{sect:comparison_studies}

The most comparable study to our work is \citet{Partmann2025}, and their study on the importance of nuclear star clusters (NSC) for MBH growth in the low-mass dwarf galaxy regime. Both their and our low-mass dwarf galaxy are very comparable in mass, but with the difference that \citet{Partmann2025} already have a pre-defined stellar disc in their initial conditions. However, despite this difference, and that fact that they are using a different hydrodynamic and physical framework, our results agree remarkably well. For instance, an initial $10^{4} \Msol$ MBH grows $\sim 30-40~\%$ of its initial mass in 600~Myr (without a NSC) in their study, which is slightly less but still very comparable to our initial $10^{4} \Msol$ MBH in the $\texttt{MBH4\_full}$ simulation (see Fig.~\ref{fig:bh_growth}). \citet{Partmann2025} also find that the gas accretion onto the MBH is episodic over time, composed of both warm and cold gas, and strongly regulated by nearby SNII explosions. Moreover, they also find that lower-mass MBHs ($10^{2-3} \Msol$) cannot grow in the absence of an NSC, and that only a higher-mass MBH ($10^{5} \Msol$) can accrete larger amount of gas. This agrees very well with our results from Sect.~\ref{sect:bh_initial_mass}. That said, we note that our initial $10^{5} \Msol$ MBH grows more than in the equivalent set-up of \citet{Partmann2025} ($> 100~\%$ compared to $\sim 10\%$ respectively). We speculate that this discrepancy could be due to higher star formation rates in the centre of the dwarf galaxy simulated by \citet{Partmann2025}, possibly due to the higher gas mass resolution ($4$ versus $20 \Msol$ in our work) and the different star formation prescription adopted, and thus a higher number of nearby SNII explosions, hindering the gas accretion onto the MBH. Regardless, the overall conclusion of both studies remain the same, that is, it is primarily the nearby gravitational potential around the MBH (either from an NSC or directly from an overmassive BH) that determines whether or not a MBH can grow via gas accretion over time (at least in an isolated low-mass dwarf galaxy environment).

\section{Conclusions}
\label{sect:conclusions}

We conduct a suite of high-resolution radiation-hydrodynamic simulations of MBH gas accretion in a low-mass dwarf galaxy, which we term the \noctua suite of simulations. These are carried out by introducing and applying the \areponoctua numerical framework, where the chemical evolution of the gas is explicitly modelled in a time-dependent non-equilibrium way. As part of this framework, we develop and apply a novel physically-motivated model for BH gas accretion, taking into account the angular momentum of the gas in the radiatively efficient regime, to estimate the gas accretion rate from the sub-grid accretion disc onto the MBH. We then study the individual and cumulative impact of stellar feedback on the growth of an initial $10^4 \Msol$ MBH, by gradually adding stellar feedback processes to the numerical set-up. This includes individually-traced type II supernova (SNII) explosions, and radiatively transferred (on-the-fly) ionising stellar radiation from OB stars. Furthermore, we explore the influence of the MBH initial mass ($10^{3-6} \Msol$) on the gas accretion, and whether or not artificially induced gas inflows (mimicking strong gas inflows potentially happening in a cosmological setting and possibly at high redshift) can `boost' the gas accretion onto the MBH. Our main findings are as follows:

\begin{enumerate}[(i)]
    \item Without any stellar feedback, a large mass fraction of the interstellar medium (ISM) reside in the cold neutral- and molecular medium (CNM+CMM) phases. This mass fraction increases as we move radially inward towards the centre of the low-mass dwarf galaxy and the MBH. There, a high cumulative net gas inflow of the CNM+CMM phase enables the MBH to grow at a steady rate, roughly doubling its initial mass after 800~Myr. 
    \item With only ISR feedback, the star formation rate (SFR) of our low-mass dwarf galaxy drops to the same level (or slightly above) as that when SNII feedback is considered. Besides an overall higher total cumulative net gas inflow, primarily driven by the CNM+CMM phase, substantial net gas inflows of the unstable- and warm neutral medium (UNM and WNM respectively) phases also contribute to the growth of the MBH. This allows the MBH to grow at a faster rate than in the no stellar feedback run, reaching a final mass of around five times its initial mass after 800~Myr. This is the fastest rate of growth found in any of our simulations of an initial $10^{4} \Msol$ MBH.
    \item With only SNII feedback, the growth of the MBH is suppressed by more than an order of magnitude compared to the no stellar feedback run. By including SNII feedback, the mass fraction of the WNM phase increases throughout the entire low-mass dwarf galaxy, at the expense of mainly the CNM+CMM phase (compared to when no SNII feedback is included). The energy injection from nearby SNII explosions disrupts the gas inflow towards the MBH, making the gravitational binding energy of the gas comparable to or smaller than that of its thermal and kinetic energy. Thus, reducing the likelihood of the inflowing gas fulfilling the accretion criteria of the MBH, that is, be gravitationally bound to the MBH and have a converging flow. Consequently, the gas inflow and therefore the MBH gas accretion rate is more episodic over time, regulated by either a few very nearby (10s of~pc) SNII explosions, or a high number relatively nearby ($\sim 100 - 200 \pc$) SNII events. 
    \item When SNII feedback is combined with ISR feedback, the growth of the MBH remains suppressed compared to the no stellar feedback run. However, compared to the SNII-only feedback run, the MBH grows slightly more rapidly. This is due to the suppression of star formation by the early radiative feedback, which leads to a lower rate of SNII explosions, enabling a higher total cumulative net gas inflow onto the MBH, and a higher probability for the inflowing gas to be gravitationally bound to the MBH and have a converging flow.
    \item Taking the $10^{4} \Msol$ MBH as a reference point, we find that an overmassive BH of $10^5 \Msol$ grows more in terms of both the total amount of mass accreted, and with respect to its initial mass. On the other hand, the opposite is found for an undermassive BH of $10^3 \Msol$, which barely is able to grow at all within a time period of 800~Myr. 
    \item Lastly, we find that artificially induced gas inflows towards the centre of the low-mass dwarf galaxy are unable to `boost' further gas accretion onto the MBH, even for a $10^6 \Msol$ MBH. This suggests that it is primarily the nearby gravitational potential around the MBH that determines how much the MBH can grow via gas accretion over time, in the presence of SNII feedback (at least in an isolated non-cosmological environment). 
\end{enumerate}

Lastly, we note that this work is the first in an upcoming series of studies, aiming to get a deeper understanding of BH growth over cosmic time. Even though our simulated low-mass dwarf galaxy is not meant to represent the conditions expected at high redshift, it still provides valuable information to what environments and/or physical processes at high redshift, that might be more or less favourable for MBH growth. With this knowledge, in future work we aim to extend the \noctua numerical framework, and conduct high-resolution cosmological zoom-in simulations of BHs in (low-mass) galaxies. We also plan to include and investigate the impact of AGN winds (Farcy et al., submitted) and radiation on MBH growth at early epochs, as well as to accurately model emission lines not only from HII regions around young star clusters, but also from resolved narrow-line regions around BHs to enable a meaningful comparison with spectroscopic observations such as from the James Webb Space Telescope \citep[going beyond previous studies of][]{Hirschmann17, Hirschmann19, Hirschmann23}.

\begin{acknowledgements}
    JP credits Patrick Hirling (Institute of Physics, EPFL) and Darwin Roduit (Department of Astronomy, University of Geneva) for the name \noctua. JP thanks Feng Yuan (Center for Astronomy and Astrophysics, Fudan University) and Yaping Li (Shanghai Astronomical Observatory, CAS) for sharing their insight on BH gas accretion theory. JP, MH and MF acknowledge funding from the Swiss National Science Foundation (SNF) via a PRIMA grant PR00P2 193577 `From cosmic dawn to high noon: the role of BHs for young galaxies'. SCOG, RSK and DJW acknowledge funding from the European Research Council via the ERC Synergy Grant ``ECOGAL'' (project ID 855130) and from the German Excellence Strategy via the Heidelberg Cluster of Excellence ``STRUCTURES'' (EXC 2181 - 390900948). RSK also thanks for support from the German Ministry for Economic Affairs and Climate Action in project ``MAINN'' (funding ID 50OO2206). RSK is grateful for computing resources provided by the Ministry of Science, Research and the Arts (MWK) of the State of Baden-W\"{u}rttemberg through bwHPC and the German Science Foundation (DFG) through grants INST 35/1134-1 FUGG and 35/1597-1 FUGG, and also for data storage at SDS@hd funded through grants INST 35/1314-1 FUGG and INST 35/1503-1 FUGG. In addition, RSK thanks the Harvard-Smithsonian Center for Astrophysics and the Radcliffe Institute for Advanced Studies for their hospitality during his sabbatical, and the 2024/25 Class of Radcliffe Fellows for highly interesting and stimulating discussions. TN acknowledges support from the Deutsche Forschungsgemeinschaft (DFG, German Research Foundation) under Germany's Excellence Strategy - EXC-2094 - 390783311 from the DFG Cluster of Excellence "ORIGINS”. DJW acknowledges support from the Programa de Becas Posdoctorales of the Direcci\'on General de Asuntos del Personal Acad\'{e}mico of the Universidad Nacional Aut\'{o}noma de M\'{e}xico (DGAPA, UNAM, Mexico).
\end{acknowledgements}

\bibliographystyle{aa}
\bibliography{references}

\begin{appendix}

\section{Variable circularisation radius}
\label{app:rcirc}

The idea for a variable circularisation radius that is dependent on the specific angular momentum of the skimmed gas is inspired by the detailed one-dimensional accretion disc sub-grid model of \citet{Tartenas2022}, where individual gas particles are placed in a sub-grid accretion disc at different radii depending on their specific angular momentum. For our study however, a simpler prescription is adopted.

Whenever an accretion event in the radiatively efficient regime (i.e. when $\dot{M}_\mathrm{BH} > 0.02 \ \dot{M}_\mathrm{Edd}$) occurs, the specific angular momentum transfer rate from the massive black hole (MBH) accretion radius $r_\mathrm{acc}$ to the gas accretion disc is estimated as

\begin{equation}
    \langle \dot{J}_\mathrm{in} \rangle = \beta_\mathrm{ff} \frac{\langle J_\mathrm{in} \rangle}{\tau_\mathrm{ff}}, 
    \label{eq:angular_momentum_transport}
\end{equation}

\noindent where $\langle J_\mathrm{in}\rangle$ is the mass-weighted mean specific angular momentum of the newly skimmed gas, $\tau_\mathrm{ff}$ is the free-fall timescale from $r_\mathrm{acc}$ down to $r_\mathrm{circ}$ (same as in Eq.~\ref{eq:effective_accretion_rate} in Sect.~\ref{sect:bh_accretion_sub_grid}), and $\beta_\mathrm{ff}=0.1$ is an efficiency factor per free-fall timescale (whose value is discussed below). The expected circularisation radius of the newly skimmed gas follows as

\begin{equation}
    r_\mathrm{circ, \ in} = \frac{(\langle \dot{J}_\mathrm{in} \rangle \Delta t)^2}{GM_\mathrm{BH}}, 
\end{equation}

\noindent with $\Delta t$ being the current timestep of the simulation. Since the MBH can already have an active accretion disc with a well-defined mass and size from a previous accretion event, the circularisation radius $r_\mathrm{circ}$ is updated by taking the mass-weighted mean value as its new value:

\begin{equation}
    r_\mathrm{circ} = \frac{M_\mathrm{disc}r_\mathrm{circ} + M_\mathrm{in}r_\mathrm{circ, \ in}}{M_\mathrm{disc} + M_\mathrm{in}}, 
\end{equation}

\noindent where $M_\mathrm{dg}$ is the mass of the accretion disc and $M_\mathrm{in}$ is the mass of the newly skimmed gas. 

\begin{figure}
    \resizebox{\hsize}{!}{\includegraphics{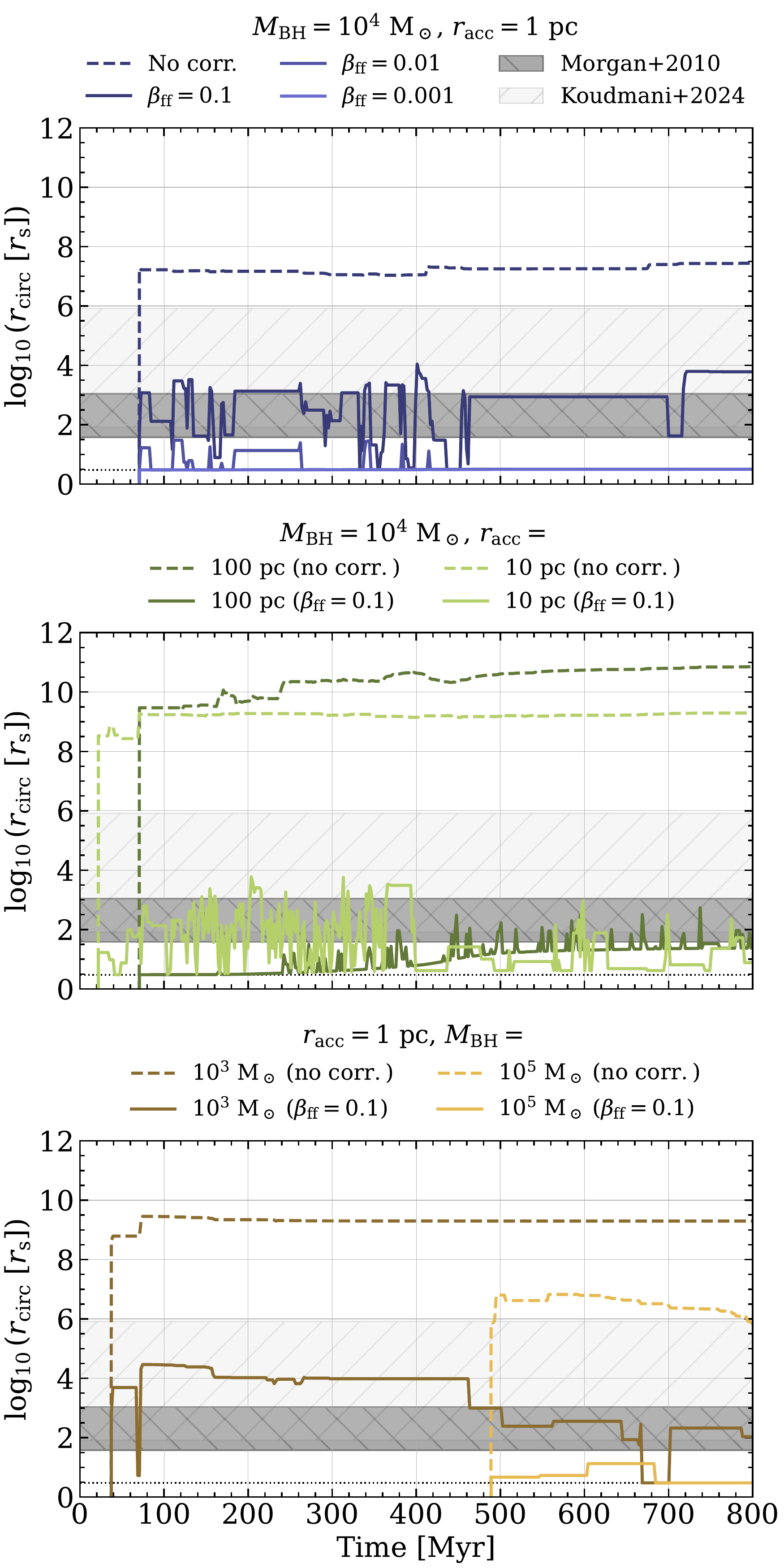}}
    \caption{Time evolution of $r_\mathrm{circ}$ in a set of simulations conducted in a low dark matter resolution set-up of our low-mass dwarf galaxy (see Table~\ref{tab:initial_conditions}). \textit{Top}: The efficiency factor $\beta_\mathrm{ff}$ (see Eq.~\ref{eq:angular_momentum_transport}) is introduced and varied between 0.1 and 0.001 for a MBH of $M_\mathrm{BH} = 10^4 \Msol$ and $r_\mathrm{acc} = 1 \pc$. The dark-grey shaded area corresponds to the observational scaling relation of \citet{Morgan2010}, while the light-grey shaded area represents the theoretical range for a $10^{1-3} \Msol$ thin disc \citep[Eq. 9 in][]{Koudmani2024}. \textit{Middle}: For a MBH of $M_\mathrm{BH} = 10^4 \Msol$, $r_\mathrm{acc}$ is varied between 100 and 10~pc, with and without $\beta_\mathrm{ff} = 0.1$. \textit{Bottom}: With $r_\mathrm{acc} = 1 \pc$, the initial mass $M_\mathrm{BH}$ of the MBH is varied between $10^{3}$ and $10^{5} \Msol$, with and without $\beta_\mathrm{ff} = 0.1$.}
    \label{fig:betacorrection}
\end{figure}

The need for an efficiency factor per free-fall time ($\beta_\mathrm{ff}$ in Eq.~\ref{eq:angular_momentum_transport}) is apparent from the left panel in Fig.~\ref{fig:betacorrection}, in which the time evolution of $r_\mathrm{circ}$ for a set of simulations\footnote{Conducted in a low dark matter resolution set-up of our low-mass dwarf galaxy system presented in Table~\ref{tab:initial_conditions}, with a MBH of $M_\mathrm{BH} = 10^4 \Msol$ and $r_\mathrm{acc} = 1 \ \mathrm{pc}$.} of different $\beta_\mathrm{ff}$ values are shown. By comparing the time evolution of $r_\mathrm{circ}$ with the observational scaling relation of \citet{Morgan2010} and the theoretical prediction for a thin disc \citep[obtained from Eq.~9 in][]{Koudmani2024}, it is evident that no $\beta_\mathrm{ff}$ correction at all overestimates $r_\mathrm{circ}$, while $\beta_\mathrm{ff} = 0.1$ agrees the most with both observational and theoretical constraints. 

By modelling the efficiency factor $\beta_\mathrm{ff}$ to be per free-fall time, the expectation is that the time evolution of $r_\mathrm{circ}$ should fall within a similar range irrespectively of $r_\mathrm{acc}$ and/or $M_\mathrm{BH}$. To assess this, the time evolution of $r_\mathrm{circ}$ for increasing values of $r_\mathrm{acc}$ (with and without $\beta_\mathrm{ff}=0.1$) is shown in the middle panel of Fig.~\ref{fig:betacorrection}. In the right panel of Fig.~\ref{fig:betacorrection}, a similar assessment is done, but for different values of $M_\mathrm{BH}$. Both panels agree with the expected outcome, however, in the case of $M_\mathrm{BH} = 10^5 \ \Msol$ (right panel), $r_\mathrm{circ}$ is able to evolve inside the upper theoretical prediction without $\beta_\mathrm{ff} = 0.1$. This suggests that as the free-fall time gets shorter (either by increasing $M_\mathrm{BH}$ and/or potentially by decreasing $r_\mathrm{acc}$), the $\beta_\mathrm{ff}$ parameter will at some point be redundant. Finding the convergence point is beyond the scope of this paper (but a potential topic for future work). Instead, based on the results of Fig.~\ref{fig:betacorrection}, a value of $\beta_\mathrm{ff}=0.1$ is concluded to be sufficient for the purpose of this study (and unless mentioned otherwise the assumed value for all of the simulations in this work). 

\section{Resolution study}
\label{app:resolution}

\begin{figure}
    \resizebox{\hsize}{!}{\includegraphics{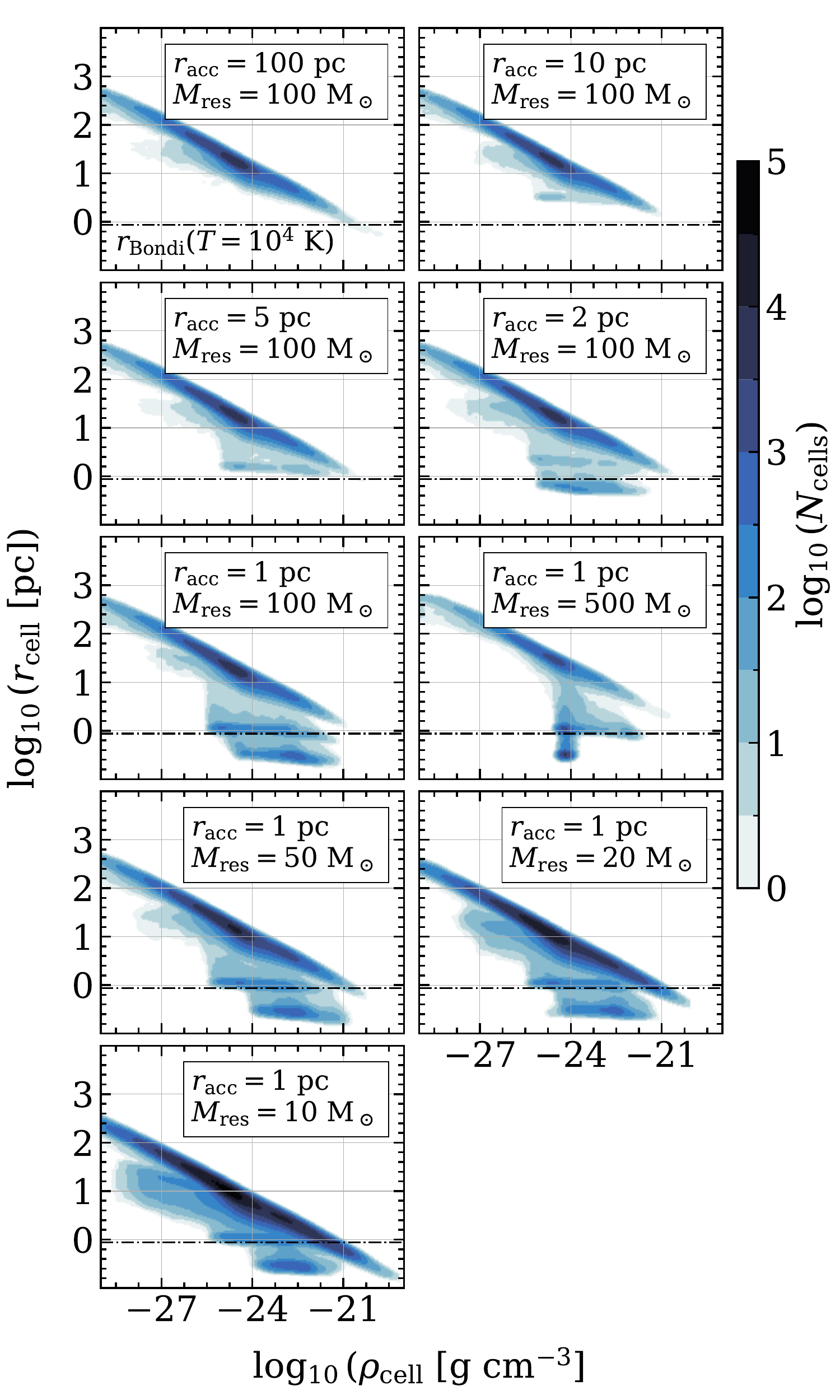}}
    \caption{Typical size of gas cells as a function of gas density (at $t \sim 300 \Myr$) in all of the simulations making up the resolution study. The characteristic length scale of Bondi-Hoyle-Lyttleton accretion $r_\mathrm{Bondi}$ for a $10^4 \Msol$ MBH in $T = 10^4 \K$ gas is indicated by the black dash-dotted line. The two horizontal regions emerging as $r_\mathrm{acc}$ decreases corresponds to the geometrical refinement strategy inside $r_\mathrm{ref}$ and $r_\mathrm{acc}$ respectively, adopted in Sect.~\ref{sect:refinement}.}
    \label{fig:resolution}
\end{figure}

With the numerical set-up described in Sect.~\ref{sect:methodology}, two parameters essentially determine the gas resolution in our simulations. These are the global target mass $M_\mathrm{res}$ of the gas cells, and the massive black hole (MBH) accretion radius $r_\mathrm{acc}$. There is also the refinement radius $r_\mathrm{ref}$ around the MBH, but since the maximum volume limit of the gas cells inside $r_\mathrm{ref}$ is directly coupled to the size of $r_\mathrm{acc}$ (see Sect.~\ref{sect:refinement} for more details), and $r_\mathrm{ref}$ is kept fixed at 10~pc in all of our simulations, we do not expected $r_\mathrm{ref}$ to be as essential for the gas accretion onto the MBH as $r_\mathrm{acc}$. To get an understanding of how influential the gas resolution is to the growth of the MBH, a resolution study is conducted. The study is made for a MBH of $M_\mathrm{BH} = 10^4 \Msol$ in a low dark matter resolution set-up of the low-mass dwarf galaxy system presented in Table~\ref{tab:initial_conditions}. For every simulation making up the study, Fig.~\ref{fig:resolution} shows how the spatial resolution of the gas cells (as function of gas density) changes with the choice of values for $r_\mathrm{acc}$ and $M_\mathrm{res}$. As \arepo favours quasi-spherical gas cells, the radius $r_\mathrm{cell}$ of a sphere is an adequate representation of the typical size of a gas cell given its volume. 

\begin{figure}
    \resizebox{\hsize}{!}{\includegraphics{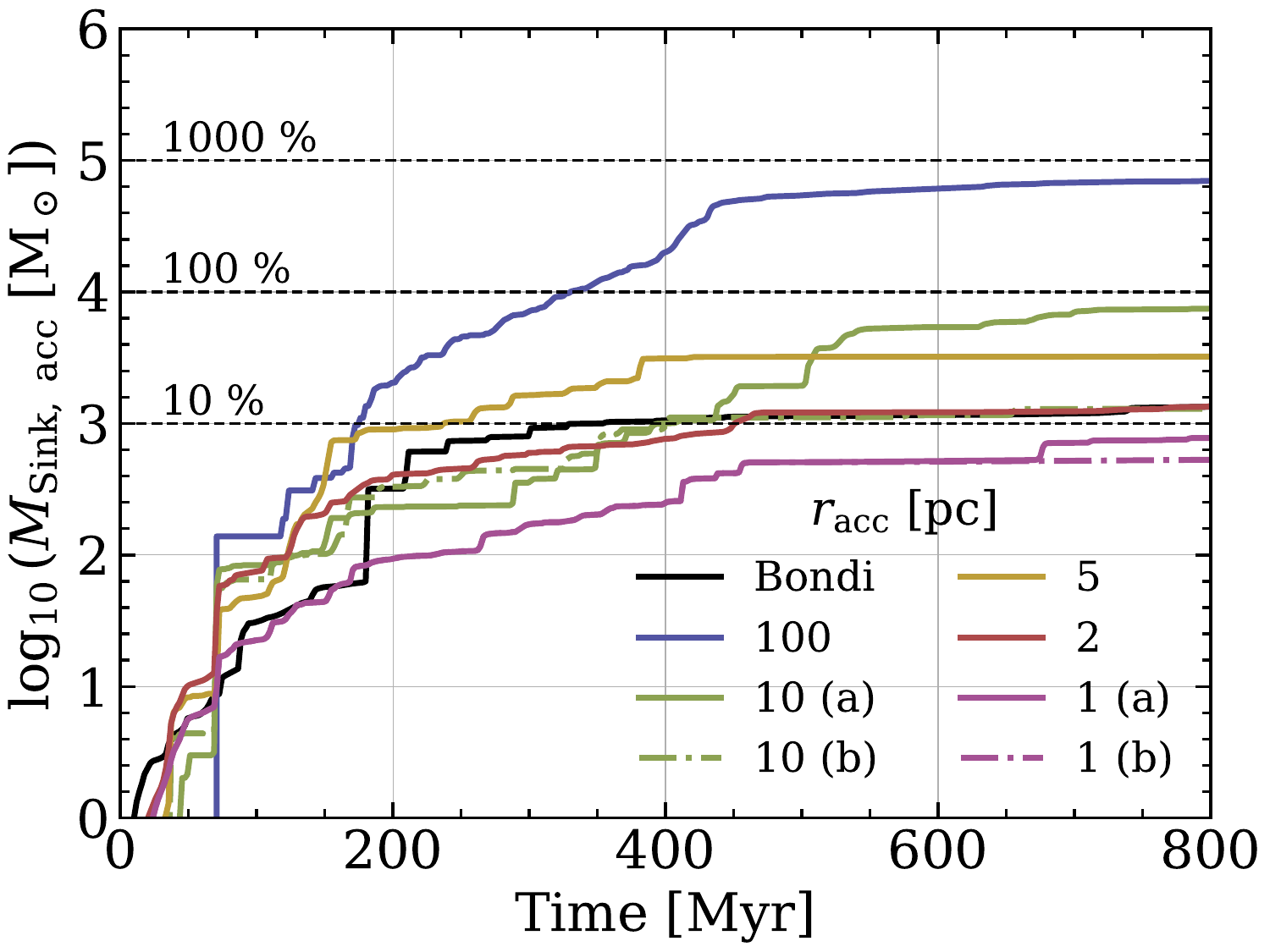}}
    \caption{Amount of mass accreted over time by an initial $10^4 \Msol$ MBH sink particle for different values of $r_\mathrm{acc}$, with two simulations run when $r_\mathrm{acc}$ is equal to 10 and 1~pc respectively (to check how volatile the growth of the MBH sink particle is to stochastic processes). A simulation of Bondi-Hoyle-Lyttleton accretion is also conducted, where $r_\mathrm{acc}=\mathrm{max}\left(1 \ \mathrm{pc}, r_\mathrm{Bondi}\right)$. Black dashed lines indicate the percentage growth of the MBH (with respect to its initial mass).}
    \label{fig:bh_growth_acc_radius}
\end{figure}

\begin{figure}
    \resizebox{\hsize}{!}{\includegraphics{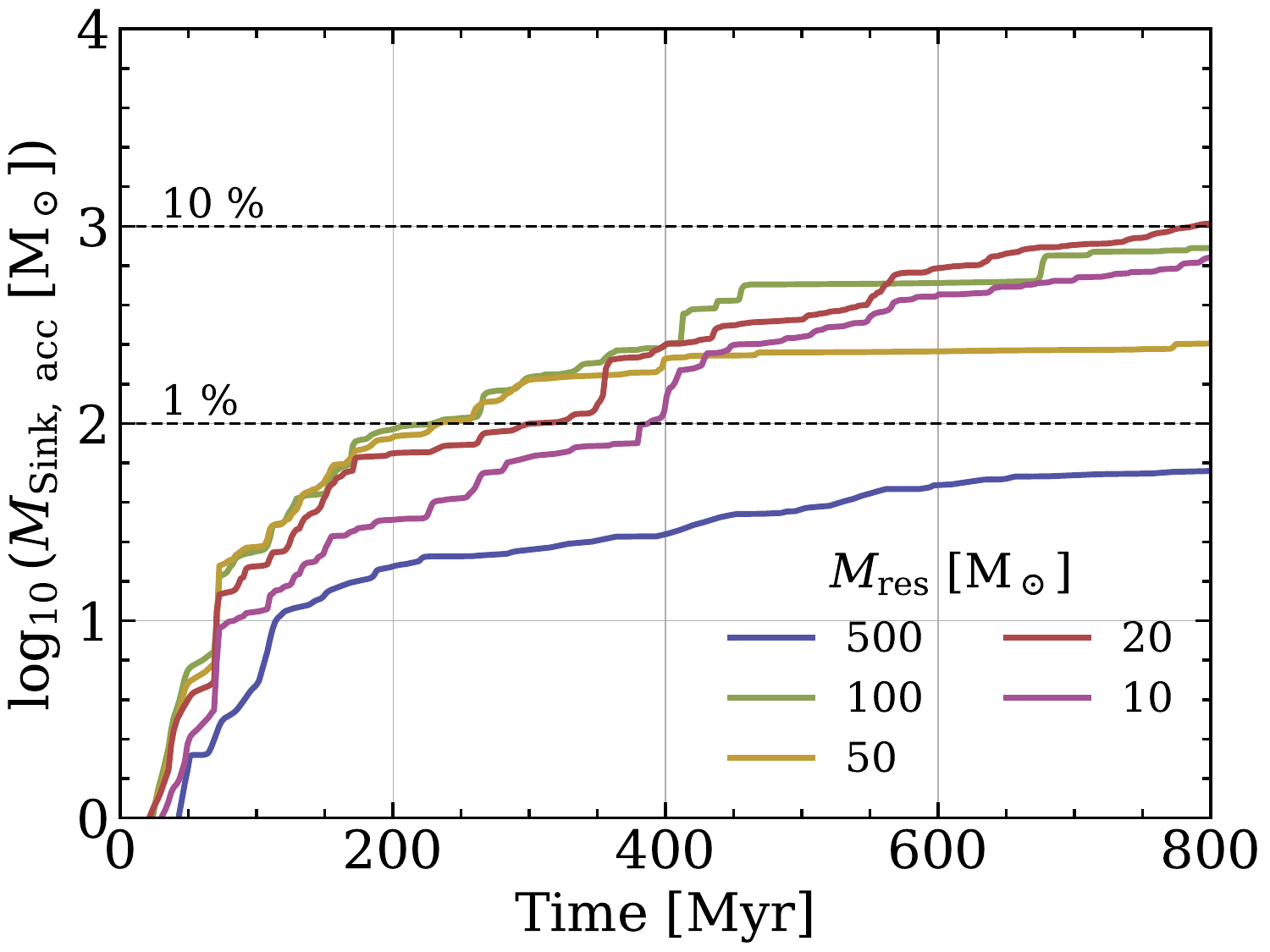}}
    \caption{Amount of mass accreted over time by an initial $10^4 \Msol$ MBH sink particle for different values of the global target mass $M_\mathrm{res}$ of gas cells. For comparison, black dashed lines mark the percentage growth of the MBH (with respect to its initial mass)}
    \label{fig:bh_growth_resolution}
\end{figure}

By gradually reducing $r_\mathrm{acc}$ from 100 to 1~pc (while $M_\mathrm{res} = 100 \Msol$), the gas resolution at the vicinity of the MBH increases (due to the refinement strategy adopted in Sect.~\ref{sect:refinement} ensuring that at least a few tens of gas cells are always resolved inside $r_\mathrm{acc}$). The two emerging vertical regions below the main bulk of gas in Fig.~\ref{fig:resolution} correspond to the geometrical refinement inside $r_\mathrm{ref}$ and $r_\mathrm{acc}$ respectively. As $r_\mathrm{acc}$ decreases, the gas cells populating these regions reduce in size, i.e. the spatial resolution increases. Figure~\ref{fig:bh_growth_acc_radius} shows the amount of mass accreted over time by the MBH sink particle\footnote{Only the amount of mass accreted on a grid level is considered (to avoid the risk of confusion when taking into account the many different steps of the sub-grid accretion model in Sect.~\ref{sect:bh_accretion_sub_grid}).} for different values of $r_\mathrm{acc}$. To get an idea how volatile the growth of the MBH sink particle is to stochastic processes (e.g. star formation), the simulations of $r_\mathrm{acc}$ being equal to 10 and 1 pc are run a second time. A simulation of Bondi-Hoyle-Lyttleton accretion \citep{Hoyle1939, Bondi1944, Bondi1952} is also conducted. Within this accretion scheme, Eq.~(\ref{eq:infalling_gas_accretion_rate}) in Sect.~\ref{sect:bh_accretion} is replaced by

\begin{equation}
    \dot{M}_\mathrm{in} = \dot{M}_\mathrm{Bondi} = \frac{4\pi G^2 M_\mathrm{BH}^2\rho}{c_s^3}, 
\end{equation}

\noindent where $\rho$ and $c_s$ are the non-weighted mean density and mean sound speed respectively of all the gas cells inside $r_\mathrm{acc}$. The recipe to skim mass of individual gas cells remains the same (see Eq.~\ref{eq:skimming_mass_of_gas_cells}), except that it is now volume-weighted. With the high spatial resolution achievable in the simulations of this work, the characteristic length scale $r_\mathrm{Bondi} = 2GM_\mathrm{BH}/c_s^2$ of Bondi-Hoyle-Lyttleton accretion is partly resolved (up to $T = 10^4 \K$; see Fig.~\ref{fig:resolution}). Therefore, $r_\mathrm{acc}$ is allowed to vary on-the-fly according to $r_\mathrm{acc} = \mathrm{max}(1 \pc,~r_\mathrm{Bondi})$ (the computational cost of a resolved accretion region when $r_\mathrm{acc}<1 \ \mathrm{pc}$ is too high). Altogether, Fig.~\ref{fig:bh_growth_acc_radius} demonstrates that the growth of the MBH decreases as $r_\mathrm{acc}$ gets reduced in size (combined with an increase in the gas resolution at the vicinity of the MBH), and is less susceptible to stochastic processes. The growth of the MBH seems to converge when $r_\mathrm{acc}$ is close to a few~pc in size, and only shows a slight divergence from our adopted Bondi-Hoyle-Lyttleton accretion scheme. This agrees well with \citet{Negri2017}, who find that Bondi-Hoyle-Lyttleton accretion converges with a flux accretion scheme when $r_\mathrm{acc}$ is comparable to $r_\mathrm{Bondi}$.

For the remaining set of simulations in Fig.~\ref{fig:resolution}, $M_\mathrm{res}$ is gradually reduced from 500 to $10 \Msol$ (while $r_\mathrm{acc} = 1 \pc$). Besides the overall higher number of gas cells, the spatial resolution increases on a global scale as $M_\mathrm{res}$ decreases. This however does not guarantee a higher spatial resolution at the vicinity of the MBH, meaning that the refinement strategy adopted in Sect.~\ref{sect:refinement} is still needed to ensure that the MBH at any given time can skim mass of nearby gas cells. The amount of mass accreted over time by the MBH for different values of $M_\mathrm{res}$ are shown in Fig.~\ref{fig:bh_growth_resolution}. The numerical set-up seem to convergence as $M_\mathrm{res}$ gets close to a few tens of $\mathrm{M}_\odot$ in size (with $M_\mathrm{res} = 50 \Msol$ being a possible outlier due to stochastic processes). 

\end{appendix}

\end{document}

%% file: header.tex
\usepackage{ifthen,xspace,ulem,makecell}
\usepackage[dvipsnames]{xcolor}

\ifthenelse{\paperversion > 0}{
    
    }{
    
}

\ifthenelse{\equal{\paperversion}{1}}{
    
    \newcommand{\draft}[1]{\begin{color}{red}\bfseries #1\end{color}}                  
    \newcommand{\idea}[1]{\begin{color}{Green}\textit{#1}\end{color}}                  
    \newcommand{\todo}[1]{\begin{color}{magenta}TODO: #1\end{color}}                   
    \newcommand{\move}[2]{\begin{color}{violet}\textbf{#1 -> MOVE TO: #2}\end{color}}  
    \newcommand{\toc}[1]{\begin{color}{orange}\textbf{#1}\end{color}}                  
    \newcommand{\trash}[1]{\begin{color}{brown}\textbf{#1}\end{color}}                 
    \newcommand{\collab}[2]{\begin{color}{Emerald}\textbf{[#1: #2]}\end{color}}        
    }{
    
    \newcommand{\draft}[1]{}
    \newcommand{\ndraft}[1]{}
    \newcommand{\idea}[1]{}
    \newcommand{\todo}[1]{}
    \newcommand{\move}[2]{}
    \newcommand{\toc}[1]{}

    \newcommand{\trash}[1]{}
    \newcommand{\collab}[2]{}
}

\DeclareRobustCommand{\orcid}[1]{%
  \begingroup\normalfont
  \href{https://orcid.org/#1}{\includegraphics[height=\fontcharht\font`\B]{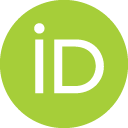}}%
  \endgroup
}

\newcommand{\Msol}{\ensuremath{~\mathrm{M}_\odot}}
\newcommand{\pc}{\ensuremath{~\mathrm{pc}}}
\newcommand{\kpc}{\ensuremath{~\mathrm{kpc}}}
\newcommand{\erg}{\ensuremath{~\mathrm{erg}}}
\newcommand{\K}{\ensuremath{~\mathrm{K}}}
\newcommand{\yr}{\ensuremath{~\mathrm{yr}}}
\newcommand{\Myr}{\ensuremath{~\mathrm{Myr}}}
\newcommand{\Gyr}{\ensuremath{~\mathrm{Gyr}}}

\newcommand{\arepo}{\textsc{Arepo}\xspace}
\newcommand{\gadgettwo}{\textsc{Gadget-2}\xspace}

\newcommand{\treecol}{\textsc{TreeCol}\xspace}

\newcommand{\noctua}{\textsc{Noctua}\xspace}

\newcommand{\sweep}{\textsc{Sweep}\xspace}

\newcommand{\areponoctua}{\textsc{ArepoNoctua}\xspace}

\newcommand{\galspec}{Institute of Physics, Laboratory for Galaxy Evolution and Spectral Modelling, Ecole Polytechnique Fédérale de Lausanne, Observatoire de Sauverny, Chemin Pegasi 51, CH-1290 Versoix, Switzerland}
\newcommand{\mpia}{Max Planck Institute for Astrophysik, Karl-Schwarzschild-Str. 1, 85748, Garching, Germany}
\newcommand{\zah}{Universit\"at Heidelberg, Zentrum f\"ur Astronomie, Institut f\"ur Theoretische Astrophysik, Albert-Ueberle-Str. 2, 69120 Heidelberg, Germany}
\newcommand{\iwr}{Universit\"at Heidelberg, Interdisziplin\"ares Zentrum f\"ur Wissenschaftliches Rechnen, Im Neuenheimer Feld 205, D-69120 Heidelberg, Germany}
\newcommand{\cca}{Center for Computational Astrophysics, Flatiron Institute, 162 5th Avenue, New York, NY 10010, USA}
\newcommand{\unam}{Universidad Nacional Aut\'onoma de M\'exico, Instituto de Radioastronom\'ia y Astrof\'isica, Antigua Carretera a P\'atzcuaro 8701, Ex-Hda. San Jos\'e de la Huerta, 58089 Morelia, Michoac\'an, M\'exico}
\newcommand{\cfa}{Harvard-Smithsonian Center for Astrophysics, 60 Garden Street, Cambridge, MA 02138, USA}
\newcommand{\radcliffe}{Elizabeth S. and Richard M. Cashin Fellow at the Radcliffe Institute for Advanced Studies at Harvard University, 10 Garden Street, Cambridge, MA 02138, USA}

%% file: aanda.bbl
\begin{thebibliography}{174}
\expandafter\ifx\csname natexlab\endcsname\relax\def\natexlab#1{#1}\fi

\bibitem[{{Agertz} {et~al.}(2020){Agertz}, {Pontzen}, {Read}, {Rey}, {Orkney},
  {Rosdahl}, {Teyssier}, {Verbeke}, {Kretschmer}, \& {Nickerson}}]{Agertz2020}
{Agertz}, O., {Pontzen}, A., {Read}, J.~I., {et~al.} 2020, \mnras, 491, 1656

\bibitem[{{Andersson} {et~al.}(2023){Andersson}, {Agertz}, {Renaud}, \&
  {Teyssier}}]{Andersson2023}
{Andersson}, E.~P., {Agertz}, O., {Renaud}, F., \& {Teyssier}, R. 2023, \mnras,
  521, 2196

\bibitem[{{Andika} {et~al.}(2020){Andika}, {Jahnke}, {Onoue}, {Ba{\~n}ados},
  {Mazzucchelli}, {Novak}, {Eilers}, {Venemans}, {Schindler}, {Walter},
  {Neeleman}, {Simcoe}, {Decarli}, {Farina}, {Marian}, {Pensabene}, {Cooper},
  \& {Rojas}}]{Andika2020}
{Andika}, I.~T., {Jahnke}, K., {Onoue}, M., {et~al.} 2020, \apj, 903, 34

\bibitem[{{Angl{\'e}s-Alc{\'a}zar} {et~al.}(2015){Angl{\'e}s-Alc{\'a}zar},
  {{\"O}zel}, {Dav{\'e}}, {Katz}, {Kollmeier}, \&
  {Oppenheimer}}]{AnglesAlcazar2015}
{Angl{\'e}s-Alc{\'a}zar}, D., {{\"O}zel}, F., {Dav{\'e}}, R., {et~al.} 2015,
  \apj, 800, 127

\bibitem[{{Asplund} {et~al.}(2009){Asplund}, {Grevesse}, {Sauval}, \&
  {Scott}}]{Asplund2009}
{Asplund}, M., {Grevesse}, N., {Sauval}, A.~J., \& {Scott}, P. 2009, \araa, 47,
  481

\bibitem[{{Ba{\~n}ados} {et~al.}(2023){Ba{\~n}ados}, {Schindler}, {Venemans},
  {Connor}, {Decarli}, {Farina}, {Mazzucchelli}, {Meyer}, {Stern}, {Walter},
  {Fan}, {Hennawi}, {Khusanova}, {Morrell}, {Nanni}, {Noirot}, {Pensabene},
  {Rix}, {Simon}, {Verdoes Kleijn}, {Xie}, {Yang}, \& {Connor}}]{Banados2023}
{Ba{\~n}ados}, E., {Schindler}, J.-T., {Venemans}, B.~P., {et~al.} 2023, \apjs,
  265, 29

\bibitem[{{Baldassare} {et~al.}(2018){Baldassare}, {Geha}, \&
  {Greene}}]{Baldassare2018}
{Baldassare}, V.~F., {Geha}, M., \& {Greene}, J. 2018, \apj, 868, 152

\bibitem[{{Baldassare} {et~al.}(2020){Baldassare}, {Geha}, \&
  {Greene}}]{Baldassare2020}
{Baldassare}, V.~F., {Geha}, M., \& {Greene}, J. 2020, \apj, 896, 10

\bibitem[{{Baldassare} {et~al.}(2017){Baldassare}, {Reines}, {Gallo}, \&
  {Greene}}]{Baldassare2017}
{Baldassare}, V.~F., {Reines}, A.~E., {Gallo}, E., \& {Greene}, J.~E. 2017,
  \apj, 836, 20

\bibitem[{{Barai} \& {de Gouveia Dal Pino}(2019)}]{Barai2019}
{Barai}, P. \& {de Gouveia Dal Pino}, E.~M. 2019, \mnras, 487, 5549

\bibitem[{{Bellovary} {et~al.}(2013){Bellovary}, {Brooks}, {Volonteri},
  {Governato}, {Quinn}, \& {Wadsley}}]{Bellovary2013}
{Bellovary}, J., {Brooks}, A., {Volonteri}, M., {et~al.} 2013, \apj, 779, 136

\bibitem[{{Bertram} {et~al.}(2015){Bertram}, {Glover}, {Clark}, \&
  {Klessen}}]{Bertram2015}
{Bertram}, E., {Glover}, S. C.~O., {Clark}, P.~C., \& {Klessen}, R.~S. 2015,
  \mnras, 451, 3679

\bibitem[{{Bondi}(1952)}]{Bondi1952}
{Bondi}, H. 1952, \mnras, 112, 195

\bibitem[{{Bondi} \& {Hoyle}(1944)}]{Bondi1944}
{Bondi}, H. \& {Hoyle}, F. 1944, \mnras, 104, 273

\bibitem[{{Booth} \& {Schaye}(2009)}]{Booth2009}
{Booth}, C.~M. \& {Schaye}, J. 2009, \mnras, 398, 53

\bibitem[{{Bu} \& {Yuan}(2014)}]{Bu2014}
{Bu}, D.-F. \& {Yuan}, F. 2014, \mnras, 442, 917

\bibitem[{{Choi} {et~al.}(2015){Choi}, {Ostriker}, {Naab}, {Oser}, \&
  {Moster}}]{Choi2015}
{Choi}, E., {Ostriker}, J.~P., {Naab}, T., {Oser}, L., \& {Moster}, B.~P. 2015,
  \mnras, 449, 4105

\bibitem[{{Ciotti} \& {Ostriker}(2007)}]{Ciotti2007}
{Ciotti}, L. \& {Ostriker}, J.~P. 2007, \apj, 665, 1038

\bibitem[{{Ciotti} {et~al.}(2009){Ciotti}, {Ostriker}, \& {Proga}}]{Ciotti2009}
{Ciotti}, L., {Ostriker}, J.~P., \& {Proga}, D. 2009, \apj, 699, 89

\bibitem[{{Ciotti} {et~al.}(2017){Ciotti}, {Pellegrini}, {Negri}, \&
  {Ostriker}}]{Ciotti2017}
{Ciotti}, L., {Pellegrini}, S., {Negri}, A., \& {Ostriker}, J.~P. 2017, \apj,
  835, 15

\bibitem[{{Clark} {et~al.}(2012){Clark}, {Glover}, \& {Klessen}}]{Clark2012}
{Clark}, P.~C., {Glover}, S. C.~O., \& {Klessen}, R.~S. 2012, \mnras, 420, 745

\bibitem[{{Deng} {et~al.}(2024){Deng}, {Li}, {Liu}, {Kannan}, {Smith}, \&
  {Bryan}}]{Deng2024}
{Deng}, Y., {Li}, H., {Liu}, B., {et~al.} 2024, \aap, 691, A231

\bibitem[{{Di Matteo} {et~al.}(2008){Di Matteo}, {Colberg}, {Springel},
  {Hernquist}, \& {Sijacki}}]{DiMatteo2008}
{Di Matteo}, T., {Colberg}, J., {Springel}, V., {Hernquist}, L., \& {Sijacki},
  D. 2008, \apj, 676, 33

\bibitem[{{Draine}(1978)}]{Draine1978}
{Draine}, B.~T. 1978, \apjs, 36, 595

\bibitem[{{Dubois} {et~al.}(2015){Dubois}, {Volonteri}, {Silk}, {Devriendt},
  {Slyz}, \& {Teyssier}}]{Dubois2015}
{Dubois}, Y., {Volonteri}, M., {Silk}, J., {et~al.} 2015, \mnras, 452, 1502

\bibitem[{{Ekstr{\"o}m} {et~al.}(2012){Ekstr{\"o}m}, {Georgy}, {Eggenberger},
  {Meynet}, {Mowlavi}, {Wyttenbach}, {Granada}, {Decressin}, {Hirschi},
  {Frischknecht}, {Charbonnel}, \& {Maeder}}]{Ekstrom2012}
{Ekstr{\"o}m}, S., {Georgy}, C., {Eggenberger}, P., {et~al.} 2012, \aap, 537,
  A146

\bibitem[{{Emerick} {et~al.}(2018){Emerick}, {Bryan}, \& {Mac
  Low}}]{Emerick2018}
{Emerick}, A., {Bryan}, G.~L., \& {Mac Low}, M.-M. 2018, \apjl, 865, L22

\bibitem[{{Emerick} {et~al.}(2019){Emerick}, {Bryan}, \& {Mac
  Low}}]{Emerick2019}
{Emerick}, A., {Bryan}, G.~L., \& {Mac Low}, M.-M. 2019, \mnras, 482, 1304

\bibitem[{{Fabian}(2012)}]{Fabian2012}
{Fabian}, A.~C. 2012, \araa, 50, 455

\bibitem[{{Farcy} {et~al.}(2022){Farcy}, {Rosdahl}, {Dubois}, {Blaizot}, \&
  {Martin-Alvarez}}]{Farcy2022}
{Farcy}, M., {Rosdahl}, J., {Dubois}, Y., {Blaizot}, J., \& {Martin-Alvarez},
  S. 2022, \mnras, 513, 5000

\bibitem[{{Ferri{\`e}re}(2001)}]{Katia2001}
{Ferri{\`e}re}, K.~M. 2001, Reviews of Modern Physics, 73, 1031

\bibitem[{{Gan} {et~al.}(2014){Gan}, {Yuan}, {Ostriker}, {Ciotti}, \&
  {Novak}}]{Gan2014}
{Gan}, Z., {Yuan}, F., {Ostriker}, J.~P., {Ciotti}, L., \& {Novak}, G.~S. 2014,
  \apj, 789, 150

\bibitem[{{Gaspari} {et~al.}(2013){Gaspari}, {Ruszkowski}, \&
  {Oh}}]{Gaspari2013}
{Gaspari}, M., {Ruszkowski}, M., \& {Oh}, S.~P. 2013, \mnras, 432, 3401

\bibitem[{{Gatto} {et~al.}(2015){Gatto}, {Walch}, {Low}, {Naab}, {Girichidis},
  {Glover}, {W{\"u}nsch}, {Klessen}, {Clark}, {Baczynski}, {Peters},
  {Ostriker}, {Ib{\'a}{\~n}ez-Mej{\'\i}a}, \& {Haid}}]{Gatto2015}
{Gatto}, A., {Walch}, S., {Low}, M. M.~M., {et~al.} 2015, \mnras, 449, 1057

\bibitem[{{Gloudemans} {et~al.}(2022){Gloudemans}, {Duncan}, {Saxena},
  {Harikane}, {Hill}, {Zeimann}, {R{\"o}ttgering}, {Yang}, {Best},
  {Ba{\~n}ados}, {Drabent}, {Hardcastle}, {Hennawi}, {Lansbury},
  {Magliocchetti}, {Miley}, {Nanni}, {Shimwell}, {Smith}, {Venemans}, \&
  {Wagenveld}}]{Gloudemans2022}
{Gloudemans}, A.~J., {Duncan}, K.~J., {Saxena}, A., {et~al.} 2022, \aap, 668,
  A27

\bibitem[{{Glover} \& {Clark}(2012)}]{Glover2012}
{Glover}, S. C.~O. \& {Clark}, P.~C. 2012, \mnras, 421, 116

\bibitem[{{Glover} \& {Mac Low}(2007{\natexlab{a}})}]{Glover2007a}
{Glover}, S. C.~O. \& {Mac Low}, M.-M. 2007{\natexlab{a}}, \apjs, 169, 239

\bibitem[{{Glover} \& {Mac Low}(2007{\natexlab{b}})}]{Glover2007b}
{Glover}, S. C.~O. \& {Mac Low}, M.-M. 2007{\natexlab{b}}, \apj, 659, 1317

\bibitem[{{Gnat} \& {Ferland}(2012)}]{Gnat2012}
{Gnat}, O. \& {Ferland}, G.~J. 2012, \apjs, 199, 20

\bibitem[{{Goldsmith} \& {Langer}(1978)}]{Goldsmith1978}
{Goldsmith}, P.~F. \& {Langer}, W.~D. 1978, \apj, 222, 881

\bibitem[{{G{\"o}ller} {et~al.}(2025){G{\"o}ller}, {Girichidis}, {Brucy},
  {Hunter}, {Kjellgren}, {Tress}, {Klessen}, {Glover}, {Hennebelle},
  {Molinari}, {Smith}, {Soler}, {Sormani}, \& {Testi}}]{Goller2025}
{G{\"o}ller}, J., {Girichidis}, P., {Brucy}, N., {et~al.} 2025, arXiv e-prints,
  arXiv:2502.02646

\bibitem[{{Graham} \& {Driver}(2007)}]{Graham2007}
{Graham}, A.~W. \& {Driver}, S.~P. 2007, \apj, 655, 77

\bibitem[{{Greene} {et~al.}(2024){Greene}, {Labbe}, {Goulding}, {Furtak},
  {Chemerynska}, {Kokorev}, {Dayal}, {Volonteri}, {Williams}, {Wang}, {Setton},
  {Burgasser}, {Bezanson}, {Atek}, {Brammer}, {Cutler}, {Feldmann}, {Fujimoto},
  {Glazebrook}, {de Graaff}, {Khullar}, {Leja}, {Marchesini}, {Maseda},
  {Matthee}, {Miller}, {Naidu}, {Nanayakkara}, {Oesch}, {Pan}, {Papovich},
  {Price}, {van Dokkum}, {Weaver}, {Whitaker}, \& {Zitrin}}]{Greene2024}
{Greene}, J.~E., {Labbe}, I., {Goulding}, A.~D., {et~al.} 2024, \apj, 964, 39

\bibitem[{{G{\"u}ltekin} {et~al.}(2009){G{\"u}ltekin}, {Richstone}, {Gebhardt},
  {Lauer}, {Tremaine}, {Aller}, {Bender}, {Dressler}, {Faber}, {Filippenko},
  {Green}, {Ho}, {Kormendy}, {Magorrian}, {Pinkney}, \&
  {Siopis}}]{Gultekin2009}
{G{\"u}ltekin}, K., {Richstone}, D.~O., {Gebhardt}, K., {et~al.} 2009, \apj,
  698, 198

\bibitem[{{Gutcke} {et~al.}(2021){Gutcke}, {Pakmor}, {Naab}, \&
  {Springel}}]{Gutcke2021}
{Gutcke}, T.~A., {Pakmor}, R., {Naab}, T., \& {Springel}, V. 2021, \mnras, 501,
  5597

\bibitem[{{Habouzit} {et~al.}(2021){Habouzit}, {Li}, {Somerville}, {Genel},
  {Pillepich}, {Volonteri}, {Dav{\'e}}, {Rosas-Guevara}, {McAlpine}, {Peirani},
  {Hernquist}, {Angl{\'e}s-Alc{\'a}zar}, {Reines}, {Bower}, {Dubois}, {Nelson},
  {Pichon}, \& {Vogelsberger}}]{Habouzit2021}
{Habouzit}, M., {Li}, Y., {Somerville}, R.~S., {et~al.} 2021, \mnras, 503, 1940

\bibitem[{{Habouzit} {et~al.}(2022){Habouzit}, {Onoue}, {Ba{\~n}ados},
  {Neeleman}, {Angl{\'e}s-Alc{\'a}zar}, {Walter}, {Pillepich}, {Dav{\'e}},
  {Jahnke}, \& {Dubois}}]{habouzit2022}
{Habouzit}, M., {Onoue}, M., {Ba{\~n}ados}, E., {et~al.} 2022, \mnras, 511,
  3751

\bibitem[{{Habouzit} {et~al.}(2017){Habouzit}, {Volonteri}, \&
  {Dubois}}]{Habouzit2017}
{Habouzit}, M., {Volonteri}, M., \& {Dubois}, Y. 2017, \mnras, 468, 3935

\bibitem[{{Harikane} {et~al.}(2023){Harikane}, {Zhang}, {Nakajima}, {Ouchi},
  {Isobe}, {Ono}, {Hatano}, {Xu}, \& {Umeda}}]{Harikane2023}
{Harikane}, Y., {Zhang}, Y., {Nakajima}, K., {et~al.} 2023, \apj, 959, 39

\bibitem[{{H{\"a}ring} \& {Rix}(2004)}]{Haring2004}
{H{\"a}ring}, N. \& {Rix}, H.-W. 2004, \apjl, 604, L89

\bibitem[{{Hernquist}(1990)}]{Hernquist1990b}
{Hernquist}, L. 1990, \apj, 356, 359

\bibitem[{{Hernquist} \& {Barnes}(1990)}]{Hernquist1990a}
{Hernquist}, L. \& {Barnes}, J.~E. 1990, \apj, 349, 562

\bibitem[{{Hirschmann} {et~al.}(2023){Hirschmann}, {Charlot}, {Feltre},
  {Curtis-Lake}, {Somerville}, {Chevallard}, {Choi}, {Nelson}, {Morisset},
  {Plat}, \& {Vidal-Garcia}}]{Hirschmann23}
{Hirschmann}, M., {Charlot}, S., {Feltre}, A., {et~al.} 2023, \mnras, 526, 3610

\bibitem[{{Hirschmann} {et~al.}(2017){Hirschmann}, {Charlot}, {Feltre}, {Naab},
  {Choi}, {Ostriker}, \& {Somerville}}]{Hirschmann17}
{Hirschmann}, M., {Charlot}, S., {Feltre}, A., {et~al.} 2017, \mnras, 472, 2468

\bibitem[{{Hirschmann} {et~al.}(2019){Hirschmann}, {Charlot}, {Feltre}, {Naab},
  {Somerville}, \& {Choi}}]{Hirschmann19}
{Hirschmann}, M., {Charlot}, S., {Feltre}, A., {et~al.} 2019, \mnras, 487, 333

\bibitem[{{Hirschmann} {et~al.}(2014){Hirschmann}, {Dolag}, {Saro}, {Bachmann},
  {Borgani}, \& {Burkert}}]{Hirschmann2014}
{Hirschmann}, M., {Dolag}, K., {Saro}, A., {et~al.} 2014, \mnras, 442, 2304

\bibitem[{{Hirschmann} {et~al.}(2010){Hirschmann}, {Khochfar}, {Burkert},
  {Naab}, {Genel}, \& {Somerville}}]{Hirschmann10}
{Hirschmann}, M., {Khochfar}, S., {Burkert}, A., {et~al.} 2010, \mnras, 407,
  1016

\bibitem[{{Hopkins} {et~al.}(2024){Hopkins}, {Grudic}, {Su}, {Wellons},
  {Angles-Alcazar}, {Steinwandel}, {Guszejnov}, {Murray}, {Faucher-Giguere},
  {Quataert}, \& {Keres}}]{Hopkins2024}
{Hopkins}, P.~F., {Grudic}, M.~Y., {Su}, K.-Y., {et~al.} 2024, The Open Journal
  of Astrophysics, 7, 18

\bibitem[{{Hoyle} \& {Lyttleton}(1939)}]{Hoyle1939}
{Hoyle}, F. \& {Lyttleton}, R.~A. 1939, Proceedings of the Cambridge
  Philosophical Society, 35, 405

\bibitem[{{Hu} {et~al.}(2017){Hu}, {Naab}, {Glover}, {Walch}, \&
  {Clark}}]{Hu2017}
{Hu}, C.-Y., {Naab}, T., {Glover}, S. C.~O., {Walch}, S., \& {Clark}, P.~C.
  2017, \mnras, 471, 2151

\bibitem[{{Hu} {et~al.}(2016){Hu}, {Naab}, {Walch}, {Glover}, \&
  {Clark}}]{Hu2016}
{Hu}, C.-Y., {Naab}, T., {Walch}, S., {Glover}, S. C.~O., \& {Clark}, P.~C.
  2016, \mnras, 458, 3528

\bibitem[{{Hu{\v{s}}ko} {et~al.}(2025){Hu{\v{s}}ko}, {Lacey}, {Roper},
  {Schaye}, {Briggs}, \& {Schaller}}]{Husko2025}
{Hu{\v{s}}ko}, F., {Lacey}, C.~G., {Roper}, W.~J., {et~al.} 2025, \mnras, 537,
  2559

\bibitem[{{Inayoshi} {et~al.}(2020){Inayoshi}, {Visbal}, \&
  {Haiman}}]{Inayoshi2020}
{Inayoshi}, K., {Visbal}, E., \& {Haiman}, Z. 2020, \araa, 58, 27

\bibitem[{{Jahnke} \& {Macci{\`o}}(2011)}]{Jahnke11}
{Jahnke}, K. \& {Macci{\`o}}, A.~V. 2011, \apj, 734, 92

\bibitem[{{Jeans}(1902)}]{Jeans1902}
{Jeans}, J.~H. 1902, Philosophical Transactions of the Royal Society of London
  Series A, 199, 1

\bibitem[{{Jiang} {et~al.}(2016){Jiang}, {McGreer}, {Fan}, {Strauss},
  {Ba{\~n}ados}, {Becker}, {Bian}, {Farnsworth}, {Shen}, {Wang}, {Wang},
  {Wang}, {White}, {Wu}, {Wu}, {Yang}, \& {Yang}}]{Jiang2016}
{Jiang}, L., {McGreer}, I.~D., {Fan}, X., {et~al.} 2016, \apj, 833, 222

\bibitem[{{Kato} {et~al.}(2008){Kato}, {Fukue}, \& {Mineshige}}]{Kato2008}
{Kato}, S., {Fukue}, J., \& {Mineshige}, S. 2008, {Black-Hole Accretion Disks
  --- Towards a New Paradigm ---}

\bibitem[{{Kim} {et~al.}(2023){Kim}, {Kim}, {Gong}, \& {Ostriker}}]{Kim2023}
{Kim}, C.-G., {Kim}, J.-G., {Gong}, M., \& {Ostriker}, E.~C. 2023, \apj, 946, 3

\bibitem[{{Kim} {et~al.}(2015){Kim}, {Im}, {Jeon}, {Kim}, {Choi}, {Hong},
  {Hyun}, {Jun}, {Karouzos}, {Kim}, {Kim}, {Kim}, {Kim}, {Lee}, {Pak}, {Park},
  {Taak}, \& {Yoon}}]{Kim2015b}
{Kim}, Y., {Im}, M., {Jeon}, Y., {et~al.} 2015, \apjl, 813, L35

\bibitem[{{King}(2003)}]{King2003}
{King}, A. 2003, \apjl, 596, L27

\bibitem[{{Klessen} \& {Glover}(2016)}]{Klessen2016}
{Klessen}, R.~S. \& {Glover}, S. C.~O. 2016, Saas-Fee Advanced Course, 43, 85

\bibitem[{{Kocevski} {et~al.}(2023){Kocevski}, {Onoue}, {Inayoshi}, {Trump},
  {Arrabal Haro}, {Grazian}, {Dickinson}, {Finkelstein}, {Kartaltepe},
  {Hirschmann}, {Aird}, {Holwerda}, {Fujimoto}, {Juneau}, {Amor{\'\i}n},
  {Backhaus}, {Bagley}, {Barro}, {Bell}, {Bisigello}, {Calabr{\`o}}, {Cleri},
  {Cooper}, {Ding}, {Grogin}, {Ho}, {Hutchison}, {Inoue}, {Jiang}, {Jones},
  {Koekemoer}, {Li}, {Li}, {McGrath}, {Molina}, {Papovich},
  {P{\'e}rez-Gonz{\'a}lez}, {Pirzkal}, {Wilkins}, {Yang}, \&
  {Yung}}]{Kocevski2023}
{Kocevski}, D.~D., {Onoue}, M., {Inayoshi}, K., {et~al.} 2023, \apjl, 954, L4

\bibitem[{{Kokorev} {et~al.}(2023){Kokorev}, {Fujimoto}, {Labbe}, {Greene},
  {Bezanson}, {Dayal}, {Nelson}, {Atek}, {Brammer}, {Caputi}, {Chemerynska},
  {Cutler}, {Feldmann}, {Fudamoto}, {Furtak}, {Goulding}, {de Graaff}, {Leja},
  {Marchesini}, {Miller}, {Nanayakkara}, {Oesch}, {Pan}, {Price}, {Setton},
  {Smit}, {Stefanon}, {Wang}, {Weaver}, {Whitaker}, {Williams}, \&
  {Zitrin}}]{Kokorev2023}
{Kokorev}, V., {Fujimoto}, S., {Labbe}, I., {et~al.} 2023, \apjl, 957, L7

\bibitem[{{Kormendy} \& {Ho}(2013)}]{Kormendy2013}
{Kormendy}, J. \& {Ho}, L.~C. 2013, \araa, 51, 511

\bibitem[{{Koudmani} {et~al.}(2021){Koudmani}, {Henden}, \&
  {Sijacki}}]{Koudmani2021}
{Koudmani}, S., {Henden}, N.~A., \& {Sijacki}, D. 2021, \mnras, 503, 3568

\bibitem[{{Koudmani} {et~al.}(2019){Koudmani}, {Sijacki}, {Bourne}, \&
  {Smith}}]{Koudmani2019}
{Koudmani}, S., {Sijacki}, D., {Bourne}, M.~A., \& {Smith}, M.~C. 2019, \mnras,
  484, 2047

\bibitem[{{Koudmani} {et~al.}(2022){Koudmani}, {Sijacki}, \&
  {Smith}}]{Koudmani2022}
{Koudmani}, S., {Sijacki}, D., \& {Smith}, M.~C. 2022, \mnras, 516, 2112

\bibitem[{{Koudmani} {et~al.}(2024){Koudmani}, {Somerville}, {Sijacki},
  {Bourne}, {Jiang}, \& {Profit}}]{Koudmani2024}
{Koudmani}, S., {Somerville}, R.~S., {Sijacki}, D., {et~al.} 2024, \mnras, 532,
  60

\bibitem[{{Kroupa}(2001)}]{Kroupa2001}
{Kroupa}, P. 2001, \mnras, 322, 231

\bibitem[{{Lah{\'e}n} {et~al.}(2020){Lah{\'e}n}, {Naab}, {Johansson},
  {Elmegreen}, {Hu}, {Walch}, {Steinwandel}, \& {Moster}}]{Lahen2020}
{Lah{\'e}n}, N., {Naab}, T., {Johansson}, P.~H., {et~al.} 2020, \apj, 891, 2

\bibitem[{{Larson} {et~al.}(2023){Larson}, {Finkelstein}, {Kocevski},
  {Hutchison}, {Trump}, {Arrabal Haro}, {Bromm}, {Cleri}, {Dickinson},
  {Fujimoto}, {Kartaltepe}, {Koekemoer}, {Papovich}, {Pirzkal}, {Tacchella},
  {Zavala}, {Bagley}, {Behroozi}, {Champagne}, {Cole}, {Jung}, {Morales},
  {Yang}, {Zhang}, {Zitrin}, {Amor{\'\i}n}, {Burgarella}, {Casey}, {Ch{\'a}vez
  Ortiz}, {Cox}, {Chworowsky}, {Fontana}, {Gawiser}, {Grazian}, {Grogin},
  {Harish}, {Hathi}, {Hirschmann}, {Holwerda}, {Juneau}, {Leung}, {Lucas},
  {McGrath}, {P{\'e}rez-Gonz{\'a}lez}, {Rigby}, {Seill{\'e}}, {Simons}, {de La
  Vega}, {Weiner}, {Wilkins}, {Yung}, \& {Ceers Team}}]{Larsson2023}
{Larson}, R.~L., {Finkelstein}, S.~L., {Kocevski}, D.~D., {et~al.} 2023, \apjl,
  953, L29

\bibitem[{{Lin} {et~al.}(2024){Lin}, {Wang}, {Fan}, {Cai}, {Champagne}, {Sun},
  {Volonteri}, {Yang}, {Hennawi}, {Ba{\~n}ados}, {Barth}, {Eilers}, {Farina},
  {Liu}, {Jin}, {Jun}, {Lupi}, {Kakiichi}, {Mazzucchelli}, {Onoue}, {Pan},
  {Pizzati}, {Rojas-Ruiz}, {Schindler}, {Trakhtenbrot}, {Shen}, {Trebitsch},
  {Zhuang}, {Endsley}, {Meyer}, {Li}, {Li}, {Pudoka}, {Tee}, {Wu}, \&
  {Zhang}}]{Lin2024}
{Lin}, X., {Wang}, F., {Fan}, X., {et~al.} 2024, \apj, 974, 147

\bibitem[{{Lupi} {et~al.}(2024){Lupi}, {Quadri}, {Volonteri}, {Colpi}, \&
  {Regan}}]{Lupi2024}
{Lupi}, A., {Quadri}, G., {Volonteri}, M., {Colpi}, M., \& {Regan}, J.~A. 2024,
  \aap, 686, A256

\bibitem[{{Lynden-Bell}(1969)}]{LyndenBell1969}
{Lynden-Bell}, D. 1969, \nat, 223, 690

\bibitem[{{Maeder}(2009)}]{Maeder2009}
{Maeder}, A. 2009, {Physics, Formation and Evolution of Rotating Stars}

\bibitem[{{Magorrian} {et~al.}(1998){Magorrian}, {Tremaine}, {Richstone},
  {Bender}, {Bower}, {Dressler}, {Faber}, {Gebhardt}, {Green}, {Grillmair},
  {Kormendy}, \& {Lauer}}]{Magorrian1998}
{Magorrian}, J., {Tremaine}, S., {Richstone}, D., {et~al.} 1998, \aj, 115, 2285

\bibitem[{{Maiolino} {et~al.}(2024){Maiolino}, {Scholtz}, {Curtis-Lake},
  {Carniani}, {Baker}, {de Graaff}, {Tacchella}, {{\"U}bler}, {D'Eugenio},
  {Witstok}, {Curti}, {Arribas}, {Bunker}, {Charlot}, {Chevallard},
  {Eisenstein}, {Egami}, {Ji}, {Jones}, {Lyu}, {Rawle}, {Robertson},
  {Rujopakarn}, {Perna}, {Sun}, {Venturi}, {Williams}, \&
  {Willott}}]{Maiolino2024}
{Maiolino}, R., {Scholtz}, J., {Curtis-Lake}, E., {et~al.} 2024, \aap, 691,
  A145

\bibitem[{{Maoz} \& {Mannucci}(2012)}]{Maoz2012}
{Maoz}, D. \& {Mannucci}, F. 2012, \pasa, 29, 447

\bibitem[{{Martizzi} {et~al.}(2015){Martizzi}, {Faucher-Gigu{\`e}re}, \&
  {Quataert}}]{Martizzi2015}
{Martizzi}, D., {Faucher-Gigu{\`e}re}, C.-A., \& {Quataert}, E. 2015, \mnras,
  450, 504

\bibitem[{{Massonneau} {et~al.}(2023){Massonneau}, {Volonteri}, {Dubois}, \&
  {Beckmann}}]{Massonneau2023}
{Massonneau}, W., {Volonteri}, M., {Dubois}, Y., \& {Beckmann}, R.~S. 2023,
  \aap, 670, A180

\bibitem[{{Matsuoka} {et~al.}(2018){Matsuoka}, {Iwasawa}, {Onoue}, {Kashikawa},
  {Strauss}, {Lee}, {Imanishi}, {Nagao}, {Akiyama}, {Asami}, {Bosch},
  {Furusawa}, {Goto}, {Gunn}, {Harikane}, {Ikeda}, {Izumi}, {Kawaguchi},
  {Kato}, {Kikuta}, {Kohno}, {Komiyama}, {Lupton}, {Minezaki}, {Miyazaki},
  {Morokuma}, {Murayama}, {Niida}, {Nishizawa}, {Oguri}, {Ono}, {Ouchi},
  {Price}, {Sameshima}, {Schulze}, {Shirakata}, {Silverman}, {Sugiyama},
  {Tait}, {Takada}, {Takata}, {Tanaka}, {Tang}, {Toba}, {Utsumi}, {Wang}, \&
  {Yamashita}}]{Matsuoka2018}
{Matsuoka}, Y., {Iwasawa}, K., {Onoue}, M., {et~al.} 2018, \apjs, 237, 5

\bibitem[{{Matsuoka} {et~al.}(2016){Matsuoka}, {Onoue}, {Kashikawa}, {Iwasawa},
  {Strauss}, {Nagao}, {Imanishi}, {Niida}, {Toba}, {Akiyama}, {Asami}, {Bosch},
  {Foucaud}, {Furusawa}, {Goto}, {Gunn}, {Harikane}, {Ikeda}, {Kawaguchi},
  {Kikuta}, {Komiyama}, {Lupton}, {Minezaki}, {Miyazaki}, {Morokuma},
  {Murayama}, {Nishizawa}, {Ono}, {Ouchi}, {Price}, {Sameshima}, {Silverman},
  {Sugiyama}, {Tait}, {Takada}, {Takata}, {Tanaka}, {Tang}, \&
  {Utsumi}}]{Matsuoka2016}
{Matsuoka}, Y., {Onoue}, M., {Kashikawa}, N., {et~al.} 2016, \apj, 828, 26

\bibitem[{{Matsuoka} {et~al.}(2019){Matsuoka}, {Onoue}, {Kashikawa}, {Strauss},
  {Iwasawa}, {Lee}, {Imanishi}, {Nagao}, {Akiyama}, {Asami}, {Bosch},
  {Furusawa}, {Goto}, {Gunn}, {Harikane}, {Ikeda}, {Izumi}, {Kawaguchi},
  {Kato}, {Kikuta}, {Kohno}, {Komiyama}, {Koyama}, {Lupton}, {Minezaki},
  {Miyazaki}, {Murayama}, {Niida}, {Nishizawa}, {Noboriguchi}, {Oguri}, {Ono},
  {Ouchi}, {Price}, {Sameshima}, {Schulze}, {Shirakata}, {Silverman},
  {Sugiyama}, {Tait}, {Takada}, {Takata}, {Tanaka}, {Tang}, {Toba}, {Utsumi},
  {Wang}, \& {Yamashita}}]{Matsuoka2019}
{Matsuoka}, Y., {Onoue}, M., {Kashikawa}, N., {et~al.} 2019, \apjl, 872, L2

\bibitem[{{Matthee} {et~al.}(2024){Matthee}, {Naidu}, {Brammer}, {Chisholm},
  {Eilers}, {Goulding}, {Greene}, {Kashino}, {Labbe}, {Lilly}, {Mackenzie},
  {Oesch}, {Weibel}, {Wuyts}, {Xiao}, {Bordoloi}, {Bouwens}, {van Dokkum},
  {Illingworth}, {Kramarenko}, {Maseda}, {Mason}, {Meyer}, {Nelson}, {Reddy},
  {Shivaei}, {Simcoe}, \& {Yue}}]{Matthee2024}
{Matthee}, J., {Naidu}, R.~P., {Brammer}, G., {et~al.} 2024, \apj, 963, 129

\bibitem[{{Mayer} \& {Bonoli}(2019)}]{Mayer2019}
{Mayer}, L. \& {Bonoli}, S. 2019, Reports on Progress in Physics, 82, 016901

\bibitem[{{Mayer} {et~al.}(2024){Mayer}, {Capelo}, {Zwick}, \& {Di
  Matteo}}]{Mayer2024}
{Mayer}, L., {Capelo}, P.~R., {Zwick}, L., \& {Di Matteo}, T. 2024, \apj, 961,
  76

\bibitem[{{McAlpine} {et~al.}(2018){McAlpine}, {Bower}, {Rosario}, {Crain},
  {Schaye}, \& {Theuns}}]{McAlpine2018}
{McAlpine}, S., {Bower}, R.~G., {Rosario}, D.~J., {et~al.} 2018, \mnras, 481,
  3118

\bibitem[{{McClintock} \& {Remillard}(2006)}]{McClintock2006}
{McClintock}, J.~E. \& {Remillard}, R.~A. 2006, in Compact stellar X-ray
  sources, Vol.~39, 157--213

\bibitem[{{McConnell} \& {Ma}(2013)}]{McConnell2013}
{McConnell}, N.~J. \& {Ma}, C.-P. 2013, \apj, 764, 184

\bibitem[{{Mehta} {et~al.}(2024){Mehta}, {Regan}, \& {Prole}}]{Mehta2024}
{Mehta}, D., {Regan}, J.~A., \& {Prole}, L. 2024, The Open Journal of
  Astrophysics, 7, 107

\bibitem[{{Mezcua}(2019)}]{Mezcua2019a}
{Mezcua}, M. 2019, Nature Astronomy, 3, 6

\bibitem[{{Mezcua} {et~al.}(2018){Mezcua}, {Civano}, {Marchesi}, {Suh},
  {Fabbiano}, \& {Volonteri}}]{Mezcua2018}
{Mezcua}, M., {Civano}, F., {Marchesi}, S., {et~al.} 2018, \mnras, 478, 2576

\bibitem[{{Mezcua} \& {Dom{\'\i}nguez S{\'a}nchez}(2020)}]{Mezcua2020}
{Mezcua}, M. \& {Dom{\'\i}nguez S{\'a}nchez}, H. 2020, \apjl, 898, L30

\bibitem[{{Mezcua} {et~al.}(2023){Mezcua}, {Siudek}, {Suh}, {Valiante},
  {Spinoso}, \& {Bonoli}}]{Mezcua2023}
{Mezcua}, M., {Siudek}, M., {Suh}, H., {et~al.} 2023, \apjl, 943, L5

\bibitem[{{Mezcua} {et~al.}(2019){Mezcua}, {Suh}, \& {Civano}}]{Mezcua2019b}
{Mezcua}, M., {Suh}, H., \& {Civano}, F. 2019, \mnras, 488, 685

\bibitem[{{Moran} {et~al.}(2014){Moran}, {Shahinyan}, {Sugarman}, {V{\'e}lez},
  \& {Eracleous}}]{Moran2014}
{Moran}, E.~C., {Shahinyan}, K., {Sugarman}, H.~R., {V{\'e}lez}, D.~O., \&
  {Eracleous}, M. 2014, \aj, 148, 136

\bibitem[{{Morgan} {et~al.}(2010){Morgan}, {Kochanek}, {Morgan}, \&
  {Falco}}]{Morgan2010}
{Morgan}, C.~W., {Kochanek}, C.~S., {Morgan}, N.~D., \& {Falco}, E.~E. 2010,
  \apj, 712, 1129

\bibitem[{{Mortlock} {et~al.}(2009){Mortlock}, {Patel}, {Warren}, {Venemans},
  {McMahon}, {Hewett}, {Simpson}, {Sharp}, {Burningham}, {Dye}, {Ellis},
  {Gonzales-Solares}, \& {Hu{\'e}lamo}}]{Mortlock2009}
{Mortlock}, D.~J., {Patel}, M., {Warren}, S.~J., {et~al.} 2009, \aap, 505, 97

\bibitem[{{Mortlock} {et~al.}(2011){Mortlock}, {Warren}, {Venemans}, {Patel},
  {Hewett}, {McMahon}, {Simpson}, {Theuns}, {Gonz{\'a}les-Solares}, {Adamson},
  {Dye}, {Hambly}, {Hirst}, {Irwin}, {Kuiper}, {Lawrence}, \&
  {R{\"o}ttgering}}]{Mortlock2011}
{Mortlock}, D.~J., {Warren}, S.~J., {Venemans}, B.~P., {et~al.} 2011, \nat,
  474, 616

\bibitem[{{Moster} {et~al.}(2013){Moster}, {Naab}, \& {White}}]{Moster2013}
{Moster}, B.~P., {Naab}, T., \& {White}, S. D.~M. 2013, \mnras, 428, 3121

\bibitem[{{Moster} {et~al.}(2010){Moster}, {Somerville}, {Maulbetsch}, {van den
  Bosch}, {Macci{\`o}}, {Naab}, \& {Oser}}]{Moster2010}
{Moster}, B.~P., {Somerville}, R.~S., {Maulbetsch}, C., {et~al.} 2010, \apj,
  710, 903

\bibitem[{{Naab} \& {Ostriker}(2017)}]{Naab2017}
{Naab}, T. \& {Ostriker}, J.~P. 2017, \araa, 55, 59

\bibitem[{{Narayan} \& {Yi}(1995)}]{Narayan1995}
{Narayan}, R. \& {Yi}, I. 1995, \apj, 452, 710

\bibitem[{{Navarro} {et~al.}(1996){Navarro}, {Frenk}, \& {White}}]{Navarro1996}
{Navarro}, J.~F., {Frenk}, C.~S., \& {White}, S. D.~M. 1996, \apj, 462, 563

\bibitem[{{Negri} \& {Volonteri}(2017)}]{Negri2017}
{Negri}, A. \& {Volonteri}, M. 2017, \mnras, 467, 3475

\bibitem[{{Nelson} {et~al.}(2019){Nelson}, {Pillepich}, {Springel}, {Pakmor},
  {Weinberger}, {Genel}, {Torrey}, {Vogelsberger}, {Marinacci}, \&
  {Hernquist}}]{Nelson2019}
{Nelson}, D., {Pillepich}, A., {Springel}, V., {et~al.} 2019, \mnras, 490, 3234

\bibitem[{{Nelson} \& {Langer}(1997)}]{Nelson1997}
{Nelson}, R.~P. \& {Langer}, W.~D. 1997, \apj, 482, 796

\bibitem[{{Nguyen} {et~al.}(2019){Nguyen}, {Seth}, {Neumayer}, {Iguchi},
  {Cappellari}, {Strader}, {Chomiuk}, {Tremou}, {Pacucci}, {Nakanishi},
  {Bahramian}, {Nguyen}, {den Brok}, {Ahn}, {Voggel}, {Kacharov}, {Tsukui},
  {Ly}, {Dumont}, \& {Pechetti}}]{Nguyen2019}
{Nguyen}, D.~D., {Seth}, A.~C., {Neumayer}, N., {et~al.} 2019, \apj, 872, 104

\bibitem[{{Novak} {et~al.}(2011){Novak}, {Ostriker}, \& {Ciotti}}]{Novak2011}
{Novak}, G.~S., {Ostriker}, J.~P., \& {Ciotti}, L. 2011, \apj, 737, 26

\bibitem[{{Onoue} {et~al.}(2023){Onoue}, {Inayoshi}, {Ding}, {Li}, {Li},
  {Molina}, {Inoue}, {Jiang}, \& {Ho}}]{Onoue2023}
{Onoue}, M., {Inayoshi}, K., {Ding}, X., {et~al.} 2023, \apjl, 942, L17

\bibitem[{{Pardo} {et~al.}(2016){Pardo}, {Goulding}, {Greene}, {Somerville},
  {Gallo}, {Hickox}, {Miller}, {Reines}, \& {Silverman}}]{Pardo2016}
{Pardo}, K., {Goulding}, A.~D., {Greene}, J.~E., {et~al.} 2016, \apj, 831, 203

\bibitem[{{Partmann} {et~al.}(2025){Partmann}, {Naab}, {Lah{\'e}n}, {Rantala},
  {Hirschmann}, {Hislop}, {Petersson}, \& {Johansson}}]{Partmann2025}
{Partmann}, C., {Naab}, T., {Lah{\'e}n}, N., {et~al.} 2025, \mnras, 537, 956

\bibitem[{{Peter} {et~al.}(2023){Peter}, {Klessen}, {Kanschat}, {Glover}, \&
  {Bastian}}]{Peter2023}
{Peter}, T., {Klessen}, R.~S., {Kanschat}, G., {Glover}, S. C.~O., \&
  {Bastian}, P. 2023, \mnras, 519, 4263

\bibitem[{{Prieto} {et~al.}(2017){Prieto}, {Escala}, {Volonteri}, \&
  {Dubois}}]{Prieto2017}
{Prieto}, J., {Escala}, A., {Volonteri}, M., \& {Dubois}, Y. 2017, \apj, 836,
  216

\bibitem[{{Reed} {et~al.}(2019){Reed}, {Banerji}, {Becker}, {Hewett},
  {Martini}, {McMahon}, {Pons}, {Rauch}, {Abbott}, {Allam}, {Annis}, {Avila},
  {Bertin}, {Brooks}, {Buckley-Geer}, {Carnero Rosell}, {Carrasco Kind},
  {Carretero}, {Castander}, {Cunha}, {D'Andrea}, {da Costa}, {De Vicente},
  {Desai}, {Diehl}, {Doel}, {Evrard}, {Flaugher}, {Frieman},
  {Garc{\'\i}a-Bellido}, {Gaztanaga}, {Gruen}, {Gschwend}, {Gutierrez},
  {Hollowood}, {Honscheid}, {Hoyle}, {James}, {Kuehn}, {Lahav}, {Lima}, {Maia},
  {Marshall}, {Miquel}, {Ogando}, {Plazas}, {Roodman}, {Sanchez}, {Scarpine},
  {Schubnell}, {Serrano}, {Sevilla-Noarbe}, {Smith}, {Smith}, {Sobreira},
  {Suchyta}, {Swanson}, {Tarle}, {Thomas}, {Tucker}, \& {Vikram}}]{Reed2019}
{Reed}, S.~L., {Banerji}, M., {Becker}, G.~D., {et~al.} 2019, \mnras, 487, 1874

\bibitem[{{Reed} {et~al.}(2017){Reed}, {McMahon}, {Martini}, {Banerji},
  {Auger}, {Hewett}, {Koposov}, {Gibbons}, {Gonzalez-Solares}, {Ostrovski},
  {Tie}, {Abdalla}, {Allam}, {Benoit-L{\'e}vy}, {Bertin}, {Brooks},
  {Buckley-Geer}, {Burke}, {Carnero Rosell}, {Carrasco Kind}, {Carretero}, {da
  Costa}, {DePoy}, {Desai}, {Diehl}, {Doel}, {Evrard}, {Finley}, {Flaugher},
  {Fosalba}, {Frieman}, {Garc{\'\i}a-Bellido}, {Gaztanaga}, {Goldstein},
  {Gruen}, {Gruendl}, {Gutierrez}, {James}, {Kuehn}, {Kuropatkin}, {Lahav},
  {Lima}, {Maia}, {Marshall}, {Melchior}, {Miller}, {Miquel}, {Nord}, {Ogando},
  {Plazas}, {Romer}, {Sanchez}, {Scarpine}, {Schubnell}, {Sevilla-Noarbe},
  {Smith}, {Sobreira}, {Suchyta}, {Swanson}, {Tarle}, {Tucker}, {Walker}, \&
  {Wester}}]{Reed2017}
{Reed}, S.~L., {McMahon}, R.~G., {Martini}, P., {et~al.} 2017, \mnras, 468,
  4702

\bibitem[{{Reines} {et~al.}(2020){Reines}, {Condon}, {Darling}, \&
  {Greene}}]{Reines2020}
{Reines}, A.~E., {Condon}, J.~J., {Darling}, J., \& {Greene}, J.~E. 2020, \apj,
  888, 36

\bibitem[{{Reines} {et~al.}(2013){Reines}, {Greene}, \& {Geha}}]{Reines2013}
{Reines}, A.~E., {Greene}, J.~E., \& {Geha}, M. 2013, \apj, 775, 116

\bibitem[{{Reines} {et~al.}(2011){Reines}, {Sivakoff}, {Johnson}, \&
  {Brogan}}]{Reines2011}
{Reines}, A.~E., {Sivakoff}, G.~R., {Johnson}, K.~E., \& {Brogan}, C.~L. 2011,
  \nat, 470, 66

\bibitem[{{Reines} \& {Volonteri}(2015)}]{Reines2015}
{Reines}, A.~E. \& {Volonteri}, M. 2015, \apj, 813, 82

\bibitem[{{Rosas-Guevara} {et~al.}(2015){Rosas-Guevara}, {Bower}, {Schaye},
  {Furlong}, {Frenk}, {Booth}, {Crain}, {Dalla Vecchia}, {Schaller}, \&
  {Theuns}}]{RosasGuevara2015}
{Rosas-Guevara}, Y.~M., {Bower}, R.~G., {Schaye}, J., {et~al.} 2015, \mnras,
  454, 1038

\bibitem[{{Rosdahl} {et~al.}(2015){Rosdahl}, {Schaye}, {Teyssier}, \&
  {Agertz}}]{Rosdahl2015}
{Rosdahl}, J., {Schaye}, J., {Teyssier}, R., \& {Agertz}, O. 2015, \mnras, 451,
  34

\bibitem[{{Sacchi} {et~al.}(2024){Sacchi}, {Bogdan}, {Chadayammuri}, \&
  {Ricarte}}]{Sacchi2024}
{Sacchi}, A., {Bogdan}, A., {Chadayammuri}, U., \& {Ricarte}, A. 2024, arXiv
  e-prints, arXiv:2406.01707

\bibitem[{{Salpeter}(1964)}]{Salpeter1964}
{Salpeter}, E.~E. 1964, \apj, 140, 796

\bibitem[{{Schaye} {et~al.}(2015){Schaye}, {Crain}, {Bower}, {Furlong},
  {Schaller}, {Theuns}, {Dalla Vecchia}, {Frenk}, {McCarthy}, {Helly},
  {Jenkins}, {Rosas-Guevara}, {White}, {Baes}, {Booth}, {Camps}, {Navarro},
  {Qu}, {Rahmati}, {Sawala}, {Thomas}, \& {Trayford}}]{Schaye2015}
{Schaye}, J., {Crain}, R.~A., {Bower}, R.~G., {et~al.} 2015, \mnras, 446, 521

\bibitem[{{Schutte} {et~al.}(2019){Schutte}, {Reines}, \&
  {Greene}}]{Schutte2019}
{Schutte}, Z., {Reines}, A.~E., \& {Greene}, J.~E. 2019, \apj, 887, 245

\bibitem[{{Shakura} \& {Sunyaev}(1973)}]{Shakura1973}
{Shakura}, N.~I. \& {Sunyaev}, R.~A. 1973, \aap, 24, 337

\bibitem[{{Sharma} {et~al.}(2020){Sharma}, {Brooks}, {Somerville}, {Tremmel},
  {Bellovary}, {Wright}, \& {Quinn}}]{Sharma2020}
{Sharma}, R.~S., {Brooks}, A.~M., {Somerville}, R.~S., {et~al.} 2020, \apj,
  897, 103

\bibitem[{{Sharma} {et~al.}(2022){Sharma}, {Brooks}, {Tremmel}, {Bellovary},
  {Ricarte}, \& {Quinn}}]{Sharma2022}
{Sharma}, R.~S., {Brooks}, A.~M., {Tremmel}, M., {et~al.} 2022, \apj, 936, 82

\bibitem[{{Shi} {et~al.}(2023){Shi}, {Kremer}, {Grudi{\'c}},
  {Gerling-Dunsmore}, \& {Hopkins}}]{Shi2023}
{Shi}, Y., {Kremer}, K., {Grudi{\'c}}, M.~Y., {Gerling-Dunsmore}, H.~J., \&
  {Hopkins}, P.~F. 2023, \mnras, 518, 3606

\bibitem[{{Sijacki} {et~al.}(2015){Sijacki}, {Vogelsberger}, {Genel},
  {Springel}, {Torrey}, {Snyder}, {Nelson}, \& {Hernquist}}]{Sijacki2015}
{Sijacki}, D., {Vogelsberger}, M., {Genel}, S., {et~al.} 2015, \mnras, 452, 575

\bibitem[{{Silk} \& {Rees}(1998)}]{Silk1998}
{Silk}, J. \& {Rees}, M.~J. 1998, \aap, 331, L1

\bibitem[{{Sivasankaran} {et~al.}(2022){Sivasankaran}, {Blecha}, {Torrey},
  {Kelley}, {Bhowmick}, {Vogelsberger}, {Losacco}, {Weinberger}, {Hernquist},
  {Marinacci}, {Sales}, \& {Qi}}]{Sivasankaran2022}
{Sivasankaran}, A., {Blecha}, L., {Torrey}, P., {et~al.} 2022, \mnras, 517,
  4752

\bibitem[{{Smith} {et~al.}(2021){Smith}, {Bryan}, {Somerville}, {Hu},
  {Teyssier}, {Burkhart}, \& {Hernquist}}]{Smith2021}
{Smith}, M.~C., {Bryan}, G.~L., {Somerville}, R.~S., {et~al.} 2021, \mnras,
  506, 3882

\bibitem[{{Smith} {et~al.}(2014){Smith}, {Glover}, {Clark}, {Klessen}, \&
  {Springel}}]{Smith2014}
{Smith}, R.~J., {Glover}, S. C.~O., {Clark}, P.~C., {Klessen}, R.~S., \&
  {Springel}, V. 2014, \mnras, 441, 1628

\bibitem[{{Somerville} \& {Dav{\'e}}(2015)}]{Somerville2015}
{Somerville}, R.~S. \& {Dav{\'e}}, R. 2015, \araa, 53, 51

\bibitem[{{Sormani} {et~al.}(2020){Sormani}, {Tress}, {Glover}, {Klessen},
  {Battersby}, {Clark}, {Hatchfield}, \& {Smith}}]{Sormani2020}
{Sormani}, M.~C., {Tress}, R.~G., {Glover}, S. C.~O., {et~al.} 2020, \mnras,
  497, 5024

\bibitem[{{Sormani} {et~al.}(2017){Sormani}, {Tre{\ss}}, {Klessen}, \&
  {Glover}}]{Sormani2017}
{Sormani}, M.~C., {Tre{\ss}}, R.~G., {Klessen}, R.~S., \& {Glover}, S. C.~O.
  2017, \mnras, 466, 407

\bibitem[{{Sormani} {et~al.}(2018){Sormani}, {Tre{\ss}}, {Ridley}, {Glover},
  {Klessen}, {Binney}, {Magorrian}, \& {Smith}}]{Sormani2018}
{Sormani}, M.~C., {Tre{\ss}}, R.~G., {Ridley}, M., {et~al.} 2018, \mnras, 475,
  2383

\bibitem[{{Springel}(2005)}]{Springel2005b}
{Springel}, V. 2005, \mnras, 364, 1105

\bibitem[{{Springel}(2010)}]{Springel2010}
{Springel}, V. 2010, \mnras, 401, 791

\bibitem[{{Springel} {et~al.}(2005){Springel}, {Di Matteo}, \&
  {Hernquist}}]{Springel2005a}
{Springel}, V., {Di Matteo}, T., \& {Hernquist}, L. 2005, \mnras, 361, 776

\bibitem[{{Springel} \& {Hernquist}(2003)}]{Springel2003}
{Springel}, V. \& {Hernquist}, L. 2003, \mnras, 339, 289

\bibitem[{{Steinborn} {et~al.}(2015){Steinborn}, {Dolag}, {Hirschmann},
  {Prieto}, \& {Remus}}]{Steinborn2015}
{Steinborn}, L.~K., {Dolag}, K., {Hirschmann}, M., {Prieto}, M.~A., \& {Remus},
  R.-S. 2015, \mnras, 448, 1504

\bibitem[{{Steinwandel} {et~al.}(2024){Steinwandel}, {Kim}, {Bryan},
  {Ostriker}, {Somerville}, \& {Fielding}}]{Steinwandel2024}
{Steinwandel}, U.~P., {Kim}, C.-G., {Bryan}, G.~L., {et~al.} 2024, \apj, 960,
  100

\bibitem[{{Tart{\.{e}}nas} \& {Zubovas}(2022)}]{Tartenas2022}
{Tart{\.{e}}nas}, M. \& {Zubovas}, K. 2022, \mnras, 516, 2522

\bibitem[{{Taylor} {et~al.}(2024){Taylor}, {Finkelstein}, {Kocevski}, {Jeon},
  {Bromm}, {Amorin}, {Arrabal Haro}, {Backhaus}, {Bagley}, {Ba{\~n}ados},
  {Bhatawdekar}, {Brooks}, {Calabro}, {Chavez Ortiz}, {Cheng}, {Cleri}, {Cole},
  {Davis}, {Dickinson}, {Donnan}, {Dunlop}, {Ellis}, {Fernandez}, {Fontana},
  {Fujimoto}, {Giavalisco}, {Grazian}, {Guo}, {Hathi}, {Holwerda},
  {Hirschmann}, {Inayoshi}, {Kartaltepe}, {Khusanova}, {Koekemoer}, {Kokorev},
  {Larson}, {Leung}, {Lucas}, {McLeod}, {Napolitano}, {Onoue}, {Pacucci},
  {Papovich}, {P{\'e}rez-Gonz{\'a}lez}, {Pirzkal}, {Somerville}, {Trump},
  {Wilkins}, {Yung}, \& {Zhang}}]{Taylor2024}
{Taylor}, A.~J., {Finkelstein}, S.~L., {Kocevski}, D.~D., {et~al.} 2024, arXiv
  e-prints, arXiv:2409.06772

\bibitem[{{Trebitsch} {et~al.}(2018){Trebitsch}, {Volonteri}, {Dubois}, \&
  {Madau}}]{Trebitsch2018}
{Trebitsch}, M., {Volonteri}, M., {Dubois}, Y., \& {Madau}, P. 2018, \mnras,
  478, 5607

\bibitem[{{Tremaine} {et~al.}(2002){Tremaine}, {Gebhardt}, {Bender}, {Bower},
  {Dressler}, {Faber}, {Filippenko}, {Green}, {Grillmair}, {Ho}, {Kormendy},
  {Lauer}, {Magorrian}, {Pinkney}, \& {Richstone}}]{Tremaine2002}
{Tremaine}, S., {Gebhardt}, K., {Bender}, R., {et~al.} 2002, \apj, 574, 740

\bibitem[{{Tress} {et~al.}(2020){Tress}, {Smith}, {Sormani}, {Glover},
  {Klessen}, {Mac Low}, \& {Clark}}]{Tress2020a}
{Tress}, R.~G., {Smith}, R.~J., {Sormani}, M.~C., {et~al.} 2020, \mnras, 492,
  2973

\bibitem[{{Tress} {et~al.}(2024){Tress}, {Sormani}, {Girichidis}, {Glover},
  {Klessen}, {Smith}, {Sobacchi}, {Armillotta}, {Barnes}, {Battersby}, {Bogue},
  {Brucy}, {Colzi}, {Federrath}, {Garc{\'\i}a}, {Ginsburg}, {G{\"o}ller},
  {Hatchfield}, {Henkel}, {Hennebelle}, {Henshaw}, {Hirschmann}, {Hu},
  {Kauffmann}, {Kruijssen}, {Lazarian}, {Lipman}, {Longmore}, {Morris},
  {Nogueras-Lara}, {Petkova}, {Pillai}, {Rivilla}, {S{\'a}nchez-Monge},
  {Soler}, {Whitworth}, \& {Zhang}}]{Tress2024}
{Tress}, R.~G., {Sormani}, M.~C., {Girichidis}, P., {et~al.} 2024, \aap, 691,
  A303

\bibitem[{{Truelove} {et~al.}(1997){Truelove}, {Klein}, {McKee}, {Holliman},
  {Howell}, \& {Greenough}}]{Truelove1997}
{Truelove}, J.~K., {Klein}, R.~I., {McKee}, C.~F., {et~al.} 1997, \apjl, 489,
  L179

\bibitem[{{{\"U}bler} {et~al.}(2023){{\"U}bler}, {Maiolino}, {Curtis-Lake},
  {P{\'e}rez-Gonz{\'a}lez}, {Curti}, {Perna}, {Arribas}, {Charlot}, {Marshall},
  {D'Eugenio}, {Scholtz}, {Bunker}, {Carniani}, {Ferruit}, {Jakobsen}, {Rix},
  {Rodr{\'\i}guez Del Pino}, {Willott}, {Boeker}, {Cresci}, {Jones}, {Kumari},
  \& {Rawle}}]{Ubler2023}
{{\"U}bler}, H., {Maiolino}, R., {Curtis-Lake}, E., {et~al.} 2023, \aap, 677,
  A145

\bibitem[{{Volonteri} {et~al.}(2016){Volonteri}, {Dubois}, {Pichon}, \&
  {Devriendt}}]{Volonteri2016}
{Volonteri}, M., {Dubois}, Y., {Pichon}, C., \& {Devriendt}, J. 2016, \mnras,
  460, 2979

\bibitem[{{Wang} {et~al.}(2021){Wang}, {Yang}, {Fan}, {Hennawi}, {Barth},
  {Banados}, {Bian}, {Boutsia}, {Connor}, {Davies}, {Decarli}, {Eilers},
  {Farina}, {Green}, {Jiang}, {Li}, {Mazzucchelli}, {Nanni}, {Schindler},
  {Venemans}, {Walter}, {Wu}, \& {Yue}}]{Wang2021}
{Wang}, F., {Yang}, J., {Fan}, X., {et~al.} 2021, \apjl, 907, L1

\bibitem[{{Weinberger} {et~al.}(2025){Weinberger}, {Bhowmick}, {Blecha},
  {Bryan}, {Buchner}, {Hernquist}, {Hlavacek-Larrondo}, \&
  {Springel}}]{Weinberger2025}
{Weinberger}, R., {Bhowmick}, A., {Blecha}, L., {et~al.} 2025, arXiv e-prints,
  arXiv:2502.13241

\bibitem[{{Weinberger} {et~al.}(2018){Weinberger}, {Springel}, {Pakmor},
  {Nelson}, {Genel}, {Pillepich}, {Vogelsberger}, {Marinacci}, {Naiman},
  {Torrey}, \& {Hernquist}}]{Weinberger2018}
{Weinberger}, R., {Springel}, V., {Pakmor}, R., {et~al.} 2018, \mnras, 479,
  4056

\bibitem[{{Whitworth} {et~al.}(2023){Whitworth}, {Smith}, {Klessen}, {Mac Low},
  {Glover}, {Tress}, {Pakmor}, \& {Soler}}]{Whitworth2023}
{Whitworth}, D.~J., {Smith}, R.~J., {Klessen}, R.~S., {et~al.} 2023, \mnras,
  520, 89

\bibitem[{{Whitworth} {et~al.}(2022){Whitworth}, {Smith}, {Tress}, {Kay},
  {Glover}, {Sormani}, \& {Klessen}}]{Whitworth2022}
{Whitworth}, D.~J., {Smith}, R.~J., {Tress}, R., {et~al.} 2022, \mnras, 510,
  4146

\bibitem[{{Willott} {et~al.}(2009){Willott}, {Delorme}, {Reyl{\'e}}, {Albert},
  {Bergeron}, {Crampton}, {Delfosse}, {Forveille}, {Hutchings}, {McLure},
  {Omont}, \& {Schade}}]{Willott2009}
{Willott}, C.~J., {Delorme}, P., {Reyl{\'e}}, C., {et~al.} 2009, \aj, 137, 3541

\bibitem[{{Willott} {et~al.}(2010){Willott}, {Delorme}, {Reyl{\'e}}, {Albert},
  {Bergeron}, {Crampton}, {Delfosse}, {Forveille}, {Hutchings}, {McLure},
  {Omont}, \& {Schade}}]{Willott2010}
{Willott}, C.~J., {Delorme}, P., {Reyl{\'e}}, C., {et~al.} 2010, \aj, 139, 906

\bibitem[{{Yang} {et~al.}(2020){Yang}, {Wang}, {Fan}, {Hennawi}, {Davies},
  {Yue}, {Banados}, {Wu}, {Venemans}, {Barth}, {Bian}, {Boutsia}, {Decarli},
  {Farina}, {Green}, {Jiang}, {Li}, {Mazzucchelli}, \& {Walter}}]{Yang2020}
{Yang}, J., {Wang}, F., {Fan}, X., {et~al.} 2020, \apjl, 897, L14

\bibitem[{{Yuan} {et~al.}(2018){Yuan}, {Yoon}, {Li}, {Gan}, {Ho}, \&
  {Guo}}]{Yuan2018}
{Yuan}, F., {Yoon}, D., {Li}, Y.-P., {et~al.} 2018, \apj, 857, 121

\bibitem[{{Zwick} {et~al.}(2023){Zwick}, {Mayer}, {Haemmerl{\'e}}, \&
  {Klessen}}]{Zwick2023}
{Zwick}, L., {Mayer}, L., {Haemmerl{\'e}}, L., \& {Klessen}, R.~S. 2023,
  \mnras, 518, 2076

\end{thebibliography}
